\RequirePackage{fix-cm}
\documentclass[smallextended]{svjour3}       

\smartqed 
\usepackage{graphicx}
\usepackage{cite}
\usepackage{fixltx2e}
\usepackage{amsmath, amsfonts}
\usepackage{algorithm}
\usepackage{algorithmic}
\usepackage{amsmath,amsfonts}
\usepackage{amssymb}

\usepackage{array}
\usepackage{balance}
 \usepackage{booktabs}
\usepackage[skip=1pt,labelfont=bf]{caption}
 \usepackage{calligra}
\usepackage{color}
\usepackage{colortbl}
\usepackage{courier}
\usepackage{csvsimple}
\usepackage{enumitem}
\usepackage{fancybox}
\usepackage{fontenc}
\usepackage{listings}
\usepackage{longtable}
\usepackage{lscape}
\usepackage{makecell}
\usepackage{moreverb}
\usepackage{multicol}
\usepackage{multirow}
\usepackage{pifont}
\usepackage{rotating}
\usepackage{setspace}
\usepackage{subfigure}
\usepackage[most]{tcolorbox}
\usepackage{threeparttable}
\usepackage{tikz}
\usepackage[normalem]{ulem}
\usepackage{url}
\usepackage{soul}
\usepackage{xspace}
\usepackage{xcolor}
\usepackage{lineno}
\usepackage{pifont}
\usepackage{natbib}
\usepackage{etoolbox}
\usepackage{bbding}
\usepackage{fontawesome5} 
 \usepackage[most]{tcolorbox}
 \usepackage[misc]{ifsym}

\usepackage{booktabs}  

\usepackage[hyperfigures,breaklinks,colorlinks,linkcolor=blue,citecolor=blue,urlcolor=blue]{hyperref}

\makeatletter

\long\def\figwindownonum[#1,#2,#3,#4] {
	\begin{window}[#1,#2,{#3},{\centering#4\par}] }
	\def\endfigwindownonum{\end{window}}

\patchcmd{\NAT@citex}
  {\@citea\NAT@hyper@{%
     \NAT@nmfmt{\NAT@nm}%
     \hyper@natlinkbreak{\NAT@aysep\NAT@spacechar}{\@citeb\@extra@b@citeb}%
     \NAT@date}}
  {\@citea\NAT@nmfmt{\NAT@nm}%
   \NAT@aysep\NAT@spacechar\NAT@hyper@{\NAT@date}}{}{}

\patchcmd{\NAT@citex}
  {\@citea\NAT@hyper@{%
     \NAT@nmfmt{\NAT@nm}%
     \hyper@natlinkbreak{\NAT@spacechar\NAT@@open\if*#1*\else#1\NAT@spacechar\fi}%
       {\@citeb\@extra@b@citeb}%
     \NAT@date}}
  {\@citea\NAT@nmfmt{\NAT@nm}%
   \NAT@spacechar\NAT@@open\if*#1*\else#1\NAT@spacechar\fi\NAT@hyper@{\NAT@date}}
  {}{}

\makeatother

\lstdefinestyle{test-smell-listing}{
    language=Java,
    basicstyle=\scriptsize\ttfamily, 
    breakatwhitespace=false,
    breaklines=true,
    captionpos=b, 
    frame=single, 
    framesep=2pt,
    framerule=0.5pt, 
    aboveskip=0pt,
    belowskip=0pt,
    xleftmargin=10pt, 
    xrightmargin=10pt,
    numbers=left, 
    numberstyle=\tiny\color{gray}, 
    stepnumber=1,
    numbersep=5pt, 
    showstringspaces=false, 
    keywordstyle=\color{blue}\bfseries, 
    commentstyle=\color{green}, 
    stringstyle=\color{red}, 
    morecomment=[s][\color{magenta}]{/**}{*/}, 
    morecomment=[s][\color{magenta}]{/*}{*/}, 
    emph={public,private,protected,static,void,int,double,float,char,boolean}, 
    emphstyle=\color{blue}\bfseries, 
}

\definecolor{darkpastelred}{rgb}{0.76, 0.23, 0.13}
\definecolor{ao(english)}{rgb}{0.0, 0.5, 0.0}
\definecolor{darkpastelred}{rgb}{0.76, 0.23, 0.13}
\definecolor{ao(english)}{rgb}{0.0, 0.5, 0.0}
\definecolor{yellow}{RGB}{255,255,153}
\definecolor{grey}{RGB}{224,224,224}

\newboolean{showcomments}
\setboolean{showcomments}{true}
\ifthenelse{\boolean{showcomments}}
 { \newcommand{\mynote}[2]{
      \fbox{\bfseries\sffamily\scriptsize#1}
        {\small$\blacktriangleright$\textsf{\emph{#2}}$\blacktriangleleft$}}}
        { \newcommand{\mynote}[2]{}}
        
\definecolor{DarkOrange}{rgb}{0.8,0.3,0.0}
\definecolor{DarkCyan}{rgb}{0.0, 0.55, 0.55}
\definecolor{DarkCyel}{rgb}{1.0, 0.49, 0.0}
\definecolor{yellow-green}{rgb}{0.6, 0.8, 0.2}

\definecolor{myblue}{RGB}{0, 0, 255}

\newcolumntype{?}{!{\vrule width 1pt}}

\newcommand{\highlight}[1]{%
  \colorbox{gray!20}{\parbox{0.98\linewidth}{#1}}%
}

\newcommand{\quotes}[1]{``#1''}

\usepackage{ulem} 

\newcolumntype{b}{>{\color{blue}}r}  
\usepackage{cancel}

\begin{document}

\title{Prompt Engineering in LLMs for Automated Unit Test Generation: A Large-Scale Study}

\author{{Wendkûuni C. Ouédraogo} \textsuperscript{1}        \and \\
        {Abdoul Kader Kaboré} \textsuperscript{1}                  \and 
        {Yinghua Li} \textsuperscript{1}                    \and \\
        {Haoye Tian} \textsuperscript{3}                    \and 
        {Anil Koyuncu} \textsuperscript{4}                  \and 
        {Jacques Klein} \textsuperscript{1}                 \and 
        {David Lo} \textsuperscript{2}                      \and \\
        {Tegawend\'e F. Bissyand\'e} \textsuperscript{1}
}        

\institute{%
    \begin{itemize}
    \item[] {Wendkûuni C. Ouédraogo} \\
              \email{wendkuuni.ouedraogo@uni.lu} \\
      \item[] {Abdoul Kader Kaboré} \\
              \email{abdoulkader.kabore@uni.lu} \\
      \item[\Letter] {Yinghua Li} \\
              \email{yinghua.li@uni.lu} \\
      \item[] {Haoye Tian} \\
              \email{haoye.tian@aalto.fi} \\
      \item[] {Anil Koyuncu} \\
              \email{anil.koyuncu@cs.bilkent.edu.tr} \\
      \item[] {Jacques Klein} \\
              \email{jacques.klein@uni.lu} \\
      \item[] {David Lo} \\
              \email{davidlo@smu.edu.sg} \\
      \item[] {Tegawend\'e F. Bissyand\'e} \\
              \email{tegawende.bissyande@uni.lu} \\
      \at
      \item[\textsuperscript{1}] SnT Centre, University of Luxembourg
      \item[\textsuperscript{2}] Singapore, Singapore Management University
      \item[\textsuperscript{3}] Finland, Aalto University
      \item[\textsuperscript{4}] Turkey, Bilkent University
    \end{itemize}
}

\date{Received: date / Accepted: date}

\maketitle
\begin{abstract}

Unit testing is essential for software reliability, yet manual test creation is time-consuming and often neglected. Although search-based software testing improves efficiency, it produces tests with poor readability and maintainability. Although LLMs show promise for test generation, existing research lacks comprehensive evaluation across execution-driven assessment, reasoning-based prompting, and real-world testing scenarios.

This study presents the first large-scale empirical evaluation of LLM-generated unit tests at the \emph{full class level}, systematically analyzing four state-of-the-art models (GPT-3.5, GPT-4, Mistral 7B, and Mixtral 8x7B) against EvoSuite across 216,300 \emph{generated} test cases targeting Defects4J, SF110, and CMD (a dataset mitigating LLM training data leakage).

We evaluate five prompting techniques—Zero-Shot Learning (ZSL), Few-Shot Learning (FSL), Chain-of-Thought (CoT), Tree-of-Thought (ToT), and Guided Tree-of-Thought (GToT)—assessing syntactic correctness, compilability, hallucination-driven failures, readability, code coverage metrics, and test smells.

Reasoning-based prompting particularly GToT significantly enhances test reliability, compilability, and structural adherence in general-purpose LLMs. However, hallucination-driven failures remain a persistent challenge, manifesting as non-existent symbol references, incorrect API calls, and fabricated dependencies, resulting in high compilation failure rates (up to 86\%). Moreover, test smell analysis reveals that while LLM-generated tests are generally more readable than those produced by traditional tools, they still suffer from recurring design issues such as Magic Number Tests and Assertion Roulette, which hinder maintainability. Overall, our findings indicate that LLMs can serve as effective assistive tools for generating readable and maintainable test suites, but hybrid approaches that combine LLM-based generation with automated validation and search-based refinement are required to achieve reliable and production-ready test generation.

\end{abstract}


\section{Introduction}
\label{intro}
Unit testing is fundamental to software quality, yet inadequate testing remains pervasive—with only 48\% of developers regularly writing tests~\citep{daka2014survey}. 
While testing ensures correctness by validating components before integration~\citep{beck2000extreme}, the labor-intensive nature of test creation leads to significant coverage gaps in practice. 
Furthermore, when tests do exist, their poor structure often creates brittle, incomprehensible test suites that hamper maintenance~\citep{panichella2020revisiting,almasi2017industrial,grano2018empirical}.

Traditional automated test generation approaches have attempted to address these challenges, but each faces significant limitations. 
Search-Based Software Testing (SBST), particularly tools like EvoSuite~\citep{fraser2011evosuite}, optimizes for code coverage but produces tests that are difficult to read~\citep{almasi2017industrial}, contain unintuitive naming~\citep{daka2015modeling}, and rely on generic assertions~\citep{panichella2020revisiting}. 
Most critically, these high-coverage tests often fail to effectively detect faults~\citep{pinto2010multi}—the fundamental purpose of testing. 
Alternative approaches like fuzzing~\citep{zeller2019fuzzing}, random testing~\citep{pacheco2007randoop}, and symbolic execution~\citep{sen2005cute} each introduce their own trade-offs between coverage, precision, and scalability.

Large Language Models (LLMs)\citep{brown2020language,fan2023large} have emerged as a promising alternative for test generation, potentially addressing traditional approaches' fundamental limitations. Initial studies suggest LLM-generated tests feature more natural naming conventions, human-like assertions, and better readability than SBST alternatives\citep{siddiq2024using,yuan2023no,tang2024chatgpt}. Moreover, LLMs can generate both functional and exploratory tests beyond mere structural coverage~\citep{wang2024software}, aligning with comprehensive testing practices. 
Yet, it remains unclear whether prompting strategies that succeed in general code generation also transfer effectively to test generation. While reasoning-based prompts (e.g., CoT, ToT) have shown strong results in functional code tasks~\citep{li2025structured,wang2024enhancing}, unit test generation involves additional challenges: importing the class under test (CUT), setting up fixtures, managing program state, and writing reliable oracles~\citep{molina2025test,hossain2024togll}. These aspects introduce domain-specific constraints absent in standard code synthesis.

In this work, we examine whether prompting hierarchies generalize to test generation, where structure, dependencies, and correctness criteria differ from conventional code synthesis. Our class-level evaluation across diverse benchmarks uncovers key limitations that must be addressed before LLM-based testing can be reliably deployed:

\vspace{-1mm}
\begin{itemize}[leftmargin=*]
    \item {\bf Impact of Prompt Engineering on Test Reliability and Usability:} 
    Prior work~\citep{tang2024chatgpt,siddiq2024using} has focused on minimal prompting strategies, leaving structured reasoning approaches (CoT, ToT, GToT) underexplored. We evaluate their effectiveness in improving test structure, compilability, and consistency.

\item {\bf Structural Correctness \& Extractability of LLM-Generated Tests:} 

    Prior work has evaluated various aspects of LLM-generated tests, including compilability and coverage~\citep{tang2024chatgpt,yang2024evaluation,siddiq2024using}, and in some cases, test smells~\citep{siddiq2024using}, providing valuable insights into the runtime viability and maintainability of generated tests. However, few studies assess whether LLM-generated outputs are structurally valid and directly usable as standard JUnit test classes. We build on this line of work by applying extractability metrics (MSR and CSR~\citep{macedo2024exploring}) to measure whether generated files conform to JUnit expectations and can be seamlessly integrated into real-world testing pipelines.

\item {\bf Readability, Maintainability \& Test Smells:} 
    LLM tests are considered more readable than SBST outputs~\citep{yuan2023no,tang2024chatgpt}, but readability doesn't ensure maintainability. Test smells reduce long-term usability~\citep{panichella2020revisiting}. We assess maintainability using readability models~\citep{scalabrino2018comprehensive,sergeyuk2024reassessing}, complexity metrics, and automated test smell detection, comparing LLMs with EvoSuite.

\item {\bf Generalizability Across Diverse Codebases:} 
    Prior studies on LLM-generated tests rely on limited benchmarks~\citep{siddiq2024using,tang2024chatgpt,yang2024evaluation}, raising concerns about generalizability. We address this by evaluating three datasets: Defects4J, featuring real-world buggy software; SF110, containing Java programs with existing test cases; and CMD, a curated dataset excluding projects released before May 2023 to mitigate LLM training data leakage.
\end{itemize}

{\bf This paper: }We systematically evaluate LLM-generated test suites and prompt engineering impact. Building on prior comparisons~\citep{tang2024chatgpt,siddiq2024using,yang2024evaluation}, surveys~\citep{wang2024software}, and AI risk studies~\citep{sallou2023breaking}, we conduct a large-scale evaluation across four LLMs and five prompting techniques against EvoSuite. 
Our study focuses on understanding prompting strategies in LLM-based test generation. We contrast LLM outputs with EvoSuite~\citep{fraser2011evosuite}, the state-of-the-art in search-based test generation (SBST) for Java, which has consistently outperformed random testing and symbolic execution~\citep{shamshiri2015automatically}. In the SBFT Tool Competition~\citep{jahangirova2023sbft}, EvoSuite consistently ranks first in line/branch coverage and mutation score, outperforming Randoop~\citep{pacheco2007randoop}, Kex~\citep{abdullin2022kex}, and UTBot\footnote{\url{https://www.utbot.org/}}, and remains the most widely adopted and validated SBST tool~\citep{ouedraogo2025enriching,almasi2017industrial,shamshiri2015automatically}. By positioning LLMs against EvoSuite, we ensure a strong, credible baseline (the de facto reference for automated Java test generation), providing a fair reference point for assessing whether LLMs, with proper prompting, can reach or surpass state-of-the-art SBST methods.
Unlike method-level approaches in prior work, our class-level evaluation aligns with real-world testing workflows.

We make the following key contributions:
\vspace{-1mm}
\begin{enumerate}
    \item[\ding{182}] {\bf Systematic Investigation of Prompt Engineering:} We expand on previous research~\citep{tang2024chatgpt} by testing five structured prompting techniques (ZSL, FSL, CoT, ToT, GToT) with general-purpose LLMs, moving beyond the code-specific models in ~\citet{yang2024evaluation}. Unlike prior method-level studies~\citep{siddiq2024using,yang2024evaluation,yuan2024evaluating,chen2024chatunitest} our class-level evaluation provides a more realistic assessment by capturing inter-method dependencies. Our results challenge claims~\citep{yang2024evaluation} that reasoning-based prompting has limited impact on test generation for code-specialized LLMs. Our experiments demonstrate GToT consistently enhances test quality—improving structure, reducing hallucinations, and increasing extractability and compilability.

    \item[\ding{183}] {\bf Systematic Structural Correctness \& Extractability Analysis:} 
    Our analysis reveals that reasoning-based prompting improves test extractability and real-world usability, addressing inconsistencies in LLM test suite generation.

    \item[\ding{185}] {\bf Systematic Differentiation Between Readability \& Maintainability:} 
    We analyze how prompt engineering impacts both readability and long-term maintainability, highlighting areas where LLMs produce cleaner or more problematic test structures.

    \item[\ding{186}] {\bf Systematic Multi-Dataset, Multi-Model Evaluation for Generalizability:} We benchmark four LLMs on Defects4J, SF110, and CMD, offering a broader dataset scope than prior studies. CMD mitigates data leakage by including projects released after May 2023.

    \item[\ding{187}] {\bf Publicly Available Benchmark \& Replication Package:} We provide a benchmark of 216,300 tests across 690 Java classes, with prompt templates, evaluation scripts, and results accessible at: \url{https://anonymous.4open.science/r/LLM4TS-0F76/}
    
\end{enumerate}

The remainder of this paper is structured as follows: Section~\ref{background} covers LLM-based test generation, Section~\ref{setup} outlines our experimental setup, Section~\ref{results} presents evaluation results, Section~\ref{discussion} discusses threats to validity, limitations and future work, Section~\ref{relatedwork} reviews related work, Section~\ref{discussion-of-findings-and-implications} discusses key findings and implications, and Section~\ref{conclusion} concludes the study.

\section{Background}
\label{background}

In this section, we introduce the core concepts underpinning our study. We outline key ideas from prompt engineering, describe metrics adapted or extended for evaluating generated tests (e.g., MSR and CSR), clarify the distinction between readability and understandability, present common test smells, and review static analysis tools used to assess code quality. These notions provide essential context for interpreting the influence of prompting strategies on LLM-based test generation.

\vspace{-6mm}
\begin{figure*}[]
	\centering
        \includegraphics[width=1\textwidth]{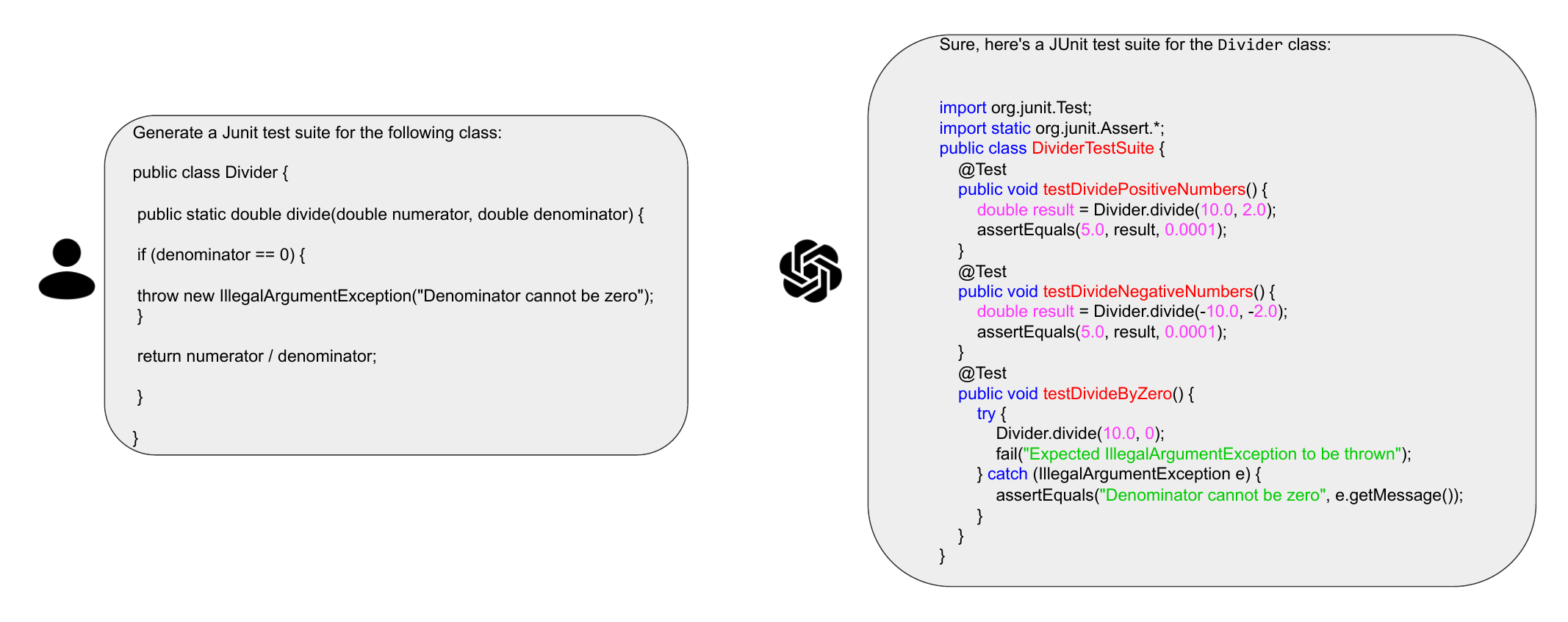}
	\caption{A Sample Use of ChatGPT in Unit test Suite generation}
	\label{fig:chatgpt-use-unit-test-generation}
    \vspace{2mm}
\end{figure*}

\subsection{Prompt Engineering}
\label{subsec:prompt-enginneering}
Prompt engineering creates effective prompts that guide LLMs in specific tasks~\citep{amatriain2024prompt,siddiq2024quality,naveed2023comprehensive,sahoo2024systematic}. This approach is valuable for complex problems where single-line reasoning is insufficient~\citep{amatriain2024prompt,kojima2022large,wei2022chain}, allowing LLMs to consider multiple possibilities before reaching conclusions. The significance of prompt engineering is twofold: specific prompts yield precise responses~\citep{si2022prompting}, and well-crafted instructions control output format, reducing post-processing needs~\citep{macedo2024exploring}. Software engineering applications typically use basic approaches like Zero-shot~\citep{kojima2022large} and Few-shot learning~\citep{brown2020language}, though recent research has expanded to include Chain-of-Thought~\citep{wei2022chain}, self-consistency~\citep{wang2022self}, and Tree-of-Thoughts~\citep{wang2022self}.

\textbf{Zero-shot learning} enables models to generalize without examples but struggles with complexity~\citep{sahoo2024systematic}. \textbf{Few-shot learning} uses examples to improve performance on novel tasks~\citep{brown2020language}. \textbf{Chain-of-Thought} (CoT)~\citep{wei2022chain} outlines intermediate reasoning steps, while \textbf{Self-consistency}~\citep{wang2022self} uses diverse reasoning paths to select consistent answers. \textbf{Tree-of-Thoughts} (ToT)~\citep{long2023large} builds on CoT by exploring a reasoning \quotes{tree} with search algorithms. This study applies these techniques to adapt LLMs for unit test generation.

\subsection{Match Success Rate (MSR) and Code Extraction Success Rate (CSR)}

Match Success Rate (MSR) and Code Extraction Success Rate (CSR), introduced by~\citet{macedo2024exploring}, evaluate the reliability of LLM-generated code. MSR measures the proportion of outputs where predefined patterns successfully extract source code, ensuring structural adherence. CSR assesses the proportion of valid test code extracted, filtering out explanatory text or formatting inconsistencies. These metrics are crucial for usability assessment, as LLM outputs often mix code with commentary (Figure~\ref{fig:chatgpt-use-unit-test-generation}).

In our setting, we adopt these metrics as a preliminary lens: rather than asking whether the code is syntactically valid or compilable, we first ask whether test code can be consistently detected and cleanly extracted from raw LLM outputs. This allows us to decouple \emph{format and extractability} from later concerns of syntax, compilation, and execution.

\paragraph{\textbf{Rationale.}}
LLM outputs often interleave code with natural language (e.g., explanations, caveats, reformatted snippets). This interleaving can bias downstream compilation- or execution-based metrics if extraction is not controlled. Following \citet{macedo2024exploring}, we decouple \emph{format/extractability} from \emph{syntax/compilation} using two format-oriented metrics tailored to JUnit tests.

\paragraph{\textbf{Definitions (adapted to test generation).}}
Let $\mathcal{O}$ be the set of all LLM responses for a given setting (model $\times$ prompt $\times$ class). Each response $o \in \mathcal{O}$ is treated as raw text. We compute:

\begin{itemize}
  \item \textbf{MSR (Match Success Rate).}
 Fraction of responses where a JUnit-oriented \emph{detector} finds a code segment that \emph{appears} to delimit a test suite. Formally,
    \begin{equation}
    \mathrm{MSR} = \frac{|\{\, o \in \mathcal{O} \mid \mathrm{detect}(o) = \mathrm{true} \,\}|}{|\mathcal{O}|}.
    \end{equation}

  Detection succeeds when either (i) explicit delimiters are respected (``\texttt{\#\#\#Test START\#\#\#}'' ... ``\texttt{\#\#\#Test END\#\#\#}''); or (ii) in their absence, fallback patterns match a plausible JUnit test file (package/imports/class declaration with \texttt{@Test} or JUnit imports). MSR is a \emph{format-compliance} signal: it tells us whether a response contains something that looks like a test suite, not whether that suite is valid or compilable.

  \item \textbf{CSR (Code Extraction Success Rate).}
  Fraction of responses where extraction yields a \emph{structured test file} that passes minimal \emph{structural} sanity checks. Formally,
  \begin{equation}
    \mathrm{CSR} = \frac{|\{\, o \in \mathcal{O} \mid \mathrm{extract}(o) \rightarrow \widehat{t} \ \wedge \ \mathrm{is\_structured}(\widehat{t}) \,\}|}{|\mathcal{O}|}.
    \end{equation}

  The extraction function isolates the best candidate $\widehat{t}$ (single file) after stripping any non-code content. $\mathrm{is\_structured}(\cdot)$ verifies that the candidate looks like a \emph{complete} JUnit test file (e.g., optional \texttt{package}, one or more \texttt{import}s including JUnit, a \texttt{public class} with at least one method or assertion). CSR is independent of syntax or compilation (both are assessed later in RQ2).
\end{itemize}

\paragraph{\textbf{Detectors and patterns.}}
We use a two-stage detector:
\begin{enumerate}
 \item \textbf{Delimiter-first.} If the prompt requested explicit boundaries (``\texttt{\#\#\#Test START\#\#\#}'' / ``\texttt{\#\#\#Test END\#\#\#}''), we extract the enclosed block.
  \item \textbf{Fallback patterns.} If delimiters are absent or malformed, we search for a \emph{single} best candidate using ordered patterns that privilege JUnit-shaped files (package line, imports, class declaration, and at least one \texttt{@Test} or assertion).
\end{enumerate}
If multiple code fences exist, we select the longest candidate that satisfies the JUnit shape. If no candidate passes detection, both MSR and CSR count that response as a failure.

\paragraph{\textbf{Structural validation for CSR.}}
$\mathrm{is\_structured}(\widehat{t})$ checks that the extracted block is not a partial snippet but a plausible file:
(i) at least one \texttt{import} or a direct reference to JUnit/assertions;
(ii) a single top-level \texttt{class} with balanced braces;
(iii) absence of obvious non-Java artifacts (Markdown fences, HTML) within the extracted region.

\paragraph{\textbf{Illustrative example (Divider).}}
Figure~\ref{fig:chatgpt-use-unit-test-generation} illustrates a typical case: the user asks for a JUnit suite for a simple \texttt{Divider} class. ChatGPT returns (i) a valid JUnit test class \texttt{DividerTestSuite}, and (ii) a natural-language summary of the scenarios. In our framework, MSR=1 because the detector identifies a plausible test file, and CSR=1 because extraction isolates a structured class with imports, assertions, and balanced braces. The trailing explanation is ignored at this stage. This example shows why MSR/CSR are necessary: without controlled detection and extraction, downstream compilation (RQ2) would be contaminated by the extraneous prose.

\paragraph{\textbf{What MSR/CSR do \emph{not} measure.}}
Neither metric judges Java syntax correctness, compilation, coverage, or fault detection. They are pre-execution format/extractability filters. As shown by \citet{macedo2024exploring}, controlling output format can substantially affect downstream compilation; we therefore isolate this step as RQ1.

\paragraph{\textbf{Synthetic corner cases.}}
\begin{enumerate}
\item \textbf{MSR = 1, CSR = 1.} Response has delimiters and encloses a full JUnit file.
\item \textbf{MSR = 1, CSR = 0.} Detector matches a region (e.g., a Markdown fence), but extracted content fails structural checks.
\item \textbf{MSR = 0, CSR = 0.} Response is purely explanatory or pseudo-code, no code region is detected.
\end{enumerate}

\paragraph{\textbf{Scope note.}}
MSR/CSR are LLM-specific. SBST tools (e.g., EvoSuite) directly emit JUnit files by construction; applying MSR/CSR would always succeed and thus be uninformative. We therefore report MSR/CSR only for LLM outputs and use syntax/compilation checks for \emph{both} paradigms in RQ2.

\subsection{Readability vs. Understandability}
Readability and understandability are distinct code quality aspects. Readability concerns visual clarity (indentation, naming, formatting), while understandability measures cognitive effort needed to comprehend logic~\citep{buse2009learning,oliveira2022systematic,grano2018empirical,winkler2024investigating}. Readability models evolved from traditional approaches~\citep{buse2009learning,daka2015modeling,posnett2011simpler,dorn2012general} to deep learning-based assessments~\citep{mi2022towards}. 
Though human evaluations are ideal, their scalability is limited by subjectivity and evaluator fatigue~\citep{kumar2024llms,zhang2024review,sun2024source,sergeyuk2024reassessing}. 

~\cite{scalabrino2018comprehensive} implemented a robust readability model integrating syntactic, structural, visual, and textual elements from Java projects. Using 104 features, it improved accuracy by 6.2\% over previous methods and better matched human assessments~\cite{sergeyuk2024reassessing}. Its training on jUnit projects makes it especially suitable for evaluating test code readability. 
Unit tests present unique challenges—poor readability obscures intent and hinders maintenance, particularly with complex fixtures~\citep{grano2018empirical,winkler2024investigating}. Our study evaluates LLM-generated tests for readability, maintainability, and practical usability.

\subsection{Test Smells}

Unit testing verifies individual components' functionality while maintaining readability. Test smells indicate poor practices~\citep{palomba2016diffusion,peruma2019distribution} that lead to fragility and tight coupling. Assertion Roulette and Magic Number Tests—documented in TsDetect\footnote{\url{https://testsmells.org/pages/testsmellexamples.html}}—commonly appear in both EvoSuite and LLM-generated tests. In Listing 1, Assertion Roulette shows multiple assertThat() calls without clarifications, obscuring failure sources. 

\vspace{3mm}
\begin{lstlisting}[style=test-smell-listing, caption= Example of Assertion Roulette Test Smell, basicstyle=\scriptsize\tiny]
@MediumTest
public void testCloneNonBareRepoFromLocalTestServer() throws Exception {
    Clone cloneOp = new Clone(false, integrationGitServerURIFor("small-repo.early.git"), helper().newFolder());
    Repository repo = executeAndWaitFor(cloneOp);
    assertThat(repo, hasGitObject("ba1f63e4430bff267d112b1e8afc1d6294db0ccc"));
    File readmeFile = new File(repo.getWorkTree(), "README");
    assertThat(readmeFile, exists());
    assertThat(readmeFile, ofLength(12));
}
\end{lstlisting}
\label{lst:assertion-roulette}
\vspace{1mm}

In Listing 2, Magic Number Test demonstrates unexplained values (15, 30) without context, compromising maintainability.

\vspace{1mm}
\begin{lstlisting}[style=test-smell-listing, caption=Example of Magic Number Test Test Smell, basicstyle=\scriptsize\tiny]
@Test
public void testGetLocalTimeAsCalendar() {
    Calendar localTime = calc.getLocalTimeAsCalendar(BigDecimal.valueOf(15.5D), Calendar.getInstance());
    assertEquals(15, localTime.get(Calendar.HOUR_OF_DAY));
    assertEquals(30, localTime.get(Calendar.MINUTE));
}
\end{lstlisting}
\vspace{2mm}
\label{lst:magic-number}

We use TsDetect~\citep{peruma2020tsdetect} to analyze how prompt engineering affects test smell prevalence in LLM-generated tests compared to EvoSuite-generated tests.

\subsection{Static Analysis Tools: Checkstyle, PMD, and SpotBugs}

To assess the structural quality of LLM-generated test suites, we rely on three widely used static analysis tools: Checkstyle\footnote{\url{https://checkstyle.sourceforge.io}}, PMD\footnote{\url{https://pmd.github.io/}}, and SpotBugs\footnote{\url{https://spotbugs.github.io/}.}

\paragraph{\textbf{Checkstyle.}}

Checkstyle is a static code analysis tool that verifies Java code conformance to coding standards. It detects formatting issues, naming convention violations, and structural inconsistencies.
In our study, we adopt both the \emph{Sun Code Conventions}\footnote{\url{https://www.oracle.com/java/technologies/javase/codeconventions-contents.html}} and \emph{Google Java Style}\footnote{\url{https://google.github.io/styleguide/javaguide.html}} profiles to evaluate coding standard adherence.

\paragraph{\textbf{PMD.}}

PMD is a source code analyzer that detects common code issues such as unused variables, empty catch blocks, and overly complex code. It provides metrics for both \textbf{cyclomatic complexity} (control flow paths) and \textbf{cognitive complexity} (human-perceived difficulty). 
Following SonarSource guidelines\footnote{\url{https://www.sonarsource.com/docs/CognitiveComplexity.pdf}}, we classify \emph{cognitive complexity} into:
\emph{Low ($<5$)}, \emph{Moderate ($5$--$10$)}, \emph{High ($11$--$20$)}, and \emph{Very High ($\geq 21$)}.
For \emph{cyclomatic complexity} (based on PMD thresholds), we use:
\emph{Low ($1$--$4$)}, \emph{Moderate ($5$--$7$)}, \emph{High ($8$--$10$)}, and \emph{Very High ($\geq 11$)}.
These thresholds allow us to assess whether generated test cases remain structurally simple and maintainable.

\paragraph{\textbf{SpotBugs.}}
SpotBugs detects potential software bugs through bytecode-level analysis. It categorizes issues based on severity into four ranked priority levels:
\begin{itemize}
  \item \textbf{Scariest}: Critical defects that threaten correctness, security, or stability.
  \item \textbf{Scary}: Severe bugs that may cause subtle but impactful failures.
  \item \textbf{Troubling}: Moderate issues, potentially leading to runtime problems if left unaddressed.
  \item \textbf{Of Concern}: Minor issues or code smells that may reduce code quality but rarely break functionality.
\end{itemize}

This categorization allows us to distinguish not only the number of detected issues, but also their potential severity and impact on code reliability. By using these severity levels, we can better characterize the trade-offs between code structure and safety in automatically generated test suites.

\vspace{-2mm}
\section{Experimental Setup}
\label{setup}
\subsection{Overview}
\begin{figure*}
  \centering
  \includegraphics[width=1.05\textwidth]{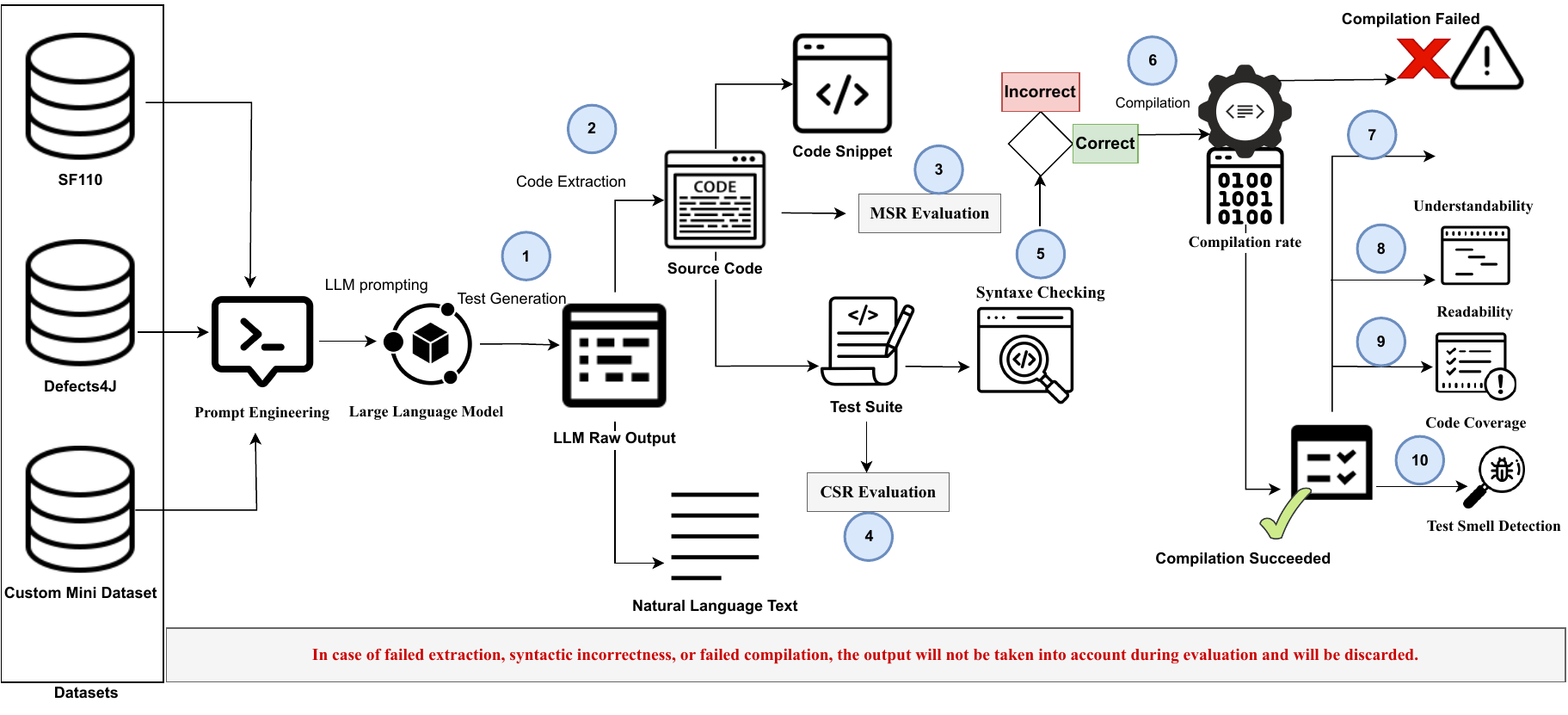}
   \caption{Overview of the pipeline used to design the experiments.}
    \label{fig:experiments-overview}
    \vspace{1mm}
\end{figure*}

Our methodology (Figure~\ref{fig:experiments-overview}) evaluates LLM-generated test suites using ZSL, FSL, CoT, ToT, and GToT prompting techniques. 
To assess prompt engineering and reliability (RQ1, RQ2), we analyze structural correctness and usability using MSR, CSR, syntax correctness, and compilability. For readability and maintainability (RQ3, RQ4), we employ automated metrics including a deep learning model~\citep{scalabrino2018comprehensive} that correlates with human assessments~\citep{sergeyuk2024reassessing}, alongside complexity metrics and coding standard adherence. For code coverage (RQ5), we compare LLM tests with EvoSuite across line, instruction, and method coverage. 
Finally, we evaluate test smell prevalence (RQ7) to determine if reasoning-based prompting improves maintainability.

\subsection{Research Questions}
This study evaluates LLMs in test generation and the impact of prompt engineering through seven research questions:

\vspace{0.2cm}\noindent\textbf{RQ0: What is the ability of SBST (EvoSuite) and LLMs to generate test code at all?}
Before analyzing correctness, readability, or maintainability, it is essential to first assess whether different paradigms can actually produce test code for a given dataset. We therefore examine the raw generation capacity of EvoSuite and LLMs across datasets, measuring how many test suites were produced out of the total expected runs. This provides a baseline for understanding the feasibility of test generation before deeper analyses.

\vspace{0.2cm}\noindent\textbf{RQ1: How do prompt engineering techniques impact the output format and extractability of LLM-generated test suites?}
We examine how different prompting strategies influence test suite structure and extractability through \ding{182} Match Success Rate (MSR), measuring detection of test code via predefined patterns, and \ding{183} Code Extraction Success Rate (CSR), evaluating reliable extraction of valid test code. These metrics determine adherence to JUnit structures and isolation from extraneous content.


\vspace{0.2cm}\noindent\textbf{RQ2: How does prompt engineering influence the generation of syntactically correct and compilable test suites by LLMs, in comparison with SBST (EvoSuite)?} 
We assess how different prompting strategies affect the ability of LLMs to generate test suites that \ding{182} adhere to syntax rules and \ding{183} successfully compile, enabling execution without ensuring functionality. 
In addition, we contrast these results with EvoSuite as a representative SBST baseline, in order to highlight similarities and differences between search-based and LLM-based test generation with respect to syntactic validity and compilability.

\vspace{0.2cm}\noindent\textbf{RQ3: Beyond syntactic correctness, how well do LLMs generate human-understandable and maintainable test suites?}
We evaluate if LLM-generated tests: \ding{182} convey test logic clearly; \ding{183} remain adaptable over time; and \ding{184} integrate seamlessly into workflows, determining if they meet practical development standards.
%
EvoSuite is not included in RQ3 analyses, as its primary goal is maximizing structural coverage rather than producing understandable or maintainable tests. This distinction has been repeatedly emphasized in the literature: \citet{rojas2015automated} show that while EvoSuite increases coverage and reduces testing time, developers consistently struggle with the readability of its outputs. Similarly, \citet{shamshiri2018automatically} report that although developers are as effective with EvoSuite-generated tests as with manual ones, they take significantly longer to maintain them and perceive them as harder to understand. These findings confirm that readability and understandability are not baselines EvoSuite provides, but rather limitations that have motivated recent efforts to refactor generated tests with LLMs~\citep{deljouyi2025leveraging,biagiola2025improving}. Therefore, RQ3 focuses exclusively on LLM-generated tests, where prompt design explicitly targets human-oriented quality dimensions such as readability and maintainability.

\vspace{0.2cm}\noindent\textbf{RQ4: How do LLM-generated test suites compare to EvoSuite and align with developer expectations in terms of readability?}
We investigate whether LLM tests meet developer expectations by: \ding{182} enhancing navigation through clear naming and structure; \ding{183} reducing ambiguity with well-organized presentation; and \ding{184} supporting collaboration. We analyze how prompting techniques influence readability using an automated model correlated with human evaluation.
%
While EvoSuite was excluded in RQ3 because static analysis tools such as Checkstyle, PMD, and SpotBugs assess 
dimensions (style compliance, cyclomatic/cognitive complexity, and bug patterns) that it was never designed to optimize, making a direct comparison unfair and potentially misleading, in RQ4 we include EvoSuite as a baseline reference. Here, the focus is on readability from a developer perspective, evaluated with an established readability model~\citep{scalabrino2018comprehensive}. EvoSuite’s widespread adoption in SBST makes it a useful empirical point of comparison to determine whether LLM-generated tests deliver tangible gains in human-oriented quality, complementing the structural strengths of SBST with improved clarity and readability.

\vspace{0.2cm}\noindent\textbf{RQ5: How Thorough Are LLM-Generated Test Suites in Terms of Code Coverage?}
We examine code coverage (line, instruction, method) of LLM-generated tests compared to EvoSuite~\citep{fraser2011evosuite}, analyzing how prompting strategies impact exploration of program behavior.

\vspace{0.2cm}\noindent\textbf{RQ6: How do LLM-generated test suites compare to EvoSuite-generated test suites in terms of test smell prevalence?}
We compare the absence of test smells in LLM-generated tests against EvoSuite to determine whether prompt engineering produces more maintainable tests with fewer design issues than traditional SBST techniques.

\subsection{Large Language Models}
\label{LargeLanguageModels}
Our study focuses on four instruction-tuned decoder-only models known for their enhanced ability to follow instructions compared to base models. These models belong to the OpenAI and MistralAI families and are detailed in Table~\ref{table:models-used}. 

\begin{table}[ht]
\centering
\caption{Summary of LLMs Used in Current Study}
\label{table:models-used}
\resizebox{0.99\columnwidth}{!}{
\begin{tabular}{@{}l c c c @{}}
\toprule
\textbf{Model Name} & \textbf{\# Parameters (Billions)} & \textbf{Context Length (Tokens)} & \textbf{Release Date} \\
\midrule
GPT 3.5-turbo~\citep{gpt35turbo} & 175 & 4\,096 & Nov, 2022 \\
GPT 4~\citep{achiam2023gpt} & $>$175 & 8\,192 & Mar, 2023 \\
Mistral 7B~\citep{jiang2023mistral} & 7 & 4\,096 & Sep, 2023 \\
Mixtral 8x7B~\citep{jiang2024mixtral} & 46.7 & 8\,192 & Dec, 2023 \\
\bottomrule
\end{tabular}
}
\end{table}

GPT-3.5-turbo has a token limit of 4,096, and exceeding this limit can hinder class context understanding, affecting test quality~\citep{tang2024chatgpt}. To ensure fairness, all models adhere to this limit, excluding inputs that exceed it. Token counts are computed using OpenAI's tiktoken\footnote{\url{https://github.com/openai/tiktoken}} library.

We select OpenAI models (GPT-3.5, GPT-4) and MistralAI models (Mistral 7B, Mixtral 8x7B) for their relevance and efficiency. OpenAI models ensure comparability with prior LLM-based test generation studies~\citep{siddiq2024using,tang2024chatgpt,yang2024evaluation}. MistralAI models, underexplored in this domain, offer cost-effective alternatives: Mistral 7B matches Llama 34B’s performance~\citep{jiang2023mistral}, while Mixtral 8x7B, leveraging Sparse Mixture-of-Experts (SMoE), surpasses Llama 2 70B with lower inference costs~\citep{jiang2024mixtral}. This selection balances established and emerging models, broadening the scope of LLM-based test generation research.

Furthermore, all models are set to a standard temperature of 0.7. Our selection criteria prioritized models that excel in code generation and possess accessible APIs, making them suitable for both academic research and industrial applications~\citep{wang2024software,fan2023large}.

\subsection{Prompt Design}
Unit test generation research emphasizes practical and structured approaches~\citep{siddiq2024using, yuan2023no, chen2023chatunitest, tang2024chatgpt}. To ensure clarity, we design prompts with varying complexity using a standardized template. Each prompt comprises: (1) a Natural Language Description (NLD) specifying the task, and (2) a Source Code placeholder for inserting the Class Under Test (CUT). 
The following sections detail these components and their alignment with unit testing practices.

\begin{itemize}[leftmargin=0.2cm]
    \item \textbf{Natural Language Description:} 
    We design Natural Language Descriptions (NLDs) that guide LLMs in test generation, including: (i) a role-playing directive simulating professional testers \citep{tang2024chatgpt,chen2023chatunitest}; (ii) a task-specific directive for JUnit 4 test file generation; and (iii) a control statement enforcing structured output with markers. We introduce delimiters (\textbf{\#\#\#Test START\#\#\#} and \textbf{\#\#\#Test END\#\#\#}) to separate test code from explanatory text.

    \item \textbf{Source Code:} 
    Aligned with prior work on learning-based unit test generation~\citep{tang2024chatgpt,chen2023chatunitest}, our prompts vary by technique, either including (i) only the class name and source code or (ii) an additional example class with its test suite to provide context and comparative guidance.
\end{itemize}

This study evaluates five prompt techniques: Zero-Shot Learning (ZSL), Few-Shot Learning (FSL), Chain-of-Thought (CoT), Tree-of-Thought (ToT), and Guided Tree-of-Thought (GToT). ZSL uses minimal context for generalization, while FSL introduces examples to improve structure. For consistency, we use the same medium-sized class across experiments, ensuring token limit compliance for FSL examples.  CoT enhances reasoning through step-by-step prompting, while ToT explores multiple reasoning paths before selecting test cases. GToT, our novel approach, simulates collaborative reasoning among three “testers” who iteratively refine tests, combining CoT's structured reasoning with ToT's diversity to improve coverage, correctness, and variability. Figure~\ref{fig:prompt-templates-cot-tot} illustrates CoT and ToT prompts, while Figure~\ref{fig:gtot-prompt} shows GToT's collaborative structure. All prompts are available in our replication package.

\begin{figure}[ht]
\centering
\scalebox{0.9}{
\begin{tcbitemize}[raster columns=1, size=fbox, boxrule=0.8pt,
                  colback=white, colframe=black, fonttitle=\bfseries,
                  title style={colframe=black, colback=gray!20}]

\tcbitem[colback=white, colframe=black]
\footnotesize
\begin{minipage}{1.01\linewidth}
    \caption*{(a) Chain-of-Thought Prompting}
    \label{fig:cot}
\begin{lstlisting}[
      basicstyle=\ttfamily\scriptsize,
      breaklines=true,
      breakindent=0pt,
      showstringspaces=false
    ]
As a professional software tester who writes Java test methods, generate Junit 4 test cases to comprehensively test all methods in the following class named {class_name} by following these steps:\n- Extract and list all the public methods including their signatures\n- For each methods, generate a basic JUnit 4 test case that checks the method's functionality\n- Given the source code of the class and the listed methods, identify potential edge cases and exception handling scenarios that should be tested\n- Generate JUnit 4 test cases that specifically test for the identified edge cases and exceptions\n- Merge all the individual test cases into a complete JUnit 4 test file ({class_name}Test.java) for the given Java class. The complete JUnit test file must start with ###Test START## and finish with ###Test END##\nHere is the {class_name} class:\n{source_code} 
\end{lstlisting}
\end{minipage}

\tcbitem[colback=white, colframe=black]
\footnotesize
\begin{minipage}{1.01\linewidth}
    \caption*{(b) Tree-of-Thoughts Prompting}
    \label{fig:tot}
\begin{lstlisting}[
      basicstyle=\ttfamily\scriptsize,
      breaklines=true,
      breakindent=0pt,
      showstringspaces=false
    ]
Imagine three different experts in software testing who are tasked with developing comprehensive JUnit 4 test cases for the following Java class. All experts will propose one test cases for each method, share it with the group, and then proceed to the next step. If any expert realizes they're wrong at any point, they leave.\nThe Java class is:\n{source_code}\nAt the end they must propose one complete (including typical use cases, edge cases, and error scenarios) JUnit 4 test file ({class_name}Test.java) for the given Java class. The complete JUnit test file must start with ###Test START## and finish with ###Test END##
\end{lstlisting}
\end{minipage}

\end{tcbitemize}
}
\caption{Examples of Chain-of-Thought and Tree-of-Thoughts for test suite generation.}
\label{fig:prompt-templates-cot-tot}
\vspace{2mm}
\end{figure}

\begin{figure}[ht]
\centering
\scalebox{0.9}{ 
\scriptsize
\begin{tcbitemize}[raster columns=1, size=fbox, boxrule=0.8pt,
                  colback=white, colframe=black, fonttitle=\bfseries,
                  title style={colframe=black, colback=gray!20}]

\tcbitem[colback=white, colframe=black]
\begin{minipage}{1.01\linewidth}
    \caption*{Guided Tree-of-Thoughts Prompting}
    \label{fig:gtot}
\begin{lstlisting}[
      basicstyle=\ttfamily\scriptsize,
      breaklines=true,
      breakindent=0pt,
      showstringspaces=false,
      columns=fullflexible
    ]
Imagine three different experts in software testing who are tasked with developing comprehensive JUnit 4 test cases for the following Java class.To comprehensively test all methods in the following class named {class_name} they must following these steps:\n- Extract and list all the public methods including their signatures\n- For each methods, generate a basic JUnit 4 test case that checks the method's functionality\n- Given the source code of the class and the listed methods, identify potential edge cases and exception handling scenarios that should be tested\n- Generate JUnit 4 test cases that specifically test for the identified edge cases and exceptions\n- Merge all the individual test cases into a complete JUnit 4 test file ({class_name}Test.java) for the given Java class.All experts will propose one test cases for each method, share it with the group, and then proceed to the next step. If any expert realizes they're wrong at any point, they leave.\nThe Java class is:\n{source_code}\nAt the end they must propose one complete (including typical use cases, edge cases, and error scenarios) JUnit 4 test file ({class_name}Test.java) for the given Java class. The complete JUnit test file must start with ###Test START## and finish with ###Test END##
\end{lstlisting}
\end{minipage}

\end{tcbitemize}
}
\caption{Example of Guided Tree-of-Thoughts (GToT) for test suite generation.}
\label{fig:gtot-prompt}
\vspace{2mm}
\end{figure}

\vspace{-5mm}
\subsection{The EvoSuite SBST tool}

For the SBST baseline, we selected EvoSuite~\citep{fraser2011evosuite} as it is the most mature and empirically validated tool for automated test generation in Java. Recent editions of the SBFT competition confirm EvoSuite’s leadership over competitors such as Randoop~\citep{pacheco2007randoop}, Kex~\citep{abdullin2022kex}, and UTBot, particularly in terms of branch and line coverage~\citep{jahangirova2023sbft}. Large-scale empirical analyses (e.g.~\citep{shamshiri2015automatically}) also show that EvoSuite can detect real faults in Defects4J more effectively than random or commercial tools. Moreover, recent research on enhancing automated test generation~\citep{ouedraogo2025enriching} continues to adopt EvoSuite as the reference point when proposing extensions or hybrid approaches. These consistent results across competitions, benchmarks, and empirical studies justify the choice of EvoSuite as the most reliable SBST comparator for our study.
We employ its default settings, including the DynaMOSA algorithm~\citep{panichella2017automated}, which optimizes branch, method, and line coverage. To mitigate the randomness of genetic algorithms, we run EvoSuite 30 times per class, following~\citep{tang2024chatgpt}. A three-minute time budget is allocated per test generation, aligning with prior studies~\citep{fraser2015does, arcuri2013parameter, shamshiri2015automatically}, which find this duration sufficient for effective test case generation. We deliberately disable EvoSuite's post-generation processing, which reduces test smells, to ensure a fair comparison with unoptimized LLM-generated tests. This approach allows for an unbiased evaluation of the baseline capabilities of both methods, highlighting their inherent strengths and limitations.

\paragraph{\textbf{Unit of analysis.}} 
Each EvoSuite run produces exactly one test suite, which corresponds to a single Java test file. 
Therefore, all counts and percentages reported in our study are based on the number of generated test files (one per suite). 
This ensures consistency with our treatment of LLM-generated outputs, where each model–prompt–iteration also produces one test suite (file).

\paragraph{\textbf{Hardware}}
For reference, EvoSuite experiments are conducted on an AMD EPYC 7552 48-Core Processor running at 3.2 GHz with 640 GB of RAM. However, it's important to note that not all available resources are necessarily utilized during the tests.

\subsection{Datasets}
\label{subsec:dataset}

To minimize selection bias and ensure generalizability, we employ three diverse datasets: SF110, Defects4J, and a Custom Mini Dataset (CMD). Both SF110 and Defects4J are used to evaluate both EvoSuite and LLM performance. The CMD, containing GitHub projects created after May 2023, is exclusively tested with LLMs due to compatibility issues with EvoSuite's Java version requirements~\footnote{\url{https://github.com/EvoSuite/evosuite/issues/433}}. This dataset focuses on well-maintained projects from the OceanBase Developer Center (ODC)~\footnote{\url{https://github.com/oceanbase/odc}} and Conductor OSS~\footnote{\url{https://github.com/conductor-oss/conductor}}.

While SF110 covers academic and industrial Java programs and Defects4J focuses on real-world bug-fixing scenarios, both datasets were filtered with token constraints so that every retained class fits all prompt engineering techniques (ZSL, FSL, CoT, ToT, GToT). This reduced SF110 from 346 to 182 classes (74 projects) and Defects4J to 16 projects (477 classes).
We built CMD to complement SF110 (academic/industrial programs) and Defects4J (bug-fixing scenarios) with contemporary, evolving software. It contains two actively maintained projects ODC (OceanBase Developer Center) and Conductor OSS, selected to assess generalization beyond legacy-style codebases. CMD is carefully curated under three constraints: (i) token limits so each class fits all prompt engineering techniques (ZSL, FSL, CoT, ToT, GToT), (ii) only classes committed after May 2023 to reflect modern Java practices and reduce potential training leakage, and (iii) a buggy–fixed pairing scheme (as in Defects4J) to enable controlled evaluation. ODC (13 releases, 527 stars, 20 contributors) is a database development tool providing SQL development, risk management, and data security; Conductor OSS (11 releases, 15.5k stars, 53 contributors) is a microservices orchestration framework originally developed at Netflix. Despite its smaller scale, CMD offers a modern, actively maintained testbed that complements SF110 and Defects4J.

The datasets encompass projects with code of varying levels of cyclomatic complexity (detailed in Table~\ref{tab:dataset-overview-complexity}), ranging from low to high. This diversity helps mitigate potential overfitting on specific complexity profiles and provides valuable insights into the potential of LLMs for real-world automated test case generation.

\vspace{2mm}
\begin{table*}[!htbp]
\centering
\caption{Overview of datasets and their complexity.}
\label{tab:dataset-overview-complexity}
\resizebox{1.03\textwidth}{!}{
\begin{tabular}{@{}l p{7cm} cc ccc@{}}
\toprule
\textbf{Dataset} & \textbf{Description} & \textbf{Number of Projects} & \textbf{Number of Classes} & \multicolumn{3}{c}{\textbf{Complexity Overview}} \\
\cmidrule(lr){5-7}
 & & & & \textbf{Low} & \textbf{Moderate} & \textbf{High} \\ 
\midrule
SF110 & 
Derived from DynaMOSA benchmark~\citep{panichella2017automated}, originally 117 projects (346 classes). After applying token constraints to ensure all prompts fit within context windows, 74 projects (182 classes) were retained.
& 74 & 182 & 170 & 138 & 120 \\ 
\midrule
Defects4J & Contains 835 reproducible bugs across 17 open-source projects. Similarly, token constraints were applied to guarantee that each selected class can accommodate all prompt engineering techniques without exceeding context limits, resulting in 16 projects (477 classes).

& 16 & 477 & 254 & 177 & 130 \\ 
\midrule
Custom mini Dataset (CMD) & Includes ODC, a collaborative project to enhance database functionalities, and Conductor OSS, an open-source project management tool aimed at efficiency and scalability. 
To ensure fairness and prevent potential training leakage, CMD was curated under three constraints: (i) token limits, so that each class can be used across all prompt engineering techniques (ZSL, FSL, CoT, ToT, GToT), (ii) only classes committed after May 2023, and (iii) a buggy–fixed pairing approach similar to Defects4J.
& 2 & 31 & 16 & 9 & 5 \\ 
\bottomrule
\end{tabular}
}
\begin{tablenotes}
\tiny
\item[1] {\bf Notice.} Reported classes correspond to those retained after applying filtering constraints (token limits and, for CMD, post-2023 commits and buggy–fixed pairing). Due to budgetary reasons, we only run  GPT-4 and Mixtral-8x7B on the smallest dataset (CMD). Other models (GPT-3.5-Turbo and Mistral-7B) are run on all datasets.  
\end{tablenotes}
\vspace{3mm}
\end{table*}

\section{Experimental Results}
\label{results}

\subsection{[RQ0]: Ability of EvoSuite and LLMs to generate test code}
\noindent\textbf{[Experiment Goal]:} 
This research question evaluates the fundamental ability of EvoSuite and LLMs to generate test suites, independently of syntactic correctness or compilability. The goal is to establish whether each paradigm can produce test code at scale for the target datasets, thereby clarifying their baseline feasibility as automated test generators.

\noindent\textbf{[Experiment Design]:}
For each dataset and each class under test, we performed 30 independent generations. EvoSuite was applied on Defects4J and SF110 but not on CMD due to version compatibility issues with Java (Section~\ref{subsec:dataset}). For LLMs, GPT-3.5-turbo and Mistral 7B were evaluated across all datasets, while GPT-4 and Mixtral 8×7B were restricted to CMD due to budget constraints (Section~\ref{subsec:dataset}). The total number of expected test files was computed as the number of classes multiplied by 30 runs. For LLMs, since five prompting strategies were applied, the expected count was further multiplied by 5. Each generation, whether by EvoSuite or by an LLM, produces exactly one test suite, which corresponds to a single Java test file. Therefore, all counts and ratios in RQ0 are calculated on the basis of test files. The outcome metric is the ratio of generated test suites (i.e., outputs containing code) over the expected total, providing a direct measure of raw generation ability.

\noindent\textbf{[Experiment Results]:}

Table~\ref{tab:rq0-generation-capacity} summarizes the raw generation capacity of EvoSuite and the LLMs across datasets.
\begin{table}[ht]
\centering
\caption{Test generation capacity of EvoSuite and LLMs across datasets.}
\label{tab:rq0-generation-capacity}
\scalebox{0.78}{
\begin{tabular}{lrrrrr}
\toprule
\textbf{Dataset} & \textbf{EvoSuite} & \textbf{GPT-3.5} & \textbf{GPT-4} & \textbf{Mistral} & \textbf{Mixtral} \\
\midrule
\textbf{Defects4J} & 11100 / 14310 & 60565 / 71550 & --             & 65180 / 71550 & --             \\
                   & (77.6\%)      & (84.7\%)       &                & (91.1\%)      &                \\
\textbf{SF110}     & 5460 / 5460   & 23048 / 27300  & --             & 25664 / 27300 & --             \\
                   & (100\%)       & (84.4\%)       &                & (94.0\%)      &                \\
\textbf{CMD}       & --            & 4633 / 4650    & 4356 / 4650    & 4226 / 4650   & 2947 / 4650    \\
                   &               & (99.6\%)       & (93.7\%)       & (90.9\%)      & (63.4\%)       \\
\bottomrule
\end{tabular}
}
\begin{tablenotes}
\tiny
\item[1] $^*$Percentages refer to successful generations.
\item[2] $^*$Mixtral = Mixtral 8×7B; Mistral = Mistral 7B; GPT-3.5 = GPT-3.5-turbo.
\end{tablenotes}
\vspace{3mm}
\end{table}
On Defects4J (477 classes), EvoSuite produced tests in 77.6\% of the expected runs (11{,}100/14{,}310). Notably, for two entire projects (\texttt{Closure} and \texttt{Mockito}), EvoSuite did not generate any tests for any class despite multiple attempts.
In contrast, the LLMs generated code more frequently in this raw sense: GPT-3.5-turbo at 84.7\% and Mistral~7B at 91.1\% of the expected outputs. 
On SF110 (182 classes), EvoSuite achieved 100\% generation (5{,}460/5{,}460), while GPT-3.5-turbo and Mistral~7B reached 84.4\% and 94.0\%, respectively.
On CMD (31 classes), only LLMs were evaluated due to Java version incompatibilities with EvoSuite; GPT-3.5-turbo approached saturation at 99.6\% (4{,}633/4{,}650), GPT-4 and Mistral~7B were slightly lower (93.7\% and 90.9\%), and Mixtral~8$\times$7B lagged at 63.4\%.

RQ0 provides a necessary preliminary step by clarifying whether different paradigms can even produce test code before deeper quality analyses. The results highlight two important observations. First, EvoSuite’s generation capacity is not uniform across datasets: while it attains perfect generation on SF110, it fails to produce tests for certain Defects4J classes, supporting concerns raised in prior work that SBST tools may not always yield outputs for all classes under test~\citep{fraser2013evosuite, arcuri2017private, ouedraogo2025enriching}. Second, LLMs show high raw generation rates across datasets, often approaching or surpassing EvoSuite in this preliminary dimension. However, generation rates differ markedly across models, as illustrated by Mixtral~8$\times$7B’s weaker performance on CMD, indicating that raw generation ability is model- and context-sensitive.

These results establish RQ0 as a necessary baseline. Even mature SBST tools like EvoSuite do not always produce tests in challenging settings such as parts of Defects4J, which has direct consequences for downstream quality and coverage analyses. At the same time, the consistently high generation rates achieved by LLMs highlight the need to scrutinize not only whether code is produced, but also \emph{what} is produced in terms of structure, syntactic correctness, compilability, readability, smells, and fault-detection capability before assessing practical viability. Finally, the divergence across models and datasets—for instance, the comparatively weaker performance of Mixtral~8$\times$7B on CMD—demonstrates that raw generation capacity is strongly model- and context-sensitive, underscoring the importance of analyzing prompting effects (RQ1) and subsequent validity dimensions (RQ2–RQ6) while carefully distinguishing performance across different models and datasets.

\vspace{0.5em}
\noindent\colorbox{gray!20}{{\parbox{0.98\linewidth}{
\textbf{Finding 1:} EvoSuite, despite being a mature SBST tool, does not always generate tests for all classes (e.g., failures on \texttt{Closure} and \texttt{Mockito} in Defects4J), achieving 77.6\% of expected outputs, while it reaches 100\% on SF110. In contrast, LLMs display high raw generation capacity across datasets (often above 90\%), though with variability across models (e.g., Mixtral~8$\times$7B at 63.4\% on CMD).
}}}
\vspace{0.5em}

\paragraph{\textbf{Relation to prior work.}}
Comparative studies with EvoSuite (e.g., \citet{siddiq2024using, tang2024chatgpt, yang2024evaluation}) typically report compilability or coverage after some code is returned, often on method-level targets or on filtered-class subsets (e.g., classes reduced to a single focal method with injected imports). In contrast, RQ0 isolates the raw question “does the tool emit an output file at all?” at the class level, where “per-class” means one generation per entire class under test, including all interdependent methods, with no import injection. Under this stricter setup across heterogeneous projects, we observe cases where EvoSuite produces no suite for some classes or projects, whereas LLMs usually emit some code output. This qualifies baseline feasibility and highlights the impact of granularity and dataset scope on raw generation.

\vspace{0.5em}
\noindent\highlight{%
Summary of \textbf{RQ0:}
Even mature SBST tools like EvoSuite do not always return test suites for all classes, especially on heterogeneous datasets such as Defects4J. In contrast, LLMs almost always emit an output file, though at this stage we do not yet establish whether the content is valid test code. RQ0 therefore provides a baseline feasibility check: it clarifies whether tools emit outputs at all, before RQ1–RQ2 investigate format adherence, extractability, syntactic correctness, and compilability.}
\vspace{0.5em}

\subsection{[RQ1]: Impact of prompt engineering on the output format and extractability.}
\noindent\textbf{[Experiment Goal]:} 
This research examines how prompt engineering affects the format and extractability of LLM-generated test suites, evaluated through Match Success Rate (MSR) for detecting test code and Code Extraction Success Rate (CSR) for extracting structured test code, regardless of its syntactic correctness or compilability.

These metrics are specific to LLMs, since unlike LLMs, EvoSuite directly produces JUnit test files without extraneous natural language or prompt-driven variability. As a result, MSR and CSR are not applicable to SBST tools and are only reported for LLM-generated outputs.

\noindent\textbf{[Experiment Design]:} 
We evaluate prompt designs using Match Success Rate (MSR) and Code Extraction Success Rate (CSR). MSR measures adherence to predefined structures, while CSR assesses extraction of structured test code excluding commentary. To address LLMs' tendency to mix code with explanations (Figure~\ref{fig:chatgpt-use-unit-test-generation}), we introduced delimiters (\#\#\#Test START\#\# and \#\#\#Test END\#\#) to improve extraction accuracy (Figures~\ref{fig:prompt-templates-cot-tot}). Following \citet{macedo2024exploring}, we used regular expressions for extraction. For CSR validity, test suites needed key components like import statements, ensuring completeness rather than isolated snippets. CSR evaluates extractability independent of syntax correctness or compilability. Experiments on SF110, Defects4J, and CMD assessed LLM performance across diverse scenarios.

\noindent\textbf{[Experiment Results]:}

Our study generated 216,300 files, with 190,619 (88.13\%) containing source code. 

Table~\ref{table:msr-csr} reports aggregated MSR and CSR scores by model and dataset, computed across all five prompting strategies (ZSL, FSL, CoT, ToT, GToT). 
This gives a global view of each model’s ability to produce extractable test code on average. 
The detailed breakdown by prompting technique is presented in Tables~\ref{table:msr-by-prompt} and~\ref{table:csr-by-prompt}, allowing fine-grained analysis of how each prompt affects format compliance and extractability.

\noindent\textbf{\underline{Dataset-Level Performance.}}
LLMs showed varying MSR and CSR scores across datasets (Table~\ref{table:msr-csr}). GPT-3.5-turbo achieved the highest MSR (99.64\%) and CSR (90.79\%) on CMD, indicating strong structured output generation. GPT-4 also performed well, reaching a CSR of 83.59\% on CMD. Mistral 7B excelled in SF110 (MSR: 94.00\%) and Defects4J (MSR: 91.10\%), demonstrating better structural adherence in these datasets. In contrast, Mixtral 8x7B struggled on CMD, with lower MSR and CSR scores.
\vspace{2mm}
\begin{table*}[!htbp]
\centering

\caption{MSR and CSR by LLM and Dataset (aggregated over prompts)}.
\label{table:msr-csr}
\resizebox{\textwidth}{!}{%
\begin{tabular}{@{}l rrr rrr rrr rrr@{}}
\toprule
\textbf{Model Name} & \multicolumn{3}{c}{\textbf{\# of Generated Files}} & \multicolumn{3}{c}{\textbf{\# of Files with code}} & \multicolumn{3}{c}{\textbf{MSR}} & \multicolumn{3}{c}{\textbf{CSR}} \\
\cmidrule(lr){2-4} \cmidrule(lr){5-7} \cmidrule(lr){8-10} \cmidrule(lr){11-13}
& \textbf{SF110} & \textbf{Defects4J} & \textbf{CMD} & \textbf{SF110} & \textbf{Defects4J} & \textbf{CMD} & \textbf{SF110} & \textbf{Defects4J} & \textbf{CMD} & \textbf{SF110} & \textbf{Defects4J} & \textbf{CMD} \\
\midrule
GPT 3.5-turbo & 27\,300 & 71\,550 & 4\,650 & 23\,048 & 60\,565 & \textbf{4\,633} & 84.42\% & 84.65\% & \textbf{99.64\%} & 78.10\% & 78.34\% & \textbf{90.79\%} \\
GPT 4$^*$\tnote{1} & -- & -- & 4\,650 & -- & -- & 4\,356 & -- & -- & 93.68\% & -- & -- & 83.59\% \\
Mistral 7B & 27\,300 & 71\,550 & 4\,650 & \textbf{25\,664} & \textbf{65\,180} & 4\,226 & \textbf{94.00\%} & \textbf{91.10\%} & 90.88\% & \textbf{79.97\%} & \textbf{80.35\%} & 83.61\% \\
Mixtral 8x7B$^*$\tnote{1} & -- & -- & 4\,650 & -- & -- & 2\,947 & -- & -- & 63.38\% & -- & -- & 53.57\% \\
\midrule
\textbf{Total/Average (\%)} & \textit{54\,600} & \textit{143\,100} & \textit{18\,600} & \textit{48\,712} & \textit{125\,745} & \textit{16\,162} & \textit{\textbf{89.21}} & \textit{87.88} & \textit{86.90} & \textit{79.04} & \textit{\textbf{79.35}} & \textit{77.89} \\
\bottomrule
\end{tabular}
}
\begin{tablenotes}
\tiny
\item[1] $^*$Note that due to budget constraints, we only paid to query GPT-4 and Mixtral 8x7B for CMD. 
\end{tablenotes}
\end{table*}

\vspace{0.5em}
\noindent\colorbox{gray!20}{{\parbox{0.98\linewidth}{
\textbf{Finding 2:}
OpenAI models (GPT-3.5, GPT-4) demonstrate superior MSR and CSR, indicating their ability to generate structured, extractable test suites. Mistral 7B also performs well, whereas Mixtral 8x7B struggles, highlighting model-specific variations.
}}}
\vspace{0.5em}

\noindent\textbf{\underline{Impact of Prompting Techniques.}}
GToT consistently achieved the highest MSR across all models and datasets(Table~\ref{table:msr-by-prompt}), reinforcing its effectiveness in enforcing structural adherence. 

For GPT-3.5-turbo, it reached \textbf{24.54\%} on Defects4J, \textbf{23.95\%} on SF110, and \textbf{20.01\%} on CMD. Mistral 7B also benefited from GToT, achieving \textbf{20.79\%} on Defects4J and \textbf{21.07\%} on SF110. GPT-4 reached \textbf{18.20\%} on CMD.  ZSL showed competitive results in certain settings, with GPT-3.5-turbo reaching \textbf{21.51\%} on Defects4J and \textbf{20.05\%} on CMD. In contrast, FSL consistently produced lower MSR scores, suggesting that additional examples do not necessarily enhance structural adherence. ToT outperformed CoT across most models, improving MSR by up to 2.51\% on Defects4J for GPT-3.5-turbo, highlighting the advantage of stepwise reasoning in structuring outputs.
\vspace{2mm}
\begin{table*}[ht]
\centering
\caption{MSR by LLM and Prompt Engineering}
\scriptsize
\label{table:msr-by-prompt}
\resizebox{\textwidth}{!}{%
\begin{tabular}{@{}l|rrrrr|rrrrr|rrrrr@{}}
\toprule
\textbf{Model Name} & \multicolumn{5}{c|}{\textbf{Defects4J (\%)}} & \multicolumn{5}{c|}{\textbf{SF110 (\%)}} & \multicolumn{5}{c}{\textbf{CMD (\%)}} \\
\cmidrule(lr){2-6} \cmidrule(lr){7-11} \cmidrule(lr){12-16}
& \textbf{ZSL} & \textbf{FSL} & \textbf{CoT} & \textbf{ToT} & \textbf{GToT} & \textbf{ZSL} & \textbf{FSL} & \textbf{CoT} & \textbf{ToT} & \textbf{GToT} & \textbf{ZSL} & \textbf{FSL} & \textbf{CoT} & \textbf{ToT} & \textbf{GToT} \\
\midrule
GPT 3.5-turbo & 21.51 & 16.43 & 17.51 & 20.02 & \textbf{24.54} & 19.95 & 17.97 & 19.40 & 18.73 & \textbf{23.95} & \textbf{20.05} & \textbf{20.05} & 19.94 & 19.94 & 20.01 \\
GPT 4         & -- & -- & -- & -- & -- & -- & -- & -- & -- & -- & 21.35 & 21.30 & 21.03 & 18.11 & 18.20 \\
Mistral 7B        & 19.72 & 19.31 & 20.07 & 20.10 & \textbf{20.79} & 19.72 & 19.39 & 19.63 & 20.19 & \textbf{21.07} & 20.04 & 19.88 & 20.04 & 19.83 & \textbf{20.21} \\
Mixtral 8x7B      & -- & -- & -- & -- & -- & -- & -- & -- & -- & -- & 20.04 & 19.88 & 20.04 & 19.83 & \textbf{20.21} \\
\bottomrule
\end{tabular}
}
\vspace{3mm}
\end{table*}

\noindent\colorbox{gray!20}{{\parbox{0.98\linewidth}{
\textbf{Finding 3:}
GToT is the most effective prompting strategy for improving adherence to follow prompt instructions (MSR), while ZSL remains a strong baseline, especially for OpenAI models. FSL underperforms, challenging assumptions that more examples lead to better structural adherence.
}}}
\vspace{0.5em}

GToT often achieved the highest CSR scores across datasets (Table~\ref{table:csr-by-prompt}), particularly for GPT‑3.5‑turbo and GPT‑4, demonstrating its effectiveness in reducing extraneous content and structuring outputs. 
For example, GPT‑3.5‑turbo reached \textbf{24.54\%} on Defects4J and \textbf{23.65\%} on SF110 with GToT. However, CSR improvements are more sensitive to model–dataset interactions than MSR, and simpler prompting strategies can sometimes match or exceed GToT. On CMD, for instance, GPT‑3.5‑turbo achieved \textbf{21.70\%} with ZSL compared to \textbf{21.06\%} with GToT, and GPT‑4 also reached over \textbf{23\%} CSR with ZSL. For Mistral 7B, CoT slightly outperformed GToT on Defects4J (20.65\% vs. 18.73\%). FSL consistently underperformed across all settings, and Mixtral 8x7B remained the weakest overall despite peaking at \textbf{23.60\%} CSR on CMD under GToT.

\vspace{2mm}
\begin{table*}[ht]
\centering
\caption{CSR by LLM and Prompt Engineering}
\scriptsize
\label{table:csr-by-prompt}
\resizebox{\textwidth}{!}{%
\begin{tabular}{@{}l|rrrrr|rrrrr|rrrrr@{}}
\toprule
\textbf{Model Name} & \multicolumn{5}{c|}{\textbf{Defects4J (\%)}} & \multicolumn{5}{c|}{\textbf{SF110 (\%)}} & \multicolumn{5}{c}{\textbf{CMD (\%)}} \\
\cmidrule(lr){2-6} \cmidrule(lr){7-11} \cmidrule(lr){12-16}
& \textbf{ZSL} & \textbf{FSL} & \textbf{CoT} & \textbf{ToT} & \textbf{GToT} & \textbf{ZSL} & \textbf{FSL} & \textbf{CoT} & \textbf{ToT} & \textbf{GToT} & \textbf{ZSL} & \textbf{FSL} & \textbf{CoT} & \textbf{ToT} & \textbf{GToT} \\
\midrule
GPT 3.5-turbo & 21.51 & 16.43 & 17.51 & 20.02 & \textbf{24.54} & 21.30 & 17.30 & 18.05 & 19.70 & \textbf{23.65} & \textbf{21.70} & 20.77 & 20.09 & 16.39 & 21.06 \\
GPT 4         & -- & -- & -- & -- & -- & -- & -- & -- & -- & -- & \textbf{23.00} & 22.90 & 20.68 & 17.60 & 15.82 \\
Mistral 7B        & 20.26 & 20.23 & \textbf{20.65} & 20.13 & 18.73 & \textbf{21.25} & 21.15 & 20.95 & 20.43 & 16.23 & \textbf{20.60} & 20.29 & 20.37 & 20.11 & 18.62 \\
Mixtral 8x7B      & -- & -- & -- & -- & -- & -- & -- & -- & -- & -- & 18.51 & 23.36 & 18.95 & 15.58 & 23.60 \\
\bottomrule
\end{tabular}
}
\end{table*}

\vspace{0.5em}
\noindent\colorbox{gray!20}{{\parbox{0.98\linewidth}{
\textbf{Finding 4:}
GToT often yields the highest CSR scores, but improvements vary depending on the model and dataset. ZSL remains competitive for GPT models sometimes even outperforming GToT while FSL consistently underperforms. Mixtral 8x7B remains the least effective overall, despite isolated gains under GToT.
}}}
\vspace{0.5em}

To address LLMs' tendency to mix test code with explanatory text, we introduced delimiters (\textbf{\#\#\#Test START\#\#\#} and \textbf{\#\#\#Test END\#\#\#}) to guide  structured output. Despite these clear boundaries, no LLM consistently follows the expected format, with MSR varying between 16\% and 25\% across models and datasets (Table~\ref{table:csr-by-prompt}). This inconsistency directly impacts CSR, which ranges from 15\% to 24\% (Table~\ref{table:csr-by-prompt}), highlighting persistent challenges in reliably extracting complete and structured test code.

\vspace{0.5em}
\noindent\colorbox{gray!20}{{\parbox{0.98\linewidth}{
\textbf{Finding 5:}
Explicit delimiters improve extraction by structuring LLM outputs, but all LLMs exhibit inconsistencies in following formatting instructions, with some failing more frequently, making automated extraction challenging.
}}}
\vspace{0.5em}

\paragraph{\textbf{Relation to prior work.}}
Existing studies on LLM-based test generation (e.g., \citet{siddiq2024using, tang2024chatgpt, yang2024evaluation, yuan2024evaluating, chen2024chatunitest}) typically report results on syntactic correctness, compilability, coverage, or fault detection once code has been extracted, without explicitly assessing the reliability of that extraction step. Their evaluation pipelines therefore proceed directly from raw responses to downstream metrics, leaving output format and extractability unexamined. In this study we adapt the MSR/CSR framework of \citet{macedo2024exploring}, originally proposed for code translation, to JUnit test generation, making format and extractability an explicit precondition before syntax and compilation analyses (RQ2).

\vspace{0.5em}
\noindent\highlight{%
Summary of \textbf{RQ1:}
Prompt engineering strongly influences extractability of LLM-generated test suites. OpenAI models (GPT-3.5-turbo, GPT-4) achieve the highest MSR and CSR, with Mistral 7B competitive and Mixtral 8×7B lagging. GToT consistently yields the highest MSR across all models and datasets, enforcing structural adherence. For CSR, GToT often attains the best results (particularly for GPT models), though simpler prompts like ZSL or CoT occasionally match it depending on model–dataset pairing. FSL underperforms across all settings. These findings extend \citet{macedo2024exploring} by showing that controlling prompt structure remains crucial in test generation: explicit reasoning and delimiter guidance substantially enhance format compliance, yet no model follows delimiters reliably. Extracting usable test code from mixed natural-language outputs remains a key challenge for practical adoption.
}
\vspace{0.5em}

\subsection{[RQ2]: Syntactical Correctness and Compilability.}
\noindent\textbf{[Experiment Goal]:}
This research question examines how prompt engineering influences test suite syntactical correctness and compilability in LLM-generated tests. Valid Java grammar and compilation without modifications are essential for automated workflow integration. We aim to identify the most effective prompting strategies for generating valid, executable test suites. 
To provide a reference point, we also report EvoSuite’s syntactic correctness and compilability, as it represents a state-of-the-art SBST baseline. This contrast highlights differences between search-based and LLM-based approaches in producing directly usable test code.

\noindent\textbf{[Experiment Design]:}  
%
We evaluate syntactical correctness using Javalang\footnote{\url{https://github.com/c2nes/javalang}} parsing and assess compilability with JVM compilation. 
A test suite is considered syntactically correct if parsing succeeds, and compilable if compilation completes without errors. For non-compilable suites, we categorize compilation errors (e.g., \texttt{Cannot Find Symbol}, \texttt{Package Does Not Exist}) to identify hallucination patterns. 
We then analyze how different prompting techniques affect these error distributions and assess their effectiveness in mitigating specific hallucinations. EvoSuite results are included to indicate the upper bound expected from an SBST tool, allowing us to contextualize LLM outputs against a mature baseline. Experiments span SF110, Defects4J, and CMD datasets to evaluate how both paradigms handle correctness, compilability, and hallucination mitigation across diverse codebases.

Each run, whether by an LLM or by EvoSuite, produces exactly one test suite, which corresponds to a single Java test file. Therefore, all counts and percentages in our study are calculated on the basis of test files, ensuring a consistent comparison between LLM-generated and SBST-generated outputs.

For the calculation of syntactic correctness and compilability percentages, we proceed as follows. 
Suppose $A$ is a given model and $B$ a given dataset. Let $X$ be the total number of syntactically correct test suites produced by $(A,B)$ across all prompting strategies. For a specific prompting strategy $p$, we denote by $Y_p$ the number of syntactically correct test suites it generates, and by $Z_p$ the number of those that also compile successfully. The syntactic correctness percentage for $(A,B,p)$ is then computed as $Y_p / X \times 100$, while the compilability percentage is computed as $Z_p / X \times 100$. For EvoSuite, which does not involve prompting strategies, syntactic correctness is measured as the fraction of syntactically correct test suites among all generated ones, and compilability is measured as the fraction of compilable test suites among those that are syntactically correct.

\noindent\textbf{[Experiment Results]:}

\noindent\textbf{\underline{Dataset-Level Analysis.}}

Across all LLM runs, we produced \textbf{216{,}300} outputs (Table~\ref{tab:rq0-generation-capacity},Table~\ref{table:msr-csr}). 
Of these, \textbf{190{,}619} (\textbf{88.13\%}) contained extractable test code, while the remaining \textbf{25{,}681} (\textbf{11.87\%}) were plain text or structurally invalid (no extractable code).
Among the code-containing outputs, \textbf{171{,}187} (\textbf{89.81\%}) were syntactically correct, while \textbf{19{,}432} (\textbf{10.19\%}) exhibited syntax errors. 
For EvoSuite, all generated test files (\textbf{16{,}560} across Defects4J and SF110) were syntactically correct. These results highlight the challenge LLMs face in consistently generating well-formed unit tests, emphasizing the need for post-generation validation.

\vspace{0.5em}
\noindent\colorbox{gray!20}{{\parbox{0.98\linewidth}{
\textbf{Finding 6:} Across datasets, \textbf{89.81\%} of LLM-generated outputs containing code are syntactically correct, while \textbf{10.19\%} include syntax errors, requiring manual intervention before integration into real-world workflows. In contrast, EvoSuite achieves \textbf{100\%} syntactic correctness on all its generated outputs.
}}}
\vspace{0.5em}

To provide full transparency, Table~\ref{tab:syntax-compile-by-prompt} reports the raw counts of syntactically correct (S) and compilable (C) test suites for each model, dataset, and prompting strategy. 
These raw values form the basis for the normalized percentages of syntactic correctness and compilability later reported in Tables~\ref{table:syntactically} and~\ref{table:compilability}.

\vspace{0.5em}
\begin{table}[ht]
\centering
\caption{Syntactic correctness (S) and compilability (C) by model, dataset, and prompt.}
\label{tab:syntax-compile-by-prompt}
\scalebox{0.7}{
\begin{tabular}{llccccc}
\toprule
\textbf{Model} & \textbf{Dataset} & \textbf{ZSL (S/C)} & \textbf{FSL (S/C)} & \textbf{CoT (S/C)} & \textbf{ToT (S/C)} & \textbf{GToT (S/C)} \\
\midrule
\multirow{3}{*}{GPT-3.5-turbo}
  & Defects4J & 12056/1656 &  9207/1244 &  9812/1808 & 11220/1268 & 13756/3789 \\
  & SF110     &  4541/1267 &  3689/897  &  3849/1164 &  4200/870  &  5042/2061 \\
  & CMD       &   916/114  &   877/103  &   848/250  &   692/113  &   889/146  \\
\midrule
\multirow{1}{*}{GPT-4}
  & CMD       &   894/69   &   890/86   &   804/103  &   684/74   &   615/93   \\
\midrule
\multirow{3}{*}{Mistral 7B}
  & Defects4J & 11649/258  & 11633/247  & 11871/153  & 11575/252  & 10766/629  \\
  & SF110     &  4639/238  &  4617/227  &  4574/199  &  4460/379  &  3543/735  \\
  & CMD       &   792/0    &   789/1    &   801/0    &   782/1    &   724/12   \\
\midrule
\multirow{1}{*}{Mixtral 8$\times$7B}
  & CMD       &   472/4    &   582/3    &   461/2    &   388/11   &   588/4    \\
\midrule
\multirow{1}{*}{EvoSuite}
  & Defects4J & \multicolumn{5}{c}{11100/10890} \\
  & SF110     & \multicolumn{5}{c}{5460/5460} \\
\bottomrule
\end{tabular}
}
\begin{flushleft}
\footnotesize
\textbf{S} = number of tests that parse successfully (syntactically correct); \textbf{C} = number of tests that compile without errors. Each cell reports \textit{S/C}.
\end{flushleft}
\end{table}

\noindent\textbf{\underline{EvoSuite Compilation Outcomes.}}  
While EvoSuite achieved full syntactic correctness on all the test files it generated, a closer inspection reveals that not all of these tests compiled successfully. In particular, we identified 210 non-compilable cases across Defects4J: two in \texttt{JacksonXml}, two in \texttt{Math}, and three in \texttt{JacksonDatabind}. This represents a small fraction of the total (210 out of 11{,}100 syntactically correct suites, i.e., 1.9\%), yet it highlights that even mature SBST tools can occasionally produce outputs that fail at compilation. On SF110, by contrast, EvoSuite compiled all 5{,}460 generated test files without error.

\vspace{0.5em}
\noindent\colorbox{gray!20}{{\parbox{0.98\linewidth}{
\textbf{Finding 7:} 
Even EvoSuite, the SBST baseline, does not guarantee full compilability: although all its generated tests were syntactically correct, 210 cases in Defects4J ($\approx1.9\%$) failed to compile, while SF110 compiled without errors. This shows that compilation failures are not unique to LLMs but can also occur in mature automated test generation tools.
}}}

\vspace{0.5em}
\noindent\textbf{\underline{LLM and Prompting Impact on Syntactical Correctness.}}
Table~\ref{table:syntactically} reveals significant variations in syntactical correctness across models and prompting techniques. 
GPT-3.5-turbo with GToT achieves the highest correctness on Defects4J (\textbf{24.54\%}) and SF110 (\textbf{23.65\%}), showing the benefit of structured reasoning for syntax. On CMD, however, it performs slightly lower than ZSL (21.06\% vs. \textbf{21.70\%}). GPT-4 also reaches high scores on CMD with ZSL (\textbf{23.00\%}). For Mistral 7B, FSL performs best on SF110 (\textbf{21.15\%}), while CoT is highest on Defects4J (\textbf{20.65\%}). These results suggest that prompting effectiveness for syntactical correctness varies across models and datasets, with no single strategy universally superior. FSL remains inconsistent overall, and Mixtral 8x7B shows the weakest performance despite isolated peaks (e.g., \textbf{23.60\%} with GToT on CMD).

\vspace{0.5em}
\noindent\colorbox{gray!20}{{\parbox{0.98\linewidth}{
\textbf{Finding 8:}
GPT models outperform other LLMs in syntactic correctness, with GPT-4 achieving the highest correctness rate (23\%) among all prompt configurations.
Prompting effectiveness, however, is dataset-dependent: GToT yields top results on Defects4J and SF110, yet ZSL slightly surpasses it on CMD. No single prompting strategy proves universally superior, while Mixtral 8x7B remains the weakest performer overall despite occasional peaks (e.g., 23.6\% with GToT on CMD).
}}}

\vspace{1mm}
\begin{table*}[ht]
\centering
\caption{Syntactic Correctness Across Datasets and Techniques$^*$\tnote{1}
}
\label{table:syntactically}
\resizebox{\textwidth}{!}{
\begin{tabular}{@{}l|rrrrr|rrrrr|rrrrr@{}}
\toprule
\textbf{Model Name} & \multicolumn{5}{c|}{\textbf{Syntactically Correct SF110 (\%)}} & \multicolumn{5}{c|}{\textbf{Syntactically Correct Defects4J (\%)}} & \multicolumn{5}{c}{\textbf{Syntactically Correct CMD (\%)}} \\
\cmidrule(lr){2-6} \cmidrule(lr){7-11} \cmidrule(lr){12-16}
& \textbf{ZSL} & \textbf{FSL} & \textbf{CoT} & \textbf{ToT} & \textbf{GToT} & \textbf{ZSL} & \textbf{FSL} & \textbf{CoT} & \textbf{ToT} & \textbf{GToT} & \textbf{ZSL} & \textbf{FSL} & \textbf{CoT} & \textbf{ToT} & \textbf{GToT} \\
\midrule
GPT 3.5-turbo & 21.30 & 17.30 & 18.05 & 19.70 & \textbf{23.65} & 21.51 & 16.43 & 17.51 & 20.02 & \textbf{24.54} & \textbf{21.70} & 20.77 & 20.09 & 16.39 & 21.06 \\
GPT 4         & -- & -- & -- & -- & -- & -- & -- & -- & -- & -- &\textbf{23.00} & 22.90 & 20.68 & 17.60 & 15.82 \\
Mistral 7B        & 21.25 & \textbf{21.15} & 20.95 & 20.43 & 16.23 & 20.26 & 20.23 & \textbf{20.65} & 20.13 & 18.73 & \textbf{20.37} & 20.29 & 20.60 & 20.11 & 18.62 \\
Mixtral 8x7B      & -- & -- & -- & -- & -- & -- & -- & -- & -- & -- &  18.85 & 23.36 & 18.51 & 15.58 & \textbf{23.60} \\
\midrule
\textbf{Average (\%)} & \textit{\textbf{21.27}} & \textit{19.22} & \textit{19.50} & \textit{20.06} & \textit{19.94} & \textit{20.89} & \textit{18.33} & \textit{19.08} & \textit{20.08} & \textit{\textbf{21.63}} & \textit{21} & \textit{\textbf{21.83}} & \textit{19.97} & \textit{17.42} & \textit{19.78} \\
\midrule
\textbf{EvoSuite} & \multicolumn{5}{c|}{\textbf{100.00}} & \multicolumn{5}{c|}{\textbf{100.00}} & \multicolumn{5}{c}{\textbf{--}} \\

\bottomrule
\end{tabular}
}
\begin{tablenotes}
\tiny
\item[1] $^*$Note that due to budget constraints, we only paid to query GPT-4 and Mixtral 8x7B for CMD. 
\end{tablenotes}
\vspace{2mm}
\end{table*}

\noindent\textbf{\underline{Compilability Rates Across Datasets and Prompting Techniques.}}
Table~\ref{table:compilability} reveals a significant gap between syntactical correctness and compilability. GPT-3.5-Turbo with GToT achieved the highest rates (9.67\% on SF110, 6.76\% on Defects4J), yet only 7.2\% of generated tests and 12\% of syntactically correct ones compiled. On CMD, GPT-3.5-Turbo with CoT surprisingly outperformed GPT-4 (5.92\% vs. 2.65\%), suggesting execution failures limit GPT-4's syntax advantage. Mixtral 8x7B showed inconsistent performance (near-zero to 20.16\% with GToT), indicating occasional success but poor overall reliability.

\vspace{2mm}
\begin{table*}[ht]
\centering
\caption{Compilability Rates of Test Suites Across Datasets and Techniques$^*$\tnote{1}
}
\label{table:compilability}
\resizebox{\textwidth}{!}{
\begin{tabular}{@{}l|rrrrr|rrrrr|rrrrr@{}}
\toprule
\textbf{Model Name} & \multicolumn{5}{c|}{\textbf{Compilable Code SF110 (\%)}} & \multicolumn{5}{c|}{\textbf{Compilable Code Defects4J (\%)}} & \multicolumn{5}{c}{\textbf{Compilable Code CMD (\%)}} \\
\cmidrule(lr){2-6} \cmidrule(lr){7-11} \cmidrule(lr){12-16}
& \textbf{ZSL} & \textbf{FSL} & \textbf{CoT} & \textbf{ToT} & \textbf{GToT} & \textbf{ZSL} & \textbf{FSL} & \textbf{CoT} & \textbf{ToT} & \textbf{GToT} & \textbf{ZSL} & \textbf{FSL} & \textbf{CoT} & \textbf{ToT} & \textbf{GToT} \\
\midrule
GPT 3.5-turbo & 5.94 & 4.21 & 5.46 & 4.08 & \textbf{9.67} & 2.95 & 2.22 & 3.23 & 2.26 & \textbf{6.76} & 2.70 & 2.44 & \textbf{5.92} & 2.68 & 3.46 \\
GPT 4         & -- & -- & -- & -- & -- & -- & -- & -- & -- & -- & 1.78 & 2.21 & \textbf{2.65} & 1.90 & 2.39 \\
Mistral 7B        & 1.09 & 1.04 & 0.91 & 1.74 & \textbf{3.37} & 0.45 & 0.43 & 0.27 & 0.44 & \textbf{1.09} & 0.00 & 0.03 & 0.00 & 0.03 & \textbf{0.31} \\
Mixtral 8x7B      & -- & -- & -- & -- & -- & -- & -- & -- & -- & -- & 0.16 & 0.12 & 0.08 & \textbf{0.44} & \textbf{20.16} \\
\midrule
\textbf{Average (\%)} & \textit{3.52} & \textit{2.62} & \textit{3.19} & \textit{2.91} & \textit{\textbf{6.52}} & \textit{1.70} & \textit{1.32} & \textit{1.75} & \textit{1.35} & \textit{\textbf{3.93}} & \text{1.16} & \text{1.20} & \textit{\textbf{2.16}} & \textit{1.26} & \textit{1.58} \\
\midrule
\textbf{EvoSuite} & \multicolumn{5}{c|}{\textbf{100.00}} & \multicolumn{5}{c|}{\textbf{98.1}} & \multicolumn{5}{c}{\textbf{--}} \\
\bottomrule
\end{tabular}
}
\begin{tablenotes}
\tiny
\item[1] $^*$ZSL: Zero-shot learning, FSL: Few-shot learning, CoT: Chain-of-Thought, ToT: Tree-of-Thoughts, GToT: Guided Tree-of-thoughts.
\end{tablenotes}
\vspace{2mm}
\end{table*}

These results contrast with prior work~\citep{siddiq2024using}, which reports substantially higher compilation rates. A key reason lies in the different experimental setups. \citet{siddiq2024using} adopt a method-level perspective, generating one test class per method under test, with all necessary imports explicitly included and post-processing heuristics applied to repair non-compilable outputs. In contrast, our study operates at the class level, where each generation targets the full class under test, often containing multiple interdependent methods and relying solely on the code available in the dataset without import injection or post-repair. Furthermore, our benchmarks (Defects4J, CMD, SF110) encompass larger and more heterogeneous projects than HumanEval or the filtered SF110 subset used by~\citet{siddiq2024using}, which further contributes to differences in compilation outcomes.

\noindent\colorbox{gray!20}{{\parbox{0.98\linewidth}{
\textbf{Finding 9:} 
The gap between syntactical correctness and compilability remains substantial, only 7.2\% of all generated test suites and 12\% of syntactically correct ones successfully compile. Compared to prior work, which reports higher compilability when generating one test class per individual method under test with injected imports and repair heuristics, our study evaluates entire classes under test (including all their interdependent methods) without import injection or post-repair. This setup highlights the additional challenges LLMs face when handling complete classes rather than isolated methods.
}}}
\vspace{0.5em}

\noindent\textbf{\underline{LLMs vs. EvoSuite on Syntactic Correctness and Compilability.}}
Tables~\ref{table:syntactically} and~\ref{table:compilability} allow a direct comparison between LLMs and EvoSuite. 
EvoSuite exhibits near-perfect reliability, with all generated test suites being syntactically correct and only 210 non-compilable cases out of 11{,}100 in Defects4J ($\approx$1.9\%), while achieving 100\% compilability on SF110. 
By contrast, LLMs display a much higher fragility: across datasets, only 89.81\% of code-containing outputs were syntactically correct, and compilability drops drastically, with success rates below 10\% in most cases. 
Reasoning-based prompting (e.g., GToT) mitigates syntax issues and improves compilability, but still falls far short of EvoSuite’s stability. This contrast underscores that while LLMs can approach EvoSuite in raw generation capacity (RQ0), their outputs are substantially less reliable when measured by structural validity and executable quality.

\vspace{0.5em}
\noindent\colorbox{gray!20}{{\parbox{0.98\linewidth}{
\textbf{Finding 10:}  
Compared to EvoSuite, LLMs remain substantially weaker in syntactic correctness and compilability: while EvoSuite compiles nearly all its generated tests (with only $\approx$1.9\% failures in Defects4J), LLMs succeed on fewer than 10\% of their syntactically correct outputs. This highlights that prompt engineering improves LLM reliability but does not yet close the gap with mature SBST tools.
}}}
\vspace{0.5em}

\noindent\textbf{\underline{Compilation Error Analysis and Hallucination Patterns.}}
Table~\ref{tab:compilation-errors-model-dataset} reveals dataset-specific hallucination patterns, with CMD showing highest rates. 'Cannot Find Symbol' (CFS) errors dominate across models: 86.47\% (GPT-3.5-Turbo), 76.99\% (GPT-4), 72.41\% (Mistral 7B), and 67.91\% (Mixtral 8x7B), indicating generation of non-existent symbols. 'Package Does Not Exist' (PDNE) errors are significant in Mixtral 8x7B (22.35\%) and Mistral 7B (23.13\%), reflecting incorrect imports.

\vspace{2mm}
\begin{table*}[ht]
\centering
\caption{Compilation Errors by Model and Dataset (All Prompting Techniques)}
\label{tab:compilation-errors-model-dataset}
\scalebox{0.37}{
\begin{tabular}{llcccccccccccccccccccccc}
\toprule
\multicolumn{1}{c}{\textbf{Model}} & \multicolumn{1}{c}{\textbf{Dataset}} & \multicolumn{1}{c}{\textbf{AR}} & \multicolumn{1}{c}{\textbf{ATM}} & \multicolumn{1}{c}{\textbf{CA}} & \multicolumn{1}{c}{\textbf{CFS}} & \multicolumn{1}{c}{\textbf{CR}} & \multicolumn{1}{c}{\textbf{CAM}} & \multicolumn{1}{c}{\textbf{DCD}} & \multicolumn{1}{c}{\textbf{ITB}} & \multicolumn{1}{c}{\textbf{IT}} & \multicolumn{1}{c}{\textbf{MAM}} & \multicolumn{1}{c}{\textbf{MRS}} & \multicolumn{1}{c}{\textbf{MNA}} & \multicolumn{1}{c}{\textbf{PDNE}} & \multicolumn{1}{c}{\textbf{PAI}} & \multicolumn{1}{c}{\textbf{PCR}} & \multicolumn{1}{c}{\textbf{SCI}} & \multicolumn{1}{c}{\textbf{UE}} & \multicolumn{1}{c}{\textbf{UCL}} & \multicolumn{1}{c}{\textbf{USL}} & \multicolumn{1}{c}{\textbf{UET}} & \multicolumn{1}{c}{\textbf{VNI}} & \multicolumn{1}{c}{\textbf{WAP}}  \\
\midrule

\multirow{3}{*}{GPT 3.5-Turbo}  
& Defects4J & 0.45 & 0.0 & 0.83 & 84.21 & 0 & 2.26 & 0.0 & 0.04 & 1.58 & 0.24 & 0.01 & 0 & 3.53 & 6.36 & 0.02 & 0 & 0.43 & 0.01 & 0 & 0.04 & 0 & 0 \\
& SF110 & 0.45 & 0.0 & 9.05 & 72.40 & 0 & 2.53 & 0.01 & 0.01 & 1.83 & 0.31 & 0.01 & 0 & 1.48 & 11.67 & 0.04 & 0.01 & 0.19 & 0 & 0 & 0 & 0 & 0 \\
& CMD & 0.21 & 0 & 0 & 86.48 & 0 & 0.46 & 0 & 0.03 & 0.51 & 1.39 & 0 & 0.00 & 7.87 & 2.96 & 0.00 & 0.00 & 0.08 & 0.00 & 0 & 0.01 & 0.00 & 0.00 \\ 
\midrule

GPT 4 & CMD & 0.21 & 0 & 0 & 76.99 & 0 & 0.24 & 0 & 0.05 & 1.08 & 3.03 & 0 & 0.00 & 16.46 & 1.50 & 0.11 & 0.01 & 0.30 & 0.01 & 0 & 0.00 & 0.01 & 0.00 \\
\midrule

\multirow{3}{*}{Mistral 7B}  
& Defects4J & 0.31 & 0.06 & 0.34 & 81.03 & 0.01 & 1.27 & 0.0 & 0.10 & 5.11 & 0.73 & 0.00 & 0.01 & 6.63 & 3.68 & 0.04 & 0.02 & 0.17 & 0.43 & 0.03 & 0.02 & 0 & 0 \\
& SF110 & 0.86 & 0.06 & 0.35 & 82.0 & 0 & 1.84 & 0.0 & 0.01 & 5.18 & 0.97 & 0.00 & 0 & 2.81 & 5.56 & 0.02 & 0.03 & 0.16 & 0.15 & 0.01 & 0 & 0 & 0 \\
& CMD & 0.29 & 0 & 0 & 72.41 & 0 & 0.71 & 0 & 0.09 & 0.86 & 1.82 & 0.02 & 23.13 & 0.61 & 0.04 & 0.00 & 0.00 & 0.00 & 0.00 & 0 & 0.00 & 0.00 & 0.01 \\ 
\midrule

Mixtral 8x7B & CMD & 0.31 & 0 & 0 & 67.91 & 0 & 0.76 & 0 & 0.38 & 1.59 & 4.77 & 0.00 & 22.35 & 1.81 & 0.04 & 0.02 & 0.05 & 0.00 & 0.00 & 0 & 0.00 & 0.00 & 0.00 \\
\bottomrule
\end{tabular}
}
\begin{tablenotes}
\tiny
\item[1]$^*$ AR: Ambiguous Reference, ATM: Array Type Mismatch, CA: Cannot Access, CFS: Cannot Find Symbol, CR: Class Redefinition, CAM: Constructor Argument Mismatch, DCD: Duplicate Class Declaration, ITB: Incompatible Type Bound, IT: Incompatible Types, MAM: Method Argument Mismatch, MRS: Missing Return Statement, MNA: Modifier Not Allowed, PDNE: Package Does Not Exist, PAI: Private Access Issue, PCR: Public Class Redefinition, SCI: Static Context Issue, UE: Unchecked Exception, UCL: Unclosed Character Literal, USL: Unclosed String Literal, UET: Unused Exception in Try, VNI: Variable Not Initialized, WAP: Weaker Access Privileges.
\end{tablenotes}
\vspace{2mm}
\end{table*}

Defects4J reveals encapsulation issues with high CFS (GPT-3.5: 80.57\%, Mistral 7B: 82.12\%) and PAI errors (GPT-3.5: 7.88\%, Mistral 7B: 4.74\%), suggesting access permission hallucinations. SF110 shows redundancy problems with prevalent CFS (GPT-3.5: 81.02\%, Mistral 7B: 82.42\%) and PAI peaking at 11.76\% with GPT-3.5 Few-shot. PCR and DCD errors indicate LLMs generate redundant test structures.
Our findings confirm ~\citet{zhang2024llm}'s conclusions that LLM hallucinations occur due to missing context and poor dependency handling, while extending their work by demonstrating these same patterns specifically in test generation contexts. The prevalence of CFS errors (up to 86.47\% in GPT-3.5-Turbo) confirms their observation about fabricated references, while PDNE errors (up to 23.13\% in Mistral 7B) extend their findings by revealing the particular challenges of dependency resolution in test suite generation.

\vspace{0.5em}
\noindent\colorbox{gray!20}{{\parbox{0.98\linewidth}{ \textbf{Finding 11:} LLM hallucinations are a major source of compilation errors. CMD is dominated by hallucinated missing symbols, Defects4J struggles with encapsulation violations, and SF110 suffers from redundant test generation. 
}}}
\vspace{0.5em}

\noindent\textbf{\underline{Prompt Engineering and Hallucination Mitigation.}}

\begin{figure*}[ht]
   \vspace{-3mm}
    \centering
    \subfigure[{\fontsize{6pt}{10pt}\selectfont By GPT 3.5-Turbo and by prompt}]{
        \includegraphics[width=5.6cm]
        {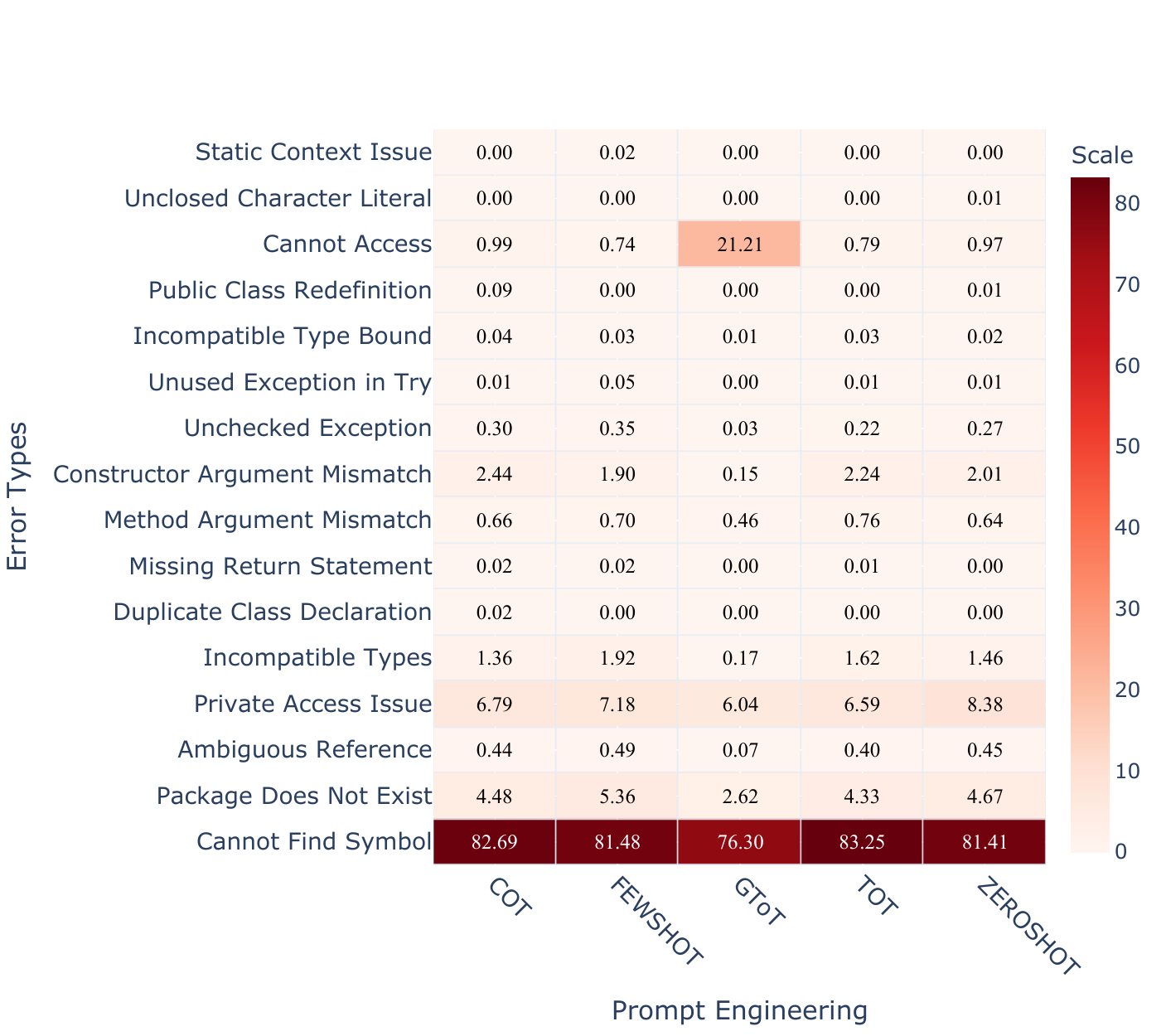}
        \label{fig:compilation-error-heatmap-gpt3}
    }
    \subfigure[{\fontsize{6pt}{10pt}\selectfont By Mistral 7B and by prompt engineering}]{
        \includegraphics[width=5.6cm]
        {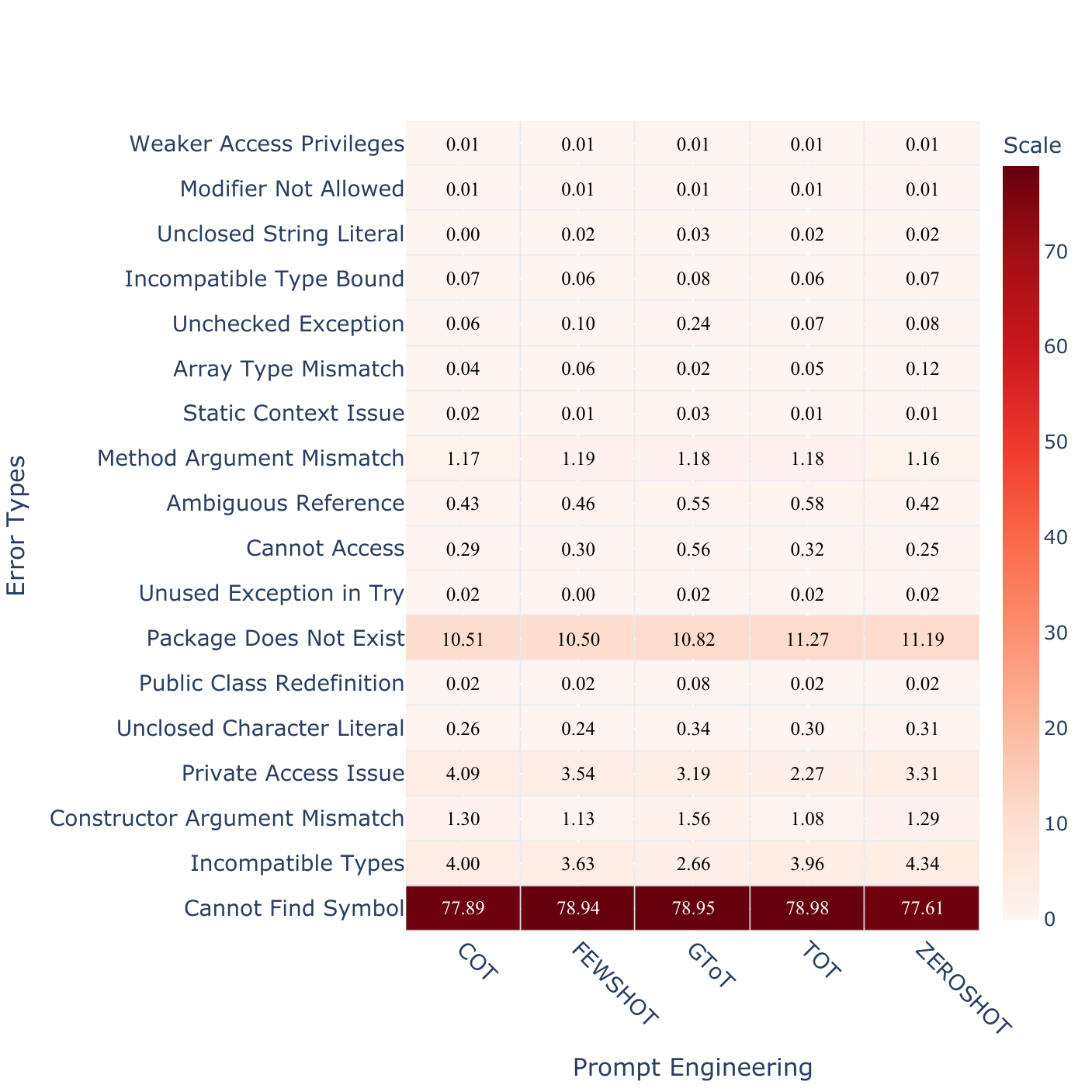}
        \label{fig:compilation-error-heatmap-mistral7b}
    }

      \vspace{-2mm}
      
    \subfigure[{\fontsize{6pt}{10pt}\selectfont By GPT 4 and by prompt}]{
        \includegraphics[width=5.6cm]
        {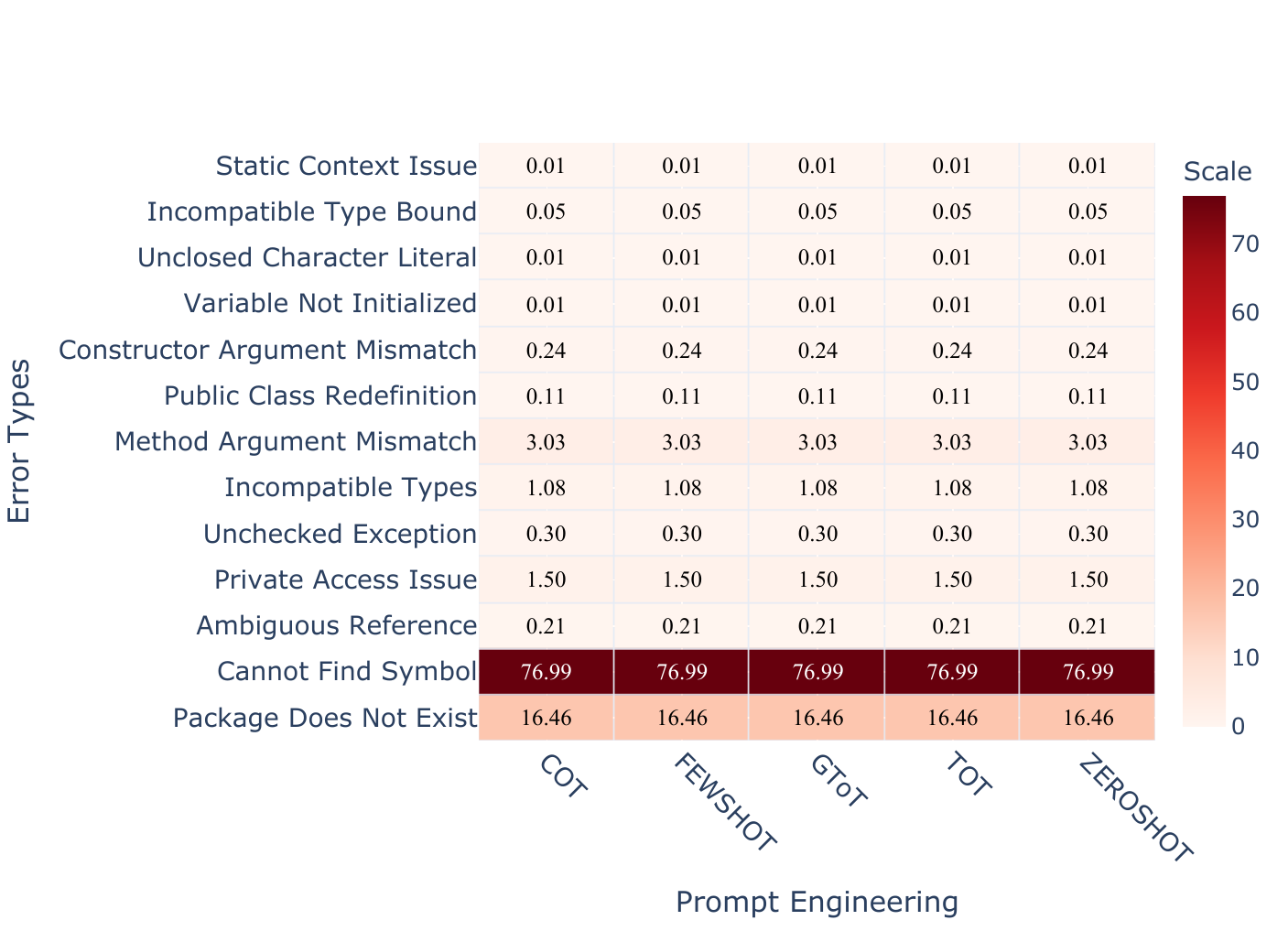}
        \label{fig:compilation-error-heatmap-gpt4}
    }  
    \subfigure[{\fontsize{6pt}{10pt}\selectfont By Mixtral 8x7B and by prompt engineering}]{
        \includegraphics[width=5.6cm]
        {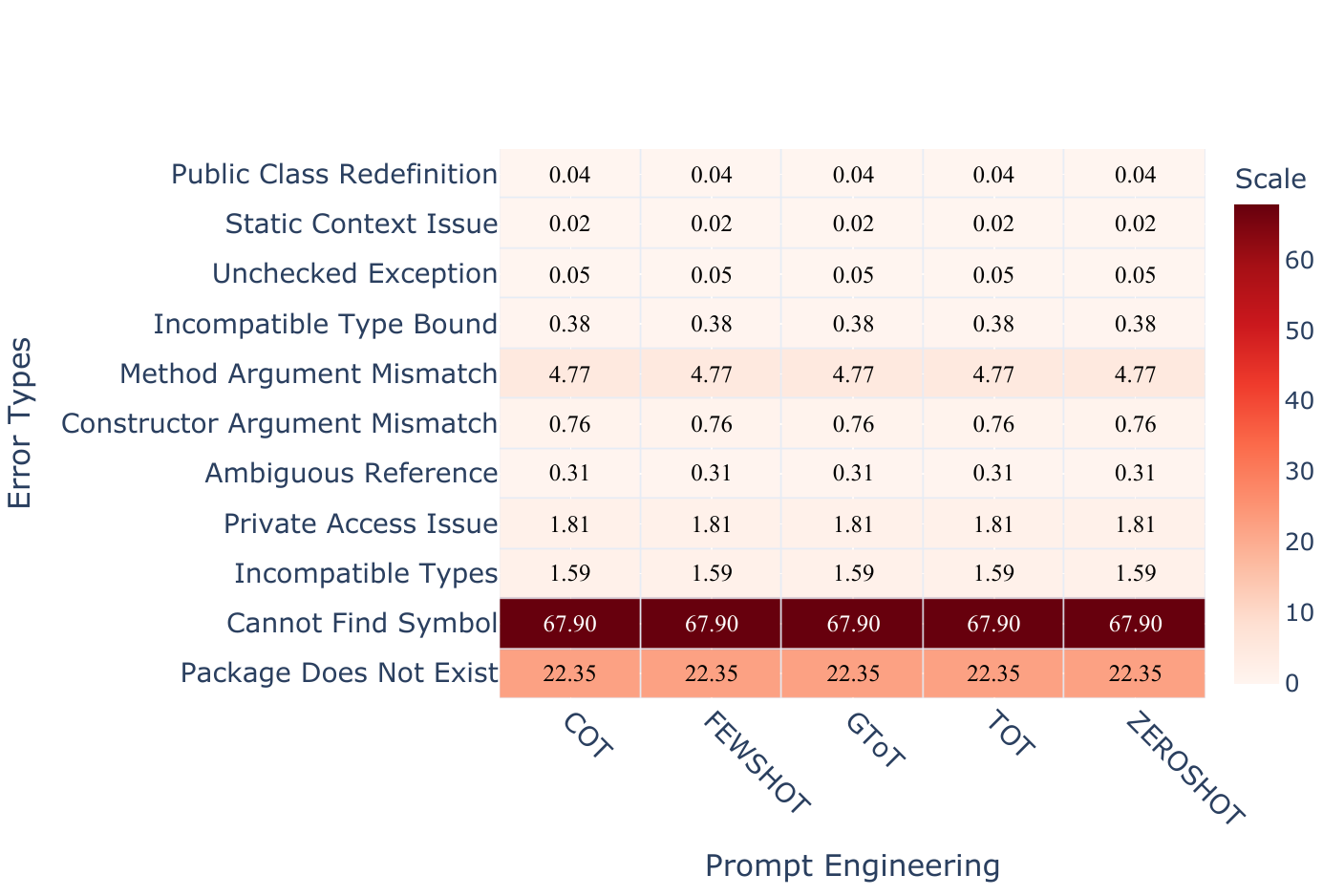}
        \label{fig:compilation-error-heatmap-mixtral}
    }  
\caption{Heatmap of compilation errors by model and prompt engineering.}
\label{fig:compilation-error-heatmap} 
\vspace{2mm}
\end{figure*}

Figure~\ref{fig:compilation-error-heatmap} reveals the distribution of non-zero error types across different machine learning models and prompt engineering techniques, with the y-axis dynamically adapting to display error types specific to each model and prompt configuration. The heatmap (Figure~\ref{fig:compilation-error-heatmap}) shows CoT and GToT reduce hallucinations more effectively than ZSL and FSL. CoT lowers 'Private Access Issue' (PAI) errors to 2.96\% (GPT-3.5) and 1.50\% (GPT-4) in CMD, improving encapsulation adherence. GToT nearly eliminates errors in Defects4J through structured dependencies. However, hallucinations in symbol resolution and redundant test generation persist despite structured reasoning.
%
%
%

FSL increases PAI errors to 11.31\% (GPT-3.5) and 5.64\% (Mistral 7B) in SF110, showing persistent access restriction misinterpretations. 'Cannot Find Symbol' (CFS) errors remain high (86.47\% in CMD, 79.16\% in SF110), indicating examples don't resolve symbol hallucinations. ZSL performs worst with CFS errors reaching 86.47\% (CMD) and 78.89\% (SF110) due to minimal context. Our findings extend ~\citet{zhang2024llm, zhang2025citywalk} by demonstrating how reasoning-based prompting like CoT and GToT can reduce certain error types—particularly PAI and CAM—by reinforcing syntactic scaffolding. While reasoning-based prompting reduces errors, challenges in reference resolution, encapsulation, and redundant test generation remain.

\vspace{0.5em}
\noindent\colorbox{gray!20}{{\parbox{0.98\linewidth}{ \textbf{Finding 12:} Reasoning-based prompting (CoT, GToT) significantly reduces errors, particularly Private Access Issues, whereas Zero-shot Learning exacerbates compilation failures by generating unstructured, non-compilable test cases. 
}}}
\vspace{0.5em}

\paragraph{\textbf{Relation to prior work.}}

Prior studies report higher LLM test compilability, but typically use configurations that simplify the task through granularity restrictions, import handling, or repair. \citet{siddiq2024using} generate one class per focal method and apply deterministic clean-up heuristics (stripping prose, fixing packages, bracket trimming), substantially inflating post-fix compilability; their raw SF110 compilability is very low with coverage under 2\% without repairs. \citet{yang2024evaluation} normalize imports and recursively prune failing test methods until classes compile, measuring coverage only on compilable residue; they attribute low validity to unresolved symbols, parameter mismatches, and abstract instantiation. \citet{tang2024chatgpt} achieve ~70\% runnable tests after light IDE fixes with a single prompt, yet EvoSuite dominates coverage. \citet{yuan2024evaluating} document modest raw ChatGPT compilability (57.9\% compilation errors) but large gains with iterative compile-feedback (ChatTester).
We evaluate full classes without import injection or post-hoc repair, yielding lower compilability but exposing true failure mode distributions—including occasional EvoSuite failures on Defects4J. Our RQ2 confirms and refines prior taxonomies from \citet{yang2024evaluation}: beyond unresolved symbols and abstract instantiation, we distinguish cannot-find-symbol, missing-package, private-access, and redundant-constructors, quantifying how their prevalence shifts across CMD, Defects4J, and SF110. Uniquely, we analyze these errors under reasoning-oriented prompting, observing that such prompts reduce specific error classes (e.g., private-access violations) while leaving others (e.g., unresolved symbols) largely unaffected. Our findings extend broader code-generation hallucination accounts to unit-test generation, showing prompt design can selectively mitigate validity failures.

\vspace{0.5em}
\noindent\highlight{%
Summary of \textbf{RQ2:}
LLMs generate a high proportion of syntactically correct tests ($\approx$90\%), yet compilability remains low ($<$10\%) due to unresolved symbols, missing packages, and access violations. EvoSuite, while more reliable, also exhibits rare compilation failures ($\approx$1.9\% in Defects4J), showing that errors are not unique to LLMs. Prompt engineering plays a critical role: reasoning-based prompts (CoT, GToT) improve structural correctness and reduce certain hallucinations, whereas FSL consistently underperforms. Compared to prior studies that operate at the method level with imports and repairs, our class-level, no-repair setup highlights the additional challenges of generating fully compilable suites directly from raw LLM outputs.
}

\vspace{-2mm}
\subsection{[RQ3]: Static Analysis of Human-Oriented Quality}
\noindent\textbf{[Experiment Goal]:} 
This research question evaluates the quality of LLM-generated test suites from a static analysis perspective. 
Unlike EvoSuite, which primarily targets structural coverage, here the focus is on human-oriented dimensions such as coding standard compliance, maintainability (via cyclomatic and cognitive complexity), and potential defects. 
By analyzing these aspects, we assess whether prompt design can improve the readability and long-term usability of automatically generated tests.

\noindent\textbf{[Experiment Design]:} 
We evaluate test quality using three approaches. Checkstyle assesses adherence to Google Java Style Checkstyle and Sun Code Conventions, measuring formatting consistency and standard violations. PMD analyzes cyclomatic complexity (independent paths) and cognitive complexity (comprehension ease)~\citep{dantas2021readability}. SpotBugs detects potential defects categorized by severity: Scariest (critical issues), Scary (high-risk errors), Troubling (moderate concerns), and Of Concern (minor warnings). We compared prompting techniques (ZSL, FSL, CoT, ToT, GToT) across SF110, Defects4J, and CMD datasets.

\noindent\textbf{[Experiment Results]:}

\begin{figure*}[ht]
\vspace{-3mm}
    \centering
    \subfigure[{\fontsize{6pt}{10pt}\selectfont Heatmap of Google Code Style Violations}]{
        \includegraphics[width=5.6cm]
        {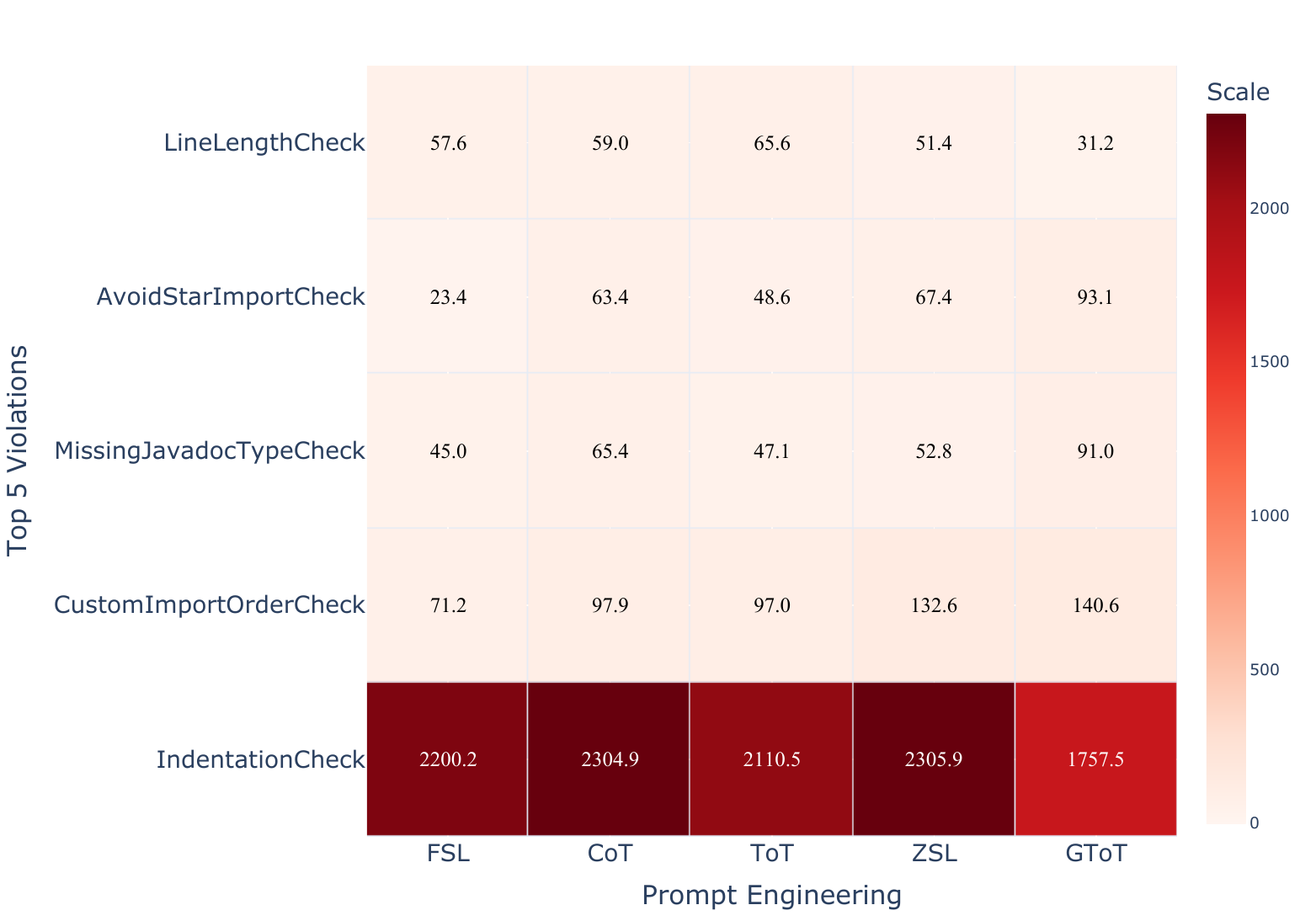}
        \label{fig:google-heatmap}
    }
    \subfigure[{\fontsize{6pt}{10pt}\selectfont Heatmap of Sun Code Style Violations}]{
        \includegraphics[width=5.6cm]
        {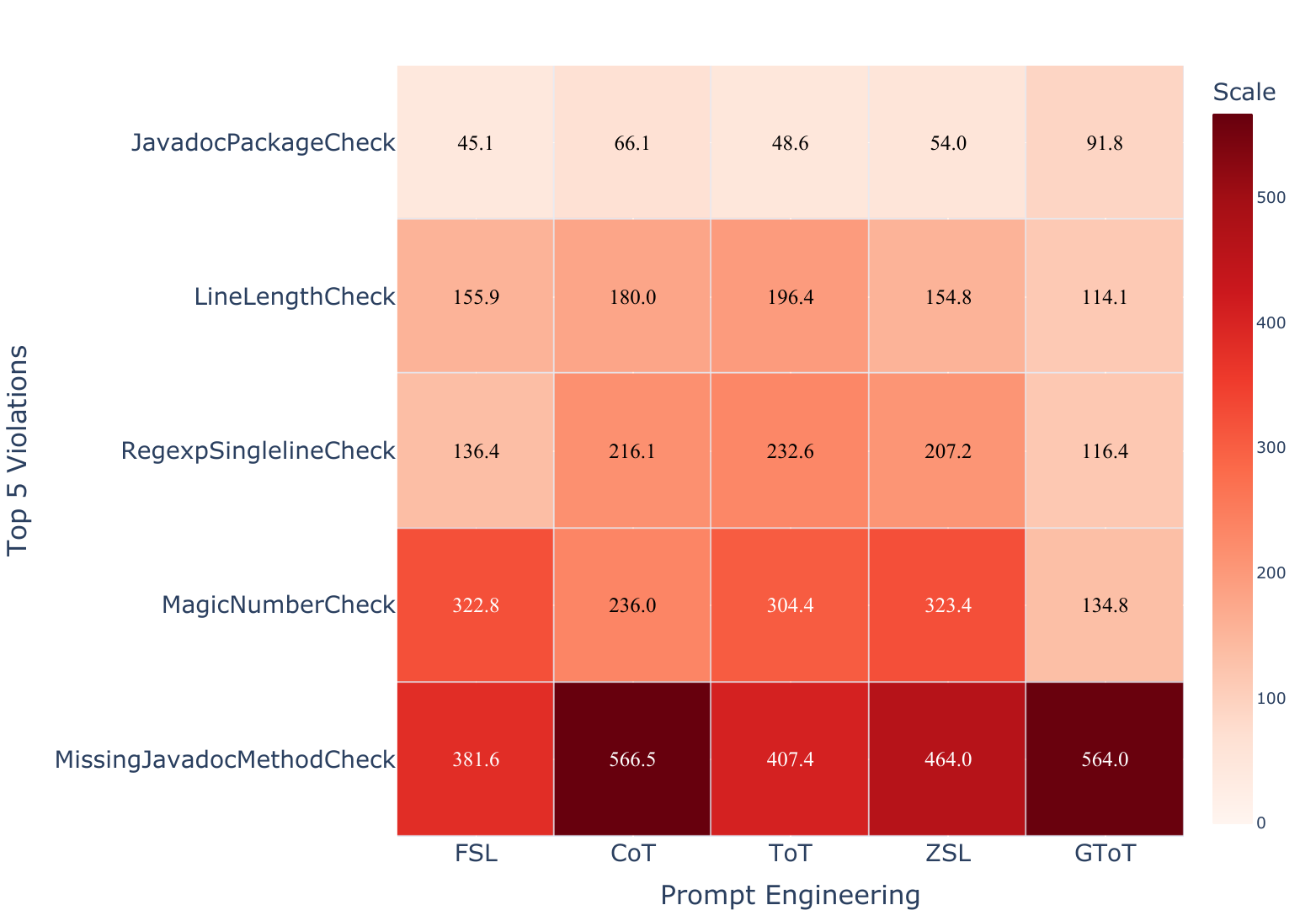}
        \label{fig:sun-heatmap}
    }  
\caption{Heatmap of the Top-5 Google and Sun Code Style Violations Across Prompt Engineering}
\label{fig:style-heatmap} 
\vspace{2mm}
\end{figure*}

\noindent\textbf{\underline{Coding Standard Compliance.}} 
LLM-generated test suites show varying coding standard adherence (Figure~\ref{fig:style-heatmap}). FSL achieves highest compliance with Google Java Style and Sun Code Conventions, producing fewer violations overall.
For Google Java Style (Figure~\ref{fig:google-heatmap}), indentation violations dominate across all prompting techniques, with GToT showing the lowest count (1757.5 vs. 2305.9 for ZSL). However, GToT increases import-related violations (140.6 CustomImportOrder and 93.1 AvoidStarImport violations) compared to FSL (71.3 and 23.4 respectively). For Sun Code Conventions (Figure~\ref{fig:sun-heatmap}), MissingJavadocMethod violations are highest with CoT (566.5) and GToT (564.0), while GToT significantly reduces MagicNumber violations (134.8 vs. 323.4 for ZSL). These patterns suggest that reasoning-based prompts (CoT, ToT, GToT) improve code structure but sacrifice documentation and import organization. FSL strikes the best balance, minimizing both structural and documentation violations. These findings partially align with ~\citet{tang2024chatgpt}, who reported high ZSL violations, though our results show FSL and CoT significantly improve compliance while ToT and GToT produce tests with better logical structure but less style adherence.

\vspace{0.5em}
\noindent\colorbox{gray!20}{{\parbox{0.98\linewidth}{
\textbf{Finding 13:}
FSL achieves the highest coding standard compliance, followed by CoT, which also reduces violations and improves adherence to style guidelines. In contrast, GToT enhances logical organization but introduces more violations, reinforcing the trade-off between structured prompting and strict adherence to coding conventions.
}}}
\vspace{0.5em}

\noindent\textbf{\underline{Code Complexity Analysis.}} 
Table~\ref{tab:understanding-complexity} shows that all prompting strategies produce test suites with very low structural complexity: more than 99.5\% of outputs fall into the lowest cyclomatic and cognitive complexity bands. GToT shows slightly higher proportions of simple cases (99.99\% cyclomatic, 100\% cognitive) and is the only strategy that avoids all moderate or high cases in cognitive complexity while nearly eliminating them in cyclomatic complexity (0.01\% moderate, 0\% high).
Although these differences are marginal in absolute terms, they point to a stabilizing effect of GToT, reducing structural anomalies rather than changing the overall average complexity.

\vspace{2mm}
\begin{table}[!h]
  \centering
  \caption{Summary of Complexity Results by Prompt Engineering Techniques\tnote{1}}
  \label{tab:understanding-complexity}
  \begin{minipage}[t]{0.48\textwidth}
    \centering
    \caption*{(a) Cyclomatic Complexity Results}
    \scalebox{0.5}{
      \begin{tabular}{lccccc}
        \toprule
        \multicolumn{1}{c}{\textbf{\begin{tabular}[c]{@{}c@{}}Cyclomatic\\ Complexity (\%)\end{tabular}}} & \textbf{ZSL} & \textbf{FSL} & \textbf{CoT} & \textbf{ToT} & \textbf{GToT} \\
        \midrule
        Low (1-4)       & 99.86 & 99.50 & 99.73 & 99.73 & 99.99 \\
        Moderate (5-7)  & 0.11  & 0.39  & 0.16  & 0.20  & 0.01  \\
        High (8-10)     & 0.03  & 0.11  & 0.11  & 0.07  & 0     \\
        Very High (11+) & 0     & 0     & 0     & 0     & 0     \\
        \bottomrule
      \end{tabular}
    }
  \end{minipage}
  \hfill
  \begin{minipage}[t]{0.48\textwidth}
    \centering
    \caption*{(b) Cognitive Complexity Results}
    \scalebox{0.5}{
      \begin{tabular}{lccccc}
        \toprule
        \multicolumn{1}{c}{\textbf{\begin{tabular}[c]{@{}c@{}}Cognitive\\ Complexity (\%)\end{tabular}}} & \textbf{ZSL} & \textbf{FSL} & \textbf{CoT} & \textbf{ToT} & \textbf{GToT} \\
        \midrule
        Low (<5)       & 99.83 & 99.76 & 99.76 & 99.90 & 100   \\
        Moderate (5-10)& 0.11  & 0.14  & 0.14  & 0.03  & 0     \\
        High (11-20)   & 0.03  & 0.10  & 0.11  & 0.07  & 0     \\
        Very High (21+)& 0.03  & 0     & 0     & 0     & 0     \\
        \bottomrule
      \end{tabular}
    }
  \end{minipage}
  \begin{tablenotes}
    \item[1] \parbox{\textwidth}{\tiny ZSL: Zero-shot learning, FSL: Few-shot learning, CoT: Chain-of-Thought, ToT: Tree-of-Thoughts, GToT: Guided Tree-of-Thoughts.}
  \end{tablenotes}
\end{table}

These findings refine the observations of \citet{tang2024chatgpt}, who showed that ZSL already produces simple tests, by highlighting that reasoning-based prompting can maintain this simplicity while reducing occasional outlier cases.

\vspace{0.5em}
\noindent\colorbox{gray!20}{{\parbox{0.98\linewidth}{
\textbf{Finding 14:}
All prompting strategies yield predominantly low-complexity test suites.
GToT achieves slightly higher proportions of simple outputs and is the only strategy with zero moderate or high cases in cognitive complexity, and nearly none in cyclomatic complexity.
While differences across strategies are small, GToT appears to offer more structural consistency by reducing complexity outliers.
}}}
\vspace{0.5em}

\noindent\textbf{\underline{Static Analysis with SpotBugs.}} 

Table~\ref{tab:spotbug-priority-level-prompt} reports the distribution of static analysis warnings (SpotBugs) across 20,530 compilable test suites. While most defects are classified as minor \quotes{Of Concern} (55.6\%), we observe notable differences across prompting strategies. FSL produces the most \quotes{Scariest} defects (e.g., 410 on SF110), while GToT yields the highest number of \quotes{Scary} (825) and \quotes{Troubling} (825) issues, particularly on SF110. These results highlight a concerning trade-off: although GToT improves structure and readability, it tends to introduce more subtle but structurally embedded defects. ZSL, while producing simpler tests, yields far fewer severe warnings. Unlike prior work~\citep{tang2024chatgpt}, which focused on ZSL with limited compilation filtering, our evaluation reveals how reasoning-oriented prompting can unintentionally lead to more severe code issues, emphasizing the need for robust post-generation validation.

\begin{table*}[!htbp]
\centering
\caption{Bug Priority Levels by Prompt Techniques}
\label{tab:spotbug-priority-level-prompt}
\scalebox{0.5}{
\begin{tabular}{ll|ccc|c|ccc|c}
\toprule
\multirow{2}{*}{\textbf{Prompt}} & \multirow{2}{*}{\textbf{Bug Level}} 
& \multicolumn{3}{c|}{\textbf{GPT-3.5}} 
& \textbf{GPT-4} 
& \multicolumn{3}{c|}{\textbf{Mistral7B}} 
& \textbf{Mixtral8x7B} \\
\cmidrule(lr){3-5} \cmidrule(lr){6-6} \cmidrule(lr){7-9} \cmidrule(lr){10-10}
 & & SF110 & D4J & CMD & CMD & SF110 & D4J & CMD & CMD \\
\midrule

\multirow{4}{*}{ZSL} 
& Of Concern   & 600 & 212 & 17 & 15 & 129 & 48 & 0 & 2 \\
& Scariest     & 0   & 0   & 0  & 0  & 129 & 48 & 0 & 0 \\
& Scary        & 0   & 0   & 0  & 0  & 129 & 0  & 0 & 0 \\
& Troubling    & 0   & 212 & 0  & 15 & 0   & 0  & 0 & 0 \\
\midrule

\multirow{4}{*}{FSL} 
& Of Concern   & 410 & 177 & 18 & 14 & 135 & 60 & 0 & 0 \\
& Scariest     & 410 & 177 & 0  & 0  & 135 & 60 & 0 & 0 \\
& Scary        & 0   & 0   & 0  & 0  & 0   & 0  & 0 & 0 \\
& Troubling    & 0   & 0   & 0  & 14 & 0   & 0  & 0 & 0 \\
\midrule

\multirow{4}{*}{CoT} 
& Of Concern   & 553 & 219 & 22 & 0  & 125 & 32 & 0 & 0 \\
& Scariest     & 0   & 0   & 0  & 0  & 0   & 32 & 0 & 0 \\
& Scary        & 0   & 0   & 0  & 0  & 125 & 0  & 0 & 0 \\
& Troubling    & 0   & 0   & 0  & 0  & 125 & 0  & 0 & 0 \\
\midrule

\multirow{4}{*}{ToT} 
& Of Concern   & 20  & 21  & 3  & 7  & 212 & 63 & 1 & 0 \\
& Scariest     & 0   & 1   & 0  & 0  & 212 & 63 & 0 & 0 \\
& Scary        & 1   & 1   & 0  & 0  & 212 & 0  & 0 & 0 \\
& Troubling    & 0   & 2   & 0  & 1  & 0   & 0  & 0 & 0 \\
\midrule

\multirow{4}{*}{GToT} 
& Of Concern   & 825 & 242 & 20 & 19 & 347 & 131 & 0 & 4 \\
& Scariest     & 0   & 0   & 0  & 0  & 0   & 0   & 0 & 0 \\
& Scary        & 825 & 0   & 0  & 0  & 0   & 0   & 0 & 0 \\
& Troubling    & 825 & 0   & 0  & 0  & 0   & 0   & 0 & 0 \\
\bottomrule
\end{tabular}
}

\tiny
\noindent$^*$ ZSL = Zero-shot Learning, FSL = Few-shot Learning, CoT = Chain-of-Thought, ToT = Tree-of-Thought, GToT = Guided Tree-of-Thought. 
SpotBugs ranks issue severity as follows: \textbf{Scariest} (most critical), \textbf{Scary}, \textbf{Troubling}, and \textbf{Of Concern} (least critical).

\vspace{1mm}

\end{table*}

\vspace{0.5em}
\noindent\colorbox{gray!20}{{\parbox{0.98\linewidth}{
\textbf{Finding 15:}
Prompting strategies impact bug prevalence differently. FSL introduces the most critical (\quotes{Scariest}) issues, while GToT yields the highest number of structurally embedded \quotes{Scary} and \quotes{Troubling} bugs, particularly on SF110. ZSL produces the fewest severe warnings, albeit with simpler outputs. These results call for stricter validation when using reasoning-based prompting.
}}}

\paragraph{\textbf{Relation to prior work.}}

\citet{tang2024chatgpt} analyze human-oriented quality through Checkstyle, PMD, and SpotBugs, showing that ChatGPT-generated tests under zero-shot prompting often violate style guidelines but remain structurally simple, with low cyclomatic and cognitive complexity. \citet{yuan2024evaluating} complement this perspective with a user study, where developers judged LLM-generated tests comparable to manual ones in terms of usability and adoption effort. By contrast, neither \citet{yang2024evaluation} nor \citet{siddiq2024using} evaluate readability or maintainability, focusing instead on correctness and coverage. Our study extends this line of work by systematically comparing prompting strategies (ZSL, FSL, CoT, ToT, GToT) and showing how they shift coding standard compliance, complexity, and defect incidence. Unlike prior studies limited to zero-shot prompting, we demonstrate that reasoning-based prompts can improve logical organization at the cost of higher defect density, while FSL maximizes style compliance. This nuanced trade-off analysis situates prompt engineering as a key determinant of human-oriented quality in LLM-generated tests, beyond what was captured in earlier evaluations.

\vspace{0.5em}
\noindent\highlight{%
Summary of \textbf{RQ3:}
LLM-generated tests generally exhibit low complexity and readable structures, consistent with prior reports. However, coding-standard compliance and defect incidence vary strongly with prompting strategies. FSL yields the cleanest style adherence, while reasoning-based prompts (CoT, ToT, GToT) enhance logical organization but introduce more subtle bugs flagged by SpotBugs. Compared to prior work limited to zero-shot prompting, our analysis shows that prompt engineering shifts the balance between readability, maintainability, and defect-proneness, underscoring its central role in shaping human-oriented quality.}

\vspace{-3mm}
\subsection{[RQ4]: Readability of LLM-Generated Test Suites: Alignment with Developer Expectations}
\noindent\textbf{[Experiment Goal]:} 
%
This experiment investigates whether LLM-generated test suites meet developer expectations of readability, focusing on clarity, structure, and ease of navigation. We use the readability model by \citet{scalabrino2018comprehensive}, which has been validated against human judgments, to quantify readability beyond surface-level formatting. EvoSuite is included here as a baseline reference: although it does not explicitly optimize for readability, its widespread use in SBST makes it a valuable point of comparison to assess whether LLM prompting strategies deliver tangible improvements in human-oriented quality.

\noindent\textbf{[Experiment Design]:}
We employ the readability model of \citet{scalabrino2018comprehensive}, trained on Java code including JUnit tests and validated against human judgments. The model evaluates syntactic structure, identifier clarity, textual coherence, and overall structural readability, thus going beyond superficial formatting checks. All analyses are restricted to compilable test suites, ensuring that readability is measured on executable outputs. 
The evaluation is conducted in three complementary stages. First, a dataset-wise analysis: we aggregate readability scores by dataset (Defects4J, SF110, CMD) for GPT-3.5-turbo and Mistral 7B. EvoSuite is also included on Defects4J and SF110, where it was executable, but excluded from CMD due to Java version incompatibilities (Section~\ref{subsec:dataset}). This view highlights how readability varies with dataset characteristics. Second, a prompt-wise analysis: we aggregate results by prompting strategy (ZSL, FSL, CoT, ToT, GToT) for GPT-3.5-turbo and Mistral 7B, and contrast them with EvoSuite on Defects4J and SF110. Although EvoSuite does not explicitly optimize for readability, its inclusion provides a widely used SBST baseline against which to assess the added value of prompting strategies.  Finally, for CMD we restrict the analysis to LLMs only, since EvoSuite could not be executed due to Java version incompatibilities. All LLMs are evaluated across the five prompting strategies, but readability scores are reported only for configurations that yielded compilable test suites.
For every configuration (dataset-wise, prompt-wise, CMD-only), readability distributions are summarized using descriptive statistics (minimum, quartiles, median, maximum, mean). Higher scores indicate more structured, navigable, and human-readable test suites.

\noindent\textbf{[Experiment Results]:}

\noindent\textbf{\underline{Dataset-Level Readability Analysis.}} 

Table~\ref{tab:dataset-gpt-mistral} shows LLM-generated test suites consistently achieve higher readability than EvoSuite across all datasets. Readability varies with codebase complexity: Defects4J exhibits highest readability scores (GPT-3.5-Turbo: 81.73\%, Mistral 7B: 69.06\% vs. EvoSuite: 57.33\%), followed by SF110 (GPT-3.5-Turbo: 75.95\%, Mistral 7B: 70.28\% vs. EvoSuite: 50.67\%). CMD presents lowest scores (GPT-3.5-Turbo: 60.93\%, Mistral 7B: 57.13\%), indicating modern, evolving codebases challenge LLMs more. While LLMs consistently outperform EvoSuite in readability, their effectiveness varies with dataset complexity.

\vspace{0.5em}
\noindent\colorbox{gray!20}{{\parbox{0.98\linewidth}{
\textbf{Finding 16:}
 LLM-generated tests outperform EvoSuite in readability. However, readability varies with dataset complexity, as shown in Table~\ref{tab:dataset-gpt-mistral}. CMD, which contains modern and evolving codebases, presents the greatest challenge, yielding the lowest readability scores across all models.
}}}
\vspace{0.5em}

\begin{table*}[ht]
\centering
\caption{Statistical Metrics from Readability Score Distribution by Model and Dataset}
\label{tab:dataset-gpt-mistral}
\scalebox{0.5}{
\begin{tabular}{llcccccc}
\toprule
\textbf{Model} & \textbf{Dataset} & \textbf{Min} & \textbf{Q1} & \textbf{Median} & \textbf{Q3} & \textbf{Max} & \textbf{Mean} \\
\midrule
\multirow{3}{*}{GPT 3.5-turbo} & Defects4J & 33.35 & 70.53 & 84.03 & 95.74 & 99.04 & 81.73 \\
                                & SF110     & 25.02 & 66.32 & 76.12 & 87.10 & 98.64 & 75.95 \\
                                & CMD       & 41.34 & 55.53 & 61.30 & 65.98 & 90.45 & 60.93 \\
\midrule
\multirow{3}{*}{Mistral7B}     & Defects4J & 15.20 & 61.23 & 68.44 & 78.03 & 97.58 & 69.06 \\
                                & SF110     & 6.58  & 60.19 & 71.04 & 80.99 & 98.59 & 70.28 \\
                                & CMD       & 13.60 & 38.68 & 52.68 & 84.24 & 95.07 & 57.13 \\
\midrule
\multirow{2}{*}{EvoSuite}      & Defects4J & 16.10 & 44.67 & 56.04 & 72.04 & 95.70 & 57.33 \\
                                & SF110     & 16.33 & 36.83 & 50.03 & 61.81 & 89.67 & 50.67 \\
\bottomrule
\end{tabular}
}
\begin{tablenotes}
\tiny
\item[1] {\bf Notice.} Values are aggregated over all prompting strategies that produced compilable test suites.
\end{tablenotes}
\vspace{3mm}
\end{table*}

\noindent\textbf{\underline{LLM vs. EvoSuite: Readability Performance.}} 

Table~\ref{tab:prompt-vs-evosuite} shows LLM-generated tests consistently surpass EvoSuite in readability. For GPT-3.5-Turbo, GToT achieves the highest score (84.75\% vs. EvoSuite's 50.77\%), followed by CoT (82.77\%) and ToT (73.32\%), while ZSL (77.93\%) and FSL (73.49\%) score lower. Similarly, for Mistral 7B, GToT leads (76.35\%), followed by ToT (69.64\%), with ZSL (63.79\%) and FSL (60.40\%) producing lowest scores. These results demonstrate reasoning-based prompting's importance for readability.

\vspace{0.5em}
\noindent\colorbox{gray!20}{{\parbox{0.98\linewidth}{
\textbf{Finding 17:}
LLM-generated test suites improve readability over EvoSuite by 21–40\%, as shown in Table~\ref{tab:prompt-vs-evosuite}. This highlights the significant advantage of LLMs in producing more comprehensible test code.
}}}

\vspace{2mm}
\begin{table}[ht]
\captionsetup{font=footnotesize, labelfont=bf}
\centering
\caption{Statistical Metrics from Readability Score Distribution (Defects4J and SF110)}
\label{tab:prompt-vs-evosuite}
\begin{minipage}{0.48\textwidth}%
    \centering
    \caption*{\textbf{(a)} GPT 3.5-turbo by Prompt vs EvoSuite}
    \scalebox{0.55}{
    \begin{tabular}{lccccccc}
    \toprule
    \textbf{Prompt} & \textbf{Min} & \textbf{Q1} & \textbf{Median} & \textbf{Q3} & \textbf{Max} & \textbf{Mean} \\
    \midrule
    ZSL       & 29.37 & 66.79 & 77.50 & 93.50 & 98.41 & 77.93 \\
    FSL       & 25.02 & 60.24 & 74.36 & 88.07 & 97.78 & 73.49 \\
    CoT       & 43.28 & 73.17 & 85.74 & 95.33 & 99.04 & 82.77 \\
    ToT       & 33.36 & 64.33 & 72.04 & 82.64 & 97.83 & 73.32 \\
    GToT      & 42.24 & 75.20 & 89.39 & 96.26 & 98.64 & 84.75 \\
    \midrule
    EvoSuite & 4.11 & 36.96  & 51.02 & 62.59 & 97.64 & 50.77 \\
    \bottomrule
    \end{tabular}
    }
\end{minipage}
\hfill
\begin{minipage}{0.48\textwidth}%
    \centering
    \caption*{\textbf{(b)} Mistral7B by Prompt vs EvoSuite}
    \scalebox{0.55}{
    \begin{tabular}{lccccccc}
    \toprule
    \textbf{Prompt} & \textbf{Min} & \textbf{Q1} & \textbf{Median} & \textbf{Q3} & \textbf{Max} & \textbf{Mean} \\
    \midrule
    ZSL      & 25.67 & 57.52 & 64.41 & 71.87 & 90.62 & 63.79 \\
    FSL      & 6.58  & 49.03 & 62.53 & 70.78 & 97.58 & 60.40 \\
    CoT      & 34.54 & 57.67 & 65.39 & 73.92 & 97.77 & 65.92 \\
    ToT      & 33.30 & 60.94 & 69.00 & 78.29 & 97.67 & 69.64 \\
    GToT     & 15.20 & 67.47 & 77.24 & 89.02 & 98.59 & 76.35 \\
    \midrule
    EvoSuite & 4.11 & 36.96 & 51.02 & 62.59 & 97.64 & 50.77 \\
    \bottomrule
    \end{tabular}
    }
\end{minipage}
\end{table}
\vspace{2mm}

\noindent\textbf{\underline{Prompting Strategies: Influence on Readability.}} 

Table~\ref{tab:cmd-prompts} shows reasoning-based prompting significantly enhances test suite clarity. GToT and CoT consistently achieve highest readability scores across datasets. For GPT-3.5-Turbo on CMD, CoT leads (67.99\%), followed by GToT (61.86\%), while ZSL (57.67\%) and FSL (55.76\%) score lower. Similarly, for GPT-4 on CMD, GToT achieves highest readability (67.71\%). These results confirm minimal prompting provides insufficient guidance for well-structured tests, while reasoning-based approaches are essential for producing maintainable test suites.

\vspace{0.1cm}
\begin{table}[ht]
\captionsetup{font=footnotesize, labelfont=bf}
\centering
\caption{Statistical Metrics from Readability Score Distribution on CMD}
\label{tab:cmd-prompts}
\begin{minipage}{0.48\textwidth}
    \centering
    \caption*{\textbf{(a)} GPT 3.5-turbo on CMD}
    \scalebox{0.6}{
    \begin{tabular}{lccccccc}
    \toprule
    \textbf{Prompt} & \textbf{Min} & \textbf{Q1} & \textbf{Median} & \textbf{Q3} & \textbf{Max} & \textbf{Mean} \\
    \midrule
    ZSL  & 41.34 & 53.19 & 57.95 & 63.26 & 69.17 & 57.67 \\
    FSL  & 42.30 & 49.14 & 55.80 & 61.70 & 66.07 & 55.76 \\
    CoT  & 45.92 & 62.59 & 66.37 & 71.95 & 90.45 & 67.99 \\
    ToT  & 47.74 & 57.77 & 62.09 & 64.61 & 71.10 & 61.38 \\
    GToT & 52.07 & 57.29 & 61.02 & 67.15 & 70.12 & 61.86 \\
    \bottomrule
    \end{tabular}
    }
\end{minipage}
\hfill
\begin{minipage}{0.48\textwidth}
    \centering
    \caption*{\textbf{(b)} GPT 4 on CMD}
    \scalebox{0.6}{
    \begin{tabular}{lccccccc}
    \toprule
    \textbf{Prompt} & \textbf{Min} & \textbf{Q1} & \textbf{Median} & \textbf{Q3} & \textbf{Max} & \textbf{Mean} \\
    \midrule
    ZSL   & 36.69 & 47.61 & 50.20 & 57.81 & 71.66 & 52.13 \\
   FSL   30.14 & 47.15 & 50.56 & 56.03 & 71.80 & 51.62 \\
    CoT   & 51.51 & 57.83 & 61.42 & 66.12 & 75.20 & 61.64 \\
    ToT   & 54.02 & 59.94 & 62.86 & 66.51 & 73.04 & 63.18 \\
    GToT  & 58.10 & 63.35 & 68.85 & 71.08 & 76.13 & 67.71 \\
    \bottomrule
    \end{tabular}
    }
\end{minipage}
\begin{minipage}{0.48\textwidth}
    \centering
    \caption*{\textbf{(c)} Mistral7B on CMD}
    \scalebox{0.6}{
    \begin{tabular}{lccccccc}
    \toprule
    \textbf{Prompt} & \textbf{Min} & \textbf{Q1} & \textbf{Median} & \textbf{Q3} & \textbf{Max} & \textbf{Mean} \\
    \midrule
    ZSL   & -- & -- & -- & -- & -- & -- \\
    FSL   & 35.18 & 35.18 & 35.18 & 35.18 & 35.18 & 35.18 \\
    
    CoT   & -- & -- & -- & -- & -- & -- \\
    ToT   & 49.18 & 49.18 & 49.18 & 49.18 & 49.18 & 49.18 \\
    GToT  & 13.60 & 45.53 & 74.89 & 93.97 & 95.07 & 64.61 \\
    \bottomrule
    \end{tabular}
    }
\end{minipage}
\hfill
\begin{minipage}{0.48\textwidth}
    \centering
    \caption*{\textbf{(d)} Mixtral8x7B on CMD}
    \scalebox{0.6}{
    \begin{tabular}{lccccccc}
    \toprule
    \textbf{Prompt} & \textbf{Min} & \textbf{Q1} & \textbf{Median} & \textbf{Q3} & \textbf{Max} & \textbf{Mean} \\
    \midrule
    ZSL   & 38.43 & 44.17 & 49.91 & 55.65 & 61.39 & 49.91 \\
    FSL   & 48.36 & 48.36 & 48.36 & 48.36 & 48.36 & 48.36 \\
    CoT   & 46.46 & 51.88 & 57.29 & 62.71 & 68.12 & 57.29 \\
    ToT   & 57.14 & 77.90 & 79.08 & 92.66 & 92.85 & 79.93 \\
    GToT  & 44.32 & 58.67 & 75.75 & 89.30 & 93.00 & 72.21 \\
    \bottomrule
    \end{tabular}
    }
\end{minipage}
\begin{tablenotes}
\item[1]$^*$ \tiny Since only compilable code was considered, there was no compilable code for Mistral 7B's CoT and ZSL prompts.
\end{tablenotes}
\vspace{2mm}
\end{table}

\vspace{0.5em}
\noindent\colorbox{gray!20}{{\parbox{0.98\linewidth}{
\textbf{Finding 18:}
CoT and GToT consistently enhance readability, even on complex datasets like CMD, where unstructured prompting results in lower scores.
}}}

\noindent\textbf{\underline{Model-Specific Readability Trends.}} 

Table~\ref{tab:cmd-prompts} reveals how different LLMs respond to reasoning-based prompting. GPT-3.5-Turbo performs best with CoT (67.99\%), while GPT-4 excels with GToT (67.71\%), confirming larger models benefit from structured reasoning. Mistral 7B shows higher variability, reaching 64.61\% with GToT, indicating sensitivity to prompt design. Mixtral 8x7B performs exceptionally with ToT (79.93\%), though its limited sample size affects stability.

\vspace{0.5em}
\noindent\colorbox{gray!20}{{\parbox{0.98\linewidth}{
\textbf{Finding 19:}
Model architecture, parameter size, and prompting strategy significantly impact readability, with larger models consistently improving when guided by reasoning-based prompting.
}}}

\paragraph{\textbf{Relation to prior work.}}
Readability from a developer perspective has been explicitly studied by \citet{yuan2024evaluating}, who conducted a user study showing that ChatGPT-generated tests were rated comparable to manual ones in terms of readability and adoption effort. Other evaluations, such as \citet{tang2024chatgpt}, focus on understandability proxies (e.g., style violations, complexity) rather than direct readability. Our study extends this line of work by systematically quantifying readability with an established model validated against human judgments, and by comparing LLMs with EvoSuite. We further show how prompting strategies (CoT, ToT, GToT) significantly enhance readability beyond zero-shot and few-shot settings, demonstrating that structured reasoning prompts align LLM outputs more closely with developer expectations.

\vspace{0.5em}
\noindent\highlight{%
Summary of \textbf{RQ4:}
LLM-generated tests consistently achieve higher readability than EvoSuite, with gains of 21–40\% across datasets. Readability varies with dataset complexity, being highest on Defects4J and lowest on CMD. Reasoning-based prompting (CoT, GToT) markedly improves clarity and navigability, while zero-shot and few-shot remain weaker. Compared to prior work, which assessed readability through targeted developer studies, our study provides a systematic, model-based evaluation showing that prompting strategies play a central role in aligning LLM outputs with developer expectations.
}

\subsection{[RQ5]: Code Coverage.}
\label{sec:RQ5}
\noindent\textbf{[Experiment Goal]:} 
This experiment evaluates LLM-generated test suite effectiveness in achieving structural coverage. We measure line, instruction, and method coverage, comparing LLMs against EvoSuite and examining how prompting strategies impact coverage. Method coverage is particularly important in our class-level evaluation to determine cross-method interaction rather than isolated functionality testing.

\noindent\textbf{[Experiment Design]:} 
We use JaCoCo\footnote{\url{https://www.jacoco.org/jacoco/}} to measure line coverage (executed source lines), instruction coverage (executed bytecode instructions), and method coverage (methods invoked at least once). Following \citet{tang2024chatgpt}, branch coverage is excluded as it focuses only on conditional branches and often yields missing values. Coverage is computed exclusively on compilable test suites (for LLMs: code that parsed and compiled; for EvoSuite: compiled outputs only).
To make comparisons both fair and informative, we report coverage along two complementary views. 
First, a prompt-wise view on the common dataset pool: for GPT-3.5-turbo and Mistral 7B, coverage distributions produced by each prompting strategy (ZSL, FSL, CoT, ToT, GToT) are aggregated over SF110 and Defects4J—the datasets where EvoSuite is available—and contrasted with the single EvoSuite baseline on the same pool. (EvoSuite has no prompting; it serves as a fixed reference distribution.)
Second, a dataset-wise view for LLMs: we provide per-dataset coverage distributions on SF110, Defects4J, and CMD to expose context sensitivity across benchmarks.
Model–dataset availability follows our experimental constraints (Section~\ref{subsec:dataset}). GPT-3.5-turbo and Mistral 7B were run on all three datasets, whereas GPT-4 and Mixtral 8$\times$7B were restricted to CMD due to budget constraints. EvoSuite could not be executed on CMD because of Java version incompatibilities. Consequently, CMD analyses are LLM-only; within CMD, we report prompt-wise results for GPT-3.5-turbo and GPT-4. Mixtral 8$\times$7B is omitted from CMD coverage summaries due to negligible compilability, yielding insufficient samples for meaningful statistics.
For every configuration (prompt-wise or dataset-wise), we summarize the resulting coverage distributions using standard descriptive statistics (minimum, quartiles, median, maximum, mean) to capture both central tendency and dispersion.

\noindent\textbf{[Experiment Results]:} 

\noindent\textbf{\underline{EvoSuite vs. LLMs: Overall Coverage Performance.}}

Tables~\ref{tab:line-cmd-prompts}, \ref{tab:instruction-cmd-prompts}, and \ref{tab:method-cmd-prompts} reveal a significant performance gap between LLMs and EvoSuite. EvoSuite consistently outperforms LLMs across all metrics, achieving median coverage of 94.34\% (line), 95.0\% (instruction), and 100\% (method) through search-based optimization. LLM-generated tests show higher variability, with reasoning-based prompting significantly improving coverage over ZSL and FSL. GPT-3.5-Turbo with CoT achieves the highest LLM median coverage (75.47\% line, 76.0\% instruction, 100\% method), though all LLM configurations fall short of EvoSuite's systematic exploration capabilities.
%
\vspace{0.1cm}
\begin{table}[ht]
\captionsetup{font=footnotesize, labelfont=bf}
\centering
\caption{Statistical metrics from Line coverage distribution.}
\label{tab:line-cmd-prompts}
\begin{minipage}{0.48\textwidth}
    \centering
    \caption*{\textbf{(a)} EvoSuite vs GPT 3.5-turbo.}
    \scalebox{0.6}{
    \begin{tabular}{lccccccc}
    \toprule
    \textbf{Prompt} & \textbf{Min} & \textbf{Q1} & \textbf{Median} & \textbf{Q3} & \textbf{Max} & \textbf{Mean} \\
    \midrule
    ZSL  & 0.0 & 1.49 & 27.68 & 100.0 & 100.0 & 46.3 \\
    FSL  & 0.0 & 1.73 & 16.14 & 100.0 & 100.0 & 41.2 \\
    CoT  & 0.0 & 3.33 & 75.47 & 100.0 & 100.0 & 55.56 \\
    ToT  & 0.0 & 1.82 & 24.39 & 93.55 & 100.0 & 41.96 \\ 
    GToT & 0.0 & 4.65 & 41.54 & 100.0 & 100.0 & 46.7 \\
    \midrule
    EvoSuite & 0.0 & 88.61 & 94.34 & 97.67 & 100.0 & 91.75 \\
    \bottomrule
    \end{tabular}
    }
\end{minipage}
\hfill
\begin{minipage}{0.48\textwidth}
    \centering
    \caption*{\textbf{(b)} EvoSuite vs Mistral 7B.}
    \scalebox{0.6}{
    \begin{tabular}{lccccccc}
    \toprule
    \textbf{Prompt} & \textbf{Min} & \textbf{Q1} & \textbf{Median} & \textbf{Q3} & \textbf{Max} & \textbf{Mean} \\
    \midrule
    ZSL  & 0.0 & 0.0 & 9.2 & 67.27 & 100.0 & 31.51 \\ 
    FSL  & 0.0 & 1.97 & 18.45 & 63.04 & 100.0 & 31.58 \\
    CoT  & 0.0 & 0.0 & 1.12 & 65.22 & 100.0 & 30.28 \\
    ToT  & 0.0 & 1.2 & 20.0 & 76.44 & 100.0 & 37.0 \\
    GToT  & 0.0 & 4.55 & 30.0 & 83.89 & 100.0 & 42.41 \\
     \midrule
    EvoSuite  & 0.0 & 88.61 & 94.34 & 97.67 & 100.0 & 91.75 \\ 
    \bottomrule
    \end{tabular}
    }
\end{minipage}
\begin{minipage}{0.48\textwidth}
    \centering
    \caption*{\textbf{(c)} GPT 3.5-turbo by datasets.}
    \scalebox{0.6}{
    \begin{tabular}{lccccccc}
    \toprule
    \textbf{Dataset} & \textbf{Min} & \textbf{Q1} & \textbf{Median} & \textbf{Q3} & \textbf{Max} & \textbf{Mean} \\
    \midrule
     Defects4J  & 0.0 & 10.22 & 82.74 & 100.0 & 100.0 & 62.61 \\
     SF110  & 0.0 & 0.0 & 3.57 & 23.24 & 100.0 & 22.72 \\ 
     CMD  & 4.0 & 53.59 & 100.0 & 100.0 & 100.0 & 76.86 \\
    \bottomrule
    \end{tabular}
    }
\end{minipage}
\hfill
\begin{minipage}{0.48\textwidth}
    \centering
    \caption*{\textbf{(d)} Mistral 7B by datasets.}
    \scalebox{0.6}{
    \begin{tabular}{lccccccc}
    \toprule
    \textbf{Dataset} & \textbf{Min} & \textbf{Q1} & \textbf{Median} & \textbf{Q3} & \textbf{Max} & \textbf{Mean} \\
    \midrule
    Defects4J  & 0.0 & 4.38 & 63.4 & 89.47 & 100.0 & 52.55 \\
    SF110  & 0.0 & 0.0 & 3.03 & 25.0 & 100.0 & 15.15 \\
    CMD  & 0.0 & 0.0 & 0.0 & 0.0 & 0.0 & 0.0 \\
    \bottomrule
    \end{tabular}
    }
\end{minipage}
\hfill
\begin{minipage}{0.48\textwidth}
    \centering
    \caption*{\textbf{(e)} GPT 3.5-turbo on CMD.}
    \scalebox{0.6}{
    \begin{tabular}{lccccccc}
    \toprule
    \textbf{Prompt} & \textbf{Min} & \textbf{Q1} & \textbf{Median} & \textbf{Q3} & \textbf{Max} & \textbf{Mean} \\
    \midrule
    ZSL  & 4.0 & 22.88 & 61.29 & 92.31 & 100.0 & 56.05 \\ 
    FSL  & 8.33 & 30.0 & 83.33 & 100.0 & 100.0 & 67.78 \\
    CoT  & 8.33 & 92.05 & 100.0 & 100.0 & 100.0 & 84.53 \\
    ToT  & 31.82 & 73.39 & 100.0 & 100.0 & 100.0 & 85.84 \\
    GToT  & 31.82 & 67.8 & 100.0 & 100.0 & 100.0 & 84.27 \\
    \bottomrule
    \end{tabular}
    }
\end{minipage}
\hfill
\begin{minipage}{0.48\textwidth}
    \centering
    \caption*{\textbf{(f)} GPT 4 on CMD.}
    \scalebox{0.6}{
    \begin{tabular}{lccccccc}
    \toprule
    \textbf{Prompt} & \textbf{Min} & \textbf{Q1} & \textbf{Median} & \textbf{Q3} & \textbf{Max} & \textbf{Mean} \\
    \midrule
    ZSL  & 0.0 & 0.0 & 8.33 & 70.97 & 89.47 & 31.88 \\ 
    FSL  & 5.56 & 12.89 & 52.51 & 85.94 & 100.0 & 52.74 \\ 
    CoT  & 0.0 & 18.68 & 65.22 & 100.0 & 100.0 & 60.35 \\
    ToT  & 0.0 & 28.25 & 72.22 & 100.0 & 100.0 & 63.82 \\
    GToT  & 0.0 & 31.61 & 72.22 & 100.0 & 100.0 & 64.01 \\
    \bottomrule
    \end{tabular}
    }
\end{minipage}
\end{table}
\vspace{2mm}

\vspace{0.5em}
\noindent\colorbox{gray!20}{{\parbox{0.98\linewidth}{
\textbf{Finding 20:}
EvoSuite outperforms all LLMs in coverage metrics. However, reasoning-based techniques (CoT, GToT, ToT) significantly enhance LLM test effectiveness, reducing the performance gap.
}}}
\vspace{0.5em}

\noindent\textbf{\underline{Dataset-Level Coverage Trends.}} 

Tables~\ref{tab:line-cmd-prompts},~\ref{tab:instruction-cmd-prompts}, and~\ref{tab:method-cmd-prompts} show coverage varies significantly across datasets. Defects4J achieves highest coverage (GPT-3.5-Turbo: 82.74\% line, 85.0\% instruction, 100\% method), suggesting its structured design facilitates test generation with clearer functional boundaries. In contrast, SF110 reports much lower coverage (3.57\% line, 2.0\% instruction, 16.67\% method) due to higher complexity and diverse architectures.
CMD reports highest median coverage (100\% for some prompts), though this likely reflects its smaller size and modular structure rather than superior LLM capabilities. The contrast between datasets highlights structural influences: Defects4J benefits from potential training exposure (\citet{sallou2023breaking}), clearer functional boundaries, and modularized code, while SF110 presents challenges through higher complexity and lower public accessibility.

\vspace{0.5em}
\noindent\colorbox{gray!20}{{\parbox{0.98\linewidth}{
\textbf{Finding 21:}
LLMs achieve higher coverage on well-documented datasets like Defects4J but struggle with complex projects like SF110, where greater structural variability makes test generation more challenging. CMD’s high coverage is likely due to its small, modular dataset rather than an inherent LLM advantage.
}}}
\vspace{0.5em}

%
\vspace{0.1cm}
\begin{table}[ht]
\captionsetup{font=footnotesize, labelfont=bf}
\centering
\caption{Statistical metrics from Instruction coverage distribution.}
\label{tab:instruction-cmd-prompts}
\begin{minipage}{0.48\textwidth}
    \centering
    \caption*{\textbf{(a)} EvoSuite vs GPT 3.5-turbo.}
    \scalebox{0.6}{
    \begin{tabular}{lccccccc}
    \toprule
    \textbf{Prompt} & \textbf{Min} & \textbf{Q1} & \textbf{Median} & \textbf{Q3} & \textbf{Max} & \textbf{Mean} \\
    \midrule
    ZSL  & 0.0 & 1.0 & 29.0 & 100.0 & 100.0 & 46.26 \\
    FSL  & 0.0 & 1.0 & 14.0 & 100.0 & 100.0 & 41.72 \\
    CoT  & 0.0 & 3.0 & 76.0 & 100.0 & 100.0 & 55.64 \\
    ToT  & 0.0 & 1.0 & 27.0 & 96.0 & 100.0 & 42.91 \\
    GToT & 0.0 & 3.5 & 45.0 & 100.0 & 100.0 & 47.61 \\
    \midrule
    EvoSuite & 0.0 & 91.0 & 95.0 & 98.0 & 100.0 & 93.35 \\
    \bottomrule
    \end{tabular}
    }
\end{minipage}
\hfill
\begin{minipage}{0.48\textwidth}
    \centering
    \caption*{\textbf{(b)} EvoSuite vs Mistral 7B.}
    \scalebox{0.6}{
    \begin{tabular}{lccccccc}
    \toprule
    \textbf{Prompt} & \textbf{Min} & \textbf{Q1} & \textbf{Median} & \textbf{Q3} & \textbf{Max} & \textbf{Mean} \\
    \midrule
    ZSL  & 0.0 & 0.0 & 10.0 & 73.25 & 100.0 & 33.63 \\
    FSL  & 0.0 & 1.0 & 18.0 & 60.5 & 100.0 & 32.48 \\
    CoT  & 0.0 & 0.0 & 0.0 & 83.0 & 100.0 & 31.19 \\
    ToT  & 0.0 & 0.0 & 17.0 & 81.0 & 100.0 & 37.49 \\
    GToT & 0.0 & 3.25 & 26.0 & 87.25 & 100.0 & 42.31 \\
     \midrule
    EvoSuite & 0.0 & 91.0 & 95.0 & 98.0 & 100.0 & 93.35 \\ 
    \bottomrule
    \end{tabular}
    }
\end{minipage}
\begin{minipage}{0.48\textwidth}
    \centering
    \caption*{\textbf{(c)} GPT 3.5-turbo by datasets.}
    \scalebox{0.6}{
    \begin{tabular}{lccccccc}
    \toprule
    \textbf{Dataset} & \textbf{Min} & \textbf{Q1} & \textbf{Median} & \textbf{Q3} & \textbf{Max} & \textbf{Mean} \\
    \midrule
     Defects4J  & 0.0 & 11.0 & 85.0 & 100.0 & 100.0 & 63.14 \\
     SF110  & 0.0 & 0.0 & 2.0 & 24.5 & 100.0 & 22.58 \\ 
     CMD  & 2.0 & 48.25 & 100.0 & 100.0 & 100.0 & 76.56 \\ 
    \bottomrule
    \end{tabular}
    }
\end{minipage}
\hfill
\begin{minipage}{0.48\textwidth}
    \centering
    \caption*{\textbf{(d)} Mistral 7B by datasets.}
    \scalebox{0.6}{
    \begin{tabular}{lccccccc}
    \toprule
    \textbf{Dataset} & \textbf{Min} & \textbf{Q1} & \textbf{Median} & \textbf{Q3} & \textbf{Max} & \textbf{Mean} \\
    \midrule
    Defects4J  & 0.0 & 2.25 & 66.0 & 93.0 & 100.0 & 53.62 \\
    SF110  & 0.0 & 0.0 & 2.0 & 22.0 & 100.0 & 15.37 \\
    CMD  & 0.0 & 0.0 & 0.0 & 0.0 & 0.0 & 0.0 \\
    \bottomrule
    \end{tabular}
    }
\end{minipage}
\hfill
\begin{minipage}{0.48\textwidth}
    \centering
    \caption*{\textbf{(e)} GPT 3.5-turbo on CMD.}
    \scalebox{0.6}{
    \begin{tabular}{lccccccc}
    \toprule
    \textbf{Prompt} & \textbf{Min} & \textbf{Q1} & \textbf{Median} & \textbf{Q3} & \textbf{Max} & \textbf{Mean} \\
    \midrule
     ZSL  & 2.0 & 22.5 & 60.0 & 91.0 & 100.0 & 55.27 \\
     FSL  & 9.0 & 31.0 & 81.0 & 100.0 & 100.0 & 67.18 \\
     CoT  & 7.0 & 93.75 & 100.0 & 100.0 & 100.0 & 84.65 \\
     ToT  & 28.0 & 77.25 & 100.0 & 100.0 & 100.0 & 85.65 \\
     GToT & 28.0 & 69.75 & 100.0 & 100.0 & 100.0 & 84.06 \\
    \bottomrule
    \end{tabular}
    }
\end{minipage}
\hfill
\begin{minipage}{0.48\textwidth}
    \centering
    \caption*{\textbf{(f)} GPT 4 on CMD.}
    \scalebox{0.6}{
    \begin{tabular}{lccccccc}
    \toprule
    \textbf{Prompt} & \textbf{Min} & \textbf{Q1} & \textbf{Median} & \textbf{Q3} & \textbf{Max} & \textbf{Mean} \\
    \midrule
    ZSL  & 0.0 & 0.0 & 6.0 & 61.0 & 95.0 & 31.46 \\
    FSL  & 3.0 & 15.5 & 51.5 & 82.25 & 100.0 & 51.92 \\
    CoT  & 0.0 & 18.0 & 61.0 & 100.0 & 100.0 & 61.05 \\
    ToT  & 0.0 & 29.0 & 79.5 & 100.0 & 100.0 & 65.19 \\
    GToT & 0.0 & 29.5 & 79.5 & 100.0 & 100.0 & 65.21 \\
    \bottomrule
    \end{tabular}
    }
\end{minipage}
\vspace{2mm}
\end{table}

While \citet{sallou2023breaking} attributes poor SF110 performance mainly to data accessibility, our analysis demonstrates SF110's inherent complexity complicates test generation (see Table~\ref{tab:dataset-overview-complexity} in  Section~\ref{subsec:dataset}). However, Defects4J's broad accessibility raises data leakage concerns, as LLMs may have encountered its test cases during pretraining. As shown in \citep{sallou2023breaking}, ChatGPT can retrieve detailed information about specific Defects4J bugs, suggesting performance may reflect memorization rather than genuine reasoning.

\vspace{0.5em}
\noindent\colorbox{gray!20}{{\parbox{0.98\linewidth}{
\textbf{Finding 22:}
Dataset characteristics influence LLM performance—Defects4J's high coverage likely stems from pretraining exposure while SF110's limited representation restricts generalization, highlighting the need for careful dataset selection to mitigate data leakage in test generation studies.
}}}
\vspace{0.5em}

\noindent\textbf{\underline{Impact of Reasoning-Based Prompting on Coverage.}} 

Tables~\ref{tab:line-cmd-prompts},~\ref{tab:instruction-cmd-prompts},~\ref{tab:method-cmd-prompts} show reasoning-based techniques (CoT, GToT, ToT) consistently improve coverage compared to ZSL and FSL.

For line coverage, GPT-3.5-Turbo with CoT achieves highest median (75.47\%), while ZSL and FSL produce significantly lower coverage (27.68\% and 16.14\%). Mistral 7B performs best with GToT (30.0\%), though still below GPT-3.5-Turbo.

For instruction coverage, EvoSuite maintains advantage (95.0\%). Among LLMs, GPT-3.5-Turbo with CoT (76.0\%) and GToT (45.0\%) lead, while Mistral 7B's best result with GToT (26.0\%) indicates smaller models' limitations.

For method coverage, GPT-3.5-Turbo with CoT reaches 100\% in some cases, while Mistral 7B improves with GToT (89.59\%) but still lags. These results demonstrate reasoning-based prompting enhances LLM performance significantly but doesn't fully bridge the gap with EvoSuite.

\vspace{0.1cm}
\begin{table}[ht]
\captionsetup{font=footnotesize, labelfont=bf}
\centering
\caption{Statistical metrics from Method coverage distribution.}
\label{tab:method-cmd-prompts}
\begin{minipage}{0.48\textwidth}
    \centering
    \caption*{\textbf{(a)} EvoSuite vs GPT 3.5-turbo.}
    \scalebox{0.6}{
    \begin{tabular}{lccccccc}
    \toprule
    \textbf{Prompt} & \textbf{Min} & \textbf{Q1} & \textbf{Median} & \textbf{Q3} & \textbf{Max} & \textbf{Mean} \\
    \midrule
    ZSL  & 0.0 & 6.25 & 83.33 & 100.0 & 100.0 & 56.95 \\
    FSL  & 0.0 & 7.14 & 66.67 & 100.0 & 100.0 & 55.31 \\
    CoT  & 0.0 & 14.29 & 100.0 & 100.0 & 100.0 & 65.18 \\
    ToT  & 0.0 & 8.33 & 66.67 & 100.0 & 100.0 & 55.31 \\
    GToT & 0.0 & 16.67 & 76.92 & 100.0 & 100.0 & 60.32 \\
    \midrule
    EvoSuite & 0.0 & 100.0 & 100.0 & 100.0 & 100.0 & 99.21 \\
    \bottomrule
    \end{tabular}
    }
\end{minipage}
\hfill
\begin{minipage}{0.48\textwidth}
    \centering
    \caption*{\textbf{(b)} EvoSuite vs Mistral 7B.}
    \scalebox{0.6}{
    \begin{tabular}{lccccccc}
    \toprule
    \textbf{Prompt} & \textbf{Min} & \textbf{Q1} & \textbf{Median} & \textbf{Q3} & \textbf{Max} & \textbf{Mean} \\
    \midrule
    ZSL  & 0.0 & 0.0 & 29.16 & 100.0 & 100.0 & 48.83 \\
    FSL  & 0.0 & 9.94 & 72.22 & 100.0 & 100.0 & 55.1 \\
    CoT  & 0.0 & 0.0 & 6.9 & 97.06 & 100.0 & 41.31 \\
    ToT  & 0.0 & 5.96 & 50.0 & 100.0 & 100.0 & 55.07 \\
    GToT & 0.0 & 17.5 & 89.59 & 100.0 & 100.0 & 62.56 \\
     \midrule
    EvoSuite & 0.0 & 100.0 & 100.0 & 100.0 & 100.0 & 99.21 \\ 
    \bottomrule
    \end{tabular}
    }
\end{minipage}
\begin{minipage}{0.48\textwidth}
    \centering
    \caption*{\textbf{(c)} GPT 3.5-turbo by datasets.}
    \scalebox{0.6}{
    \begin{tabular}{lccccccc}
    \toprule
    \textbf{Dataset} & \textbf{Min} & \textbf{Q1} & \textbf{Median} & \textbf{Q3} & \textbf{Max} & \textbf{Mean} \\
    \midrule
     Defects4J  & 0.0 & 33.33 & 100.0 & 100.0 & 100.0 & 73.86 \\
     SF110  & 0.0 & 0.0 & 16.67 & 89.44 & 100.0 & 35.17 \\
     CMD  & 20.0 & 100.0 & 100.0 & 100.0 & 100.0 & 89.35 \\
    \bottomrule
    \end{tabular}
    }
\end{minipage}
\hfill
\begin{minipage}{0.48\textwidth}
    \centering
    \caption*{\textbf{(d)} Mistral 7B by datasets.}
    \scalebox{0.6}{
    \begin{tabular}{lccccccc}
    \toprule
    \textbf{Dataset} & \textbf{Min} & \textbf{Q1} & \textbf{Median} & \textbf{Q3} & \textbf{Max} & \textbf{Mean} \\
    \midrule
    Defects4J  & 0.0 & 17.05 & 100.0 & 100.0 & 100.0 & 70.1 \\
    SF110  & 0.0 & 0.0 & 20.0 & 50.0 & 100.0 & 34.7 \\
    CMD  & 0.0 & 0.0 & 0.0 & 0.0 & 0.0 & 0.0 \\
    \bottomrule
    \end{tabular}
    }
\end{minipage}
\hfill
\begin{minipage}{0.48\textwidth}
    \centering
    \caption*{\textbf{(e)} GPT 3.5-turbo on CMD.}
    \scalebox{0.6}{
    \begin{tabular}{lccccccc}
    \toprule
    \textbf{Prompt} & \textbf{Min} & \textbf{Q1} & \textbf{Median} & \textbf{Q3} & \textbf{Max} & \textbf{Mean} \\
    \midrule
     ZSL  & 20.0 & 35.41 & 90.91 & 100.0 & 100.0 & 71.29 \\
     FSL  & 25.0 & 88.89 & 100.0 & 100.0 & 100.0 & 84.15 \\
     CoT  & 20.0 & 100.0 & 100.0 & 100.0 & 100.0 & 91.0 \\
     ToT  & 75.0 & 100.0 & 100.0 & 100.0 & 100.0 & 98.04 \\
     GToT & 75.0 & 100.0 & 100.0 & 100.0 & 100.0 & 97.82 \\
    \bottomrule
    \end{tabular}
    }
\end{minipage}
\hfill
\begin{minipage}{0.48\textwidth}
    \centering
    \caption*{\textbf{(f)} GPT 4 on CMD.}
    \scalebox{0.6}{
    \begin{tabular}{lccccccc}
    \toprule
    \textbf{Prompt} & \textbf{Min} & \textbf{Q1} & \textbf{Median} & \textbf{Q3} & \textbf{Max} & \textbf{Mean} \\
    \midrule
    ZSL  & 0.0 & 0.0 & 40.0 & 75.0 & 100.0 & 43.56 \\
    FSL  & 16.67 & 38.33 & 65.0 & 100.0 & 100.0 & 64.86 \\
    CoT  & 0.0 & 34.72 & 100.0 & 100.0 & 100.0 & 69.8 \\
    ToT  & 0.0 & 62.5 & 100.0 & 100.0 & 100.0 & 79.59 \\
    GToT & 0.0 & 70.46 & 100.0 & 100.0 & 100.0 & 80.25 \\
    \bottomrule
    \end{tabular}
    }
\end{minipage}
\vspace{2mm}
\end{table}

\vspace{0.5em}
\noindent\colorbox{gray!20}{{\parbox{0.98\linewidth}{
\textbf{Finding 23:}
Reasoning-based prompting significantly enhances LLM coverage across all datasets and models. However, smaller models like Mistral 7B exhibit greater limitations even with reasoning-based prompting.
}}}
\vspace{0.5em}

\paragraph{\textbf{Relation to prior work.}}

Coverage is central in prior LLM-for-testing evaluations. \citet{siddiq2024using} report high HumanEval coverage but extremely low SF110 coverage ($<\!2\%$), with gains contingent on strong post-fix heuristics. \citet{tang2024chatgpt} find ChatGPT achieves $\approx55\%$ statement coverage on a DynaMOSA-derived benchmark, trailing EvoSuite's $\approx74\%$ (30 runs). \citet{yang2024evaluation} confirm GPT-4 lags behind EvoSuite on Defects4J, arguing hallucination-driven invalidity is the primary barrier; moreover, directly adapting In-Context Learning  (ICL) techniques (CoT/RAG) from other tasks does not improve effectiveness and can reduce it for code-specialized LLMs. \citet{yuan2024evaluating} show iterative compile-feedback (ChatTester) substantially raises compilability/executability, narrowing the gap to manually written tests.

Our results both corroborate and extend these findings. First, evaluating at the class level across Defects4J, SF110, and CMD, we show dataset characteristics strongly shape LLM coverage: relatively higher on Defects4J (potential pretraining exposure), very low on SF110 (structural complexity), and inflated on CMD (small, modular projects). Second, we systematically study prompting strategies and observe reasoning-oriented prompts (e.g., CoT/ToT/GToT) consistently improve coverage—yet still fall short of EvoSuite's systematic exploration. This contrasts with \citet{yang2024evaluation}, where CoT and RAG—applied to code-specialized LLMs with a retrieval corpus not tailored to unit tests—are ineffective or detrimental. A plausible explanation: our prompts target developer-facing reasoning on general-purpose LLMs, whereas \citet{yang2024evaluation} adapt In-Context Learning methods from code-generation settings to code-specialized models. Overall, our results indicate coverage improvements from ICL are model- and prompt-design dependent: effective when reasoning aligns with test-generation needs, ineffective when naively ported.

\vspace{0.5em}
\noindent\highlight{%
Summary of \textbf{RQ5:}
EvoSuite consistently outperforms LLMs in structural coverage, achieving near-complete line, instruction, and method coverage. LLMs, however, display higher variability: while Defects4J shows competitive coverage (up to 82\% line coverage with GPT-3.5), SF110 highlights their limitations ($<5\%$) due to architectural complexity. CMD yields artificially high scores, reflecting dataset modularity rather than genuine reasoning ability. Reasoning-based prompting (CoT, ToT, GToT) narrows the gap by substantially improving LLM coverage, but systematic exploration remains a unique strength of SBST approaches like EvoSuite.}

\subsection{[RQ6]: Test Smell detection.}
\noindent\textbf{[Experiment Goal]:} Our aim is to evaluate the extent to which the test suites generated by LLMs exhibit test smells compared with those produced by an SBST method like EvoSuite.

\noindent\textbf{[Experiment Design]:} 
This experiment assesses test smell prevalence using TsDetect~\citep{peruma2020tsdetect}, which identifies 20 different smell types~\citep{peruma2019distribution,peruma2020tsdetect}. 
We record the proportion of test suites containing at least one instance of each smell type. These percentages reflect smell distribution rather than absolute counts, enabling comparative analysis of structural quality and maintainability.

\noindent\textbf{[Experiment Results]:} 

\noindent\textbf{\underline{Test Smell Distribution Across LLMs and EvoSuite.}}

Table~\ref{tab:test-smells} shows that Assertion Roulette (AR), Conditional Logic Test (CLT), and Magic Number Test (MNT) are the most prevalent smells in our setting. The near-ubiquity of MNT across models is consistent with \citet{siddiq2024using}, who analyze multiple smells over two benchmarks (HumanEval and SF110). Our AR variability (e.g., 73.15\% with GPT-3.5/FSL on Defects4J vs.\ 19.86\% with GToT on CMD) also falls within the ranges they report. For Exception Handling (EH), we observe higher rates in EvoSuite than in LLM outputs; this differs from patterns noted by \citet{siddiq2024using} and suggests that reasoning prompts, as part of our prompt engineering (CoT, ToT, GToT), can elicit more explicit exception modeling. We do not observe Default Test, Verbose Test, or Dependent Test in our corpus, consistent with the broader trend that LLM-generated tests tend to be structurally simpler.
Across prompt techniques, we see two systematic effects. First, under reasoning prompts (CoT/ToT/GToT), Empty Test (EM) rates increase on CMD and Defects4J, indicating that more structured prompting may improve formatting and organization yet coincide with shallower test bodies (cf.\ Finding~20). Second, EH tends to increase with reasoning prompts, pointing to more explicit exception scenarios. These patterns persist across model families and benchmarks; notably, MNT remains consistently dominant, with AR and CLT next in prevalence irrespective of the prompt technique.

\begin{table*}[]
\caption{Test smells distribution by Model, Prompt and by Datasets.}
\label{tab:test-smells}
\scalebox{0.44}
{
\begin{tabular}{llllllllllllllllllllllllcccccccccccccccccc}
\toprule
\multirow{2}{*}{\textbf{Model}}       & \multirow{2}{*}{\textbf{Dataset}}   & \multirow{2}{*}{\textbf{Prompt}} & \multicolumn{18}{c}{\textbf{Test Smell (\%)}} \\ 
\cmidrule{4-7} \cmidrule{8-21}
                             &                            &                         & \multicolumn{1}{l}{AR}     & \multicolumn{1}{l}{CLT}     & \multicolumn{1}{l}{CI}     & \multicolumn{1}{l}{EM} & \multicolumn{1}{l}{EH} & \multicolumn{1}{l}{GF} & \multicolumn{1}{l}{MG} & \multicolumn{1}{l}{RP} & \multicolumn{1}{l}{RA} & \multicolumn{1}{l}{SE} & \multicolumn{1}{l}{ST} & \multicolumn{1}{l}{EA} & \multicolumn{1}{l}{LT} & \multicolumn{1}{l}{DA} & \multicolumn{1}{l}{UT} & \multicolumn{1}{l}{IT} & \multicolumn{1}{l}{RO} & \multicolumn{1}{l}{MNT} \\ \hline
\multirow{15}{*}{GPT 3}      & \multirow{5}{*}{SF110}                      & ZSL                     & \multicolumn{1}{c}{46.17} & \multicolumn{1}{c}{2.68}   & \multicolumn{1}{c}{0.16}   & \multicolumn{1}{c}{12.71}   & 13.02 & 11.13 & 2.92 & 0.0 & 0.79 & 23.52 & 0.39 & 48.38 & 53.75 & 9.31 & 27.70 & 0.0 & 3.24 & 99.84 \\
                             &                            & FSL                     & \multicolumn{1}{c}{74.44} & \multicolumn{1}{c}{3.91} & \multicolumn{1}{c}{0.11}   & \multicolumn{1}{c}{7.14}   & 14.51 & 6.70 & 3.57 & 0.0 & 3.13 & 33.15 & 0.45 & 70.65 & 53.91 & 16.52 & 16.29 & 0.0 & 4.13 & 100 \\
                             &                            & CoT                     & \multicolumn{1}{c}{40.98} & \multicolumn{1}{c}{2.15}   & \multicolumn{1}{c}{0.0}   & \multicolumn{1}{c}{22.25}   & 11.60 & 5.93 & 1.98 & 0.0 & 0.95 & 17.53 & 0.09 & 42.44 & 50.26 & 6.53 & 34.54 & 0.0 & 1.89 & 99.05 \\
                             &                            & ToT                     & \multicolumn{1}{c}{59.77}  & \multicolumn{1}{c}{3.22}   & \multicolumn{1}{c}{0.11}   & \multicolumn{1}{c}{4.94}   & 19.20 & 13.45 & 2.41 & 0.34 & 3.56 & 28.51 & 0.92 & 56.90 & 76.21 & 13.33 & 15.06 & 0.11 & 2.64 & 100 \\
                             &                            & GToT                    & \multicolumn{1}{c}{38.23} & \multicolumn{1}{c}{0.58}   & \multicolumn{1}{c}{0.0} & \multicolumn{1}{c}{28.97} & 10.24 & 2.77 & 1.99 & 0.0 & 2.13 & 11.11 & 0.05 & 42.84 & 40.03 & 3.30 & 37.17 & 0.0 & 2.23 & 100 \\ \cline{2-21} 
                             & \multirow{5}{*}{Defects4J} & ZSL                     & \multicolumn{1}{c}{57.01} & \multicolumn{1}{c}{4.05}   & \multicolumn{1}{c}{0.06}   & \multicolumn{1}{c}{15.72} & 24.30 & 7.50 & 1.09 & 0.0 & 0.91 & 26.90 & 0.79 & 48.55 & 60.70 & 9.92 & 19.65 & 0.0 & 1.09 & 99.64 \\
                             &                            & FSL                     & \multicolumn{1}{c}{73.15} & \multicolumn{1}{c}{4.90} & \multicolumn{1}{c}{0.24}   & \multicolumn{1}{c}{13.18}   & 24.68 & 3.94 & 0.88 & 0.0 & 2.09 & 31.19 & 1.61 & 64.95 & 57.64 & 17.36 & 15.19 & 0.0 & 0.96 & 100 \\
                             &                            & CoT                     & \multicolumn{1}{c}{40.06} & \multicolumn{1}{c}{2.83}   & \multicolumn{1}{c}{0.11}   & \multicolumn{1}{c}{34.40}   & 16.79 & 2.99 & 0.61 & 0.0 & 0.66 & 17.73 & 0.61 & 36.45 & 46.81 & 5.71 & 41.72 & 0.0 & 0.66 & 99.28 \\
                             &                            & ToT                     & \multicolumn{1}{c}{69.56}  & \multicolumn{1}{c}{5.52}   & \multicolumn{1}{c}{0.32}   & \multicolumn{1}{c}{4.57}   & 30.60 & 11.20 & 1.03 & 0.0 & 0.79 & 29.81 & 0.79 & 52.92 & 75.95 & 11.20 & 6.23 & 0.0 & 1.0 & 99.97 \\
                             &                            & GToT                    & \multicolumn{1}{c}{29.64} & \multicolumn{1}{c}{2.06}   & \multicolumn{1}{c}{0.0}   & \multicolumn{1}{c}{52.39}   & 12.67 & 2.03 & 0.90 & 0.0 & 1.11 & 7.97 & 0.24 & 28.42 & 32.22 & 1.45 & 55.05 & 0.0 & 1.0 & 99.97 \\ \cline{2-21} 
                             & \multirow{5}{*}{CMD}       & ZSL                     & \multicolumn{1}{c}{37.72}  & \multicolumn{1}{c}{0.0}   & \multicolumn{1}{c}{0.0}   & \multicolumn{1}{c}{33.33}   & 15.79 & 35.96 & 0.0 & 0.0 & 0.0 & 2.63 & 0.0 & 30.70 & 27.19 & 0.0 & 54.39 & 0.0 & 0.0 & 100 \\
                             &                            & FSL                     & \multicolumn{1}{c}{18.45}  & \multicolumn{1}{c}{0.97}   & \multicolumn{1}{c}{0.0}   & \multicolumn{1}{c}{49.51}   & 25.24 & 14.56 & 0.0 & 0.0 & 0.0 & 0.0 & 0.0 & 18.45 & 16.50 & 0.0 & 77.67 & 0.0 & 0.0 & 100 \\
                             &                            & CoT                     & \multicolumn{1}{c}{6.4}  & \multicolumn{1}{c}{0.0}   & \multicolumn{1}{c}{0.0}   & \multicolumn{1}{c}{78.80}   & 7.20 & 2.80 & 0.0 & 0.0 & 0.40 & 0.0 & 0.0 & 6.40 & 7.20 & 0.0 & 89.20 & 0.0 & 0.0 & 100 \\
                             &                            & ToT                     & \multicolumn{1}{c}{15.04}   & \multicolumn{1}{c}{0.0}   & \multicolumn{1}{c}{0.0}   & \multicolumn{1}{c}{71.68}   & 7.96 & 18.58 & 0.0 & 0.0 & 0.0 & 0.88 & 0.0 & 17.70 & 15.04 & 0.88 & 76.99 & 0.0 & 0.0 & 100 \\
                             &                            & GToT                    & \multicolumn{1}{c}{19.86}  & \multicolumn{1}{c}{0.0}   & \multicolumn{1}{c}{0.0}   & \multicolumn{1}{c}{58.90}   & 15.07 & 17.81 & 0.0 & 0.0 & 0.68 & 1.37 & 0.0 & 18.49 & 21.23 & 0.0 & 74.66 & 0.0 & 0.0 & 100 \\ \hline
\multirow{5}{*}{GPT 4}       & \multirow{5}{*}{CMD}       & ZSL                     & \multicolumn{1}{c}{30.43}  & \multicolumn{1}{c}{1.45}   & \multicolumn{1}{c}{0.0}   & \multicolumn{1}{c}{11.59}   & 14.49 & 20.29 & 0.0 & 0.0 & 1.45 & 2.90 & 0.0 & 36.23 & 36.23 & 0.0 & 34.78 & 0.0 & 0.0 & 98.55 \\
                             &                            & FSL                     & \multicolumn{1}{c}{41.18}  & \multicolumn{1}{c}{1.18}   & \multicolumn{1}{c}{0.0}   & \multicolumn{1}{c}{10.59}   & 31.76 & 20.0 & 0.0 & 1.18 & 0.0 & 2.35 & 0.0 & 48.24 & 45.88 & 2.35 & 36.47 & 0.0 & 0.0 & 100 \\
                             &                            & CoT                     & \multicolumn{1}{c}{14.56}  & \multicolumn{1}{c}{0.0}   & \multicolumn{1}{c}{0.0}   & \multicolumn{1}{c}{28.16}   & 18.45 & 17.48 & 0.0 & 0.0 & 0.0 & 0.97 & 0.0 & 18.45 & 35.92 & 0.0 & 54.37 & 0.0 & 0.0 & 98.06 \\
                             &                            & ToT                     & \multicolumn{1}{c}{16.90}   & \multicolumn{1}{c}{0.0}   & \multicolumn{1}{c}{0.0}   & \multicolumn{1}{c}{33.80}   & 16.90 & 15.49 & 0.0 & 0.0 & 0.0 & 0.0 & 0.0 & 23.94 & 22.54 & 0.0 & 53.52 & 0.0 & 0.0 & 97.18 \\
                             &                            & GToT                    & \multicolumn{1}{c}{13.98}  & \multicolumn{1}{c}{0.0}   & \multicolumn{1}{c}{0.0}   & \multicolumn{1}{c}{49.46}   & 17.20 & 19.35 & 0.0 & 0.0 & 1.08 & 0.0 & 0.0 & 16.13 & 21.51 & 0.0 & 69.89 & 1.08 & 0.0 & 97.85 \\ \hline
\multirow{15}{*}{Mistral 7B}      & \multirow{5}{*}{SF110}                      & ZSL                     & \multicolumn{1}{c}{63.45} & \multicolumn{1}{c}{9.66}   & \multicolumn{1}{c}{0.0}   & \multicolumn{1}{c}{0.0}   & 14.71 & 7.98 & 0.42 & 1.26 & 1.26 & 31.51 & 0.84 & 65.13 & 80.25 & 10.08 & 3.36 & 0.42 & 0.42 & 100 \\
                             &                            & FSL                     & \multicolumn{1}{c}{72.69} & \multicolumn{1}{c}{11.89} & \multicolumn{1}{c}{0.0}   & \multicolumn{1}{c}{0.88}   & 7.05 & 0.44 & 0.0 & 1.32 & 6.17 & 27.75 & 0.0 & 68.28 & 69.60 & 15.86 & 5.29 & 0.0 & 0.0 & 100 \\
                             &                            & CoT                     & \multicolumn{1}{c}{52.26} & \multicolumn{1}{c}{7.04}   & \multicolumn{1}{c}{0.0}   & \multicolumn{1}{c}{2.51}   & 19.10 & 2.01 & 0.5 & 1.51 & 1.51 & 22.61 & 0.5 & 53.27 & 84.92 & 11.56 & 5.53 & 0.0 & 0.5 & 94.47 \\
                             &                            & ToT                     & \multicolumn{1}{c}{42.59}  & \multicolumn{1}{c}{2.65}   & \multicolumn{1}{c}{0.0}   & \multicolumn{1}{c}{8.47}   & 12.70 & 2.38 & 0.26 & 0.26 & 1.59 & 10.58 & 0.0 & 43.12 & 42.59 & 8.99 & 11.11 & 0.0 & 0.26 & 92.59 \\
                             &                            & GToT                    & \multicolumn{1}{c}{21.20} & \multicolumn{1}{c}{0.82}   & \multicolumn{1}{c}{0.14} & \multicolumn{1}{c}{13.54} & 6.57 & 0.82 & 0.14 & 0.0 & 0.55 & 3.42 & 0.0 & 27.91 & 16.01 & 2.74 & 21.34 & 0.0 & 0.14 & 85.64 \\ \cline{2-21} 
                             & \multirow{5}{*}{Defects4J} & ZSL                     & \multicolumn{1}{c}{73.64} & \multicolumn{1}{c}{14.34}   & \multicolumn{1}{c}{0.39}   & \multicolumn{1}{c}{0.0} & 17.44 & 1.55 & 0.78 & 0.78 & 1.55 & 13.18 & 1.55 & 48.06 & 58.14 & 32.17 & 2.33 & 0.0 & 0.39 & 100 \\
                             &                            & FSL                     & \multicolumn{1}{c}{79.35} & \multicolumn{1}{c}{14.17} & \multicolumn{1}{c}{0.4}   & \multicolumn{1}{c}{1.62}   & 8.5 & 0.81 & 0.4 & 0.0 & 1.21 & 25.51 & 0.81 & 62.75 & 78.95 & 21.46 & 2.02 & 0.4 & 0.4 & 100 \\
                             &                            & CoT                     & \multicolumn{1}{c}{65.13} & \multicolumn{1}{c}{13.82}   & \multicolumn{1}{c}{0.0}   & \multicolumn{1}{c}{0.66}   & 21.05 & 0.66 & 0.0 & 0.0 & 1.32 & 10.53 & 1.32 & 41.45 & 81.58 & 31.58 & 0.66 & 0.0 & 0.0 & 95.39 \\
                             &                            & ToT                     & \multicolumn{1}{c}{69.84}  & \multicolumn{1}{c}{7.54}   & \multicolumn{1}{c}{0.4}   & \multicolumn{1}{c}{1.59}   & 22.62 & 2.38 & 0.0 & 0.0 & 1.59 & 22.22 & 1.59 & 64.29 & 78.97 & 23.41 & 3.17 & 0.0 & 0.0 & 98.41 \\
                             &                            & GToT                    & \multicolumn{1}{c}{27.97} & \multicolumn{1}{c}{2.89}   & \multicolumn{1}{c}{0.0}   & \multicolumn{1}{c}{12.86}   & 13.67 & 0.0 & 0.0 & 0.0 & 0.16 & 4.82 & 0.16 & 28.94 & 30.55 & 4.18 & 16.08 & 0.0 & 0.0 & 84.41 \\ \cline{2-21} 
                             & \multirow{5}{*}{CMD}       & ZSL                     & \multicolumn{1}{c}{-}  & \multicolumn{1}{c}{-}   & \multicolumn{1}{c}{-}   & \multicolumn{1}{c}{-}   & - & - & - & - & - & - & - & - & - & - & - & - & - & - \\
                             &                            & FSL                     & \multicolumn{1}{c}{0.0}  & \multicolumn{1}{c}{0.0}   & \multicolumn{1}{c}{0.0}   & \multicolumn{1}{c}{0.0}   & 0.0 & 100 & 0.0 & 0.0 & 0.0 & 0.0 & 0.0 & 0.0 & 0.0 & 0.0 & 100 & 0.0 & 0.0 & 100 \\
                             &                            & CoT                     & \multicolumn{1}{c}{-}  & \multicolumn{1}{c}{-}   & \multicolumn{1}{c}{-}   & \multicolumn{1}{c}{-}   & - & - & - & - & - & - & - & - & - & - & - & - & - & - \\
                             &                            & ToT                     & \multicolumn{1}{c}{0.0}   & \multicolumn{1}{c}{0.0}   & \multicolumn{1}{c}{0.0}   & \multicolumn{1}{c}{0.0}   & 100 & 0.0 & 0.0 & 0.0 & 0.0 & 0.0 & 0.0 & 0.0 & 100 & 0.0 & 100 & 0.0 & 0.0 & 100 \\
                             &                            & GToT                    & \multicolumn{1}{c}{16.67}  & \multicolumn{1}{c}{0.0}   & \multicolumn{1}{c}{0.0}   & \multicolumn{1}{c}{33.33}   & 0.0 & 0.0 & 0.0 & 0.0 & 0.0 & 0.0 & 0.0 & 16.67 & 0.0 & 0.0 & 33.33 & 0.0 & 0.0 & 66.67 \\ \hline
\multirow{5}{*}{Mixtral8x7B}       & \multirow{5}{*}{CMD}       & ZSL                     & \multicolumn{1}{c}{75}  & \multicolumn{1}{c}{0.0}   & \multicolumn{1}{c}{0.0}   & \multicolumn{1}{c}{0.0}   & 25.0 & 75.0 & 0.0 & 0.0 & 0.0 & 0.0 & 0.0 & 25.0 & 25.0 & 0.0 & 0.0 & 0.0 & 0.0 & 100 \\
                             &                            & FSL                     & \multicolumn{1}{c}{0.0}  & \multicolumn{1}{c}{0.0}   & \multicolumn{1}{c}{0.0}   & \multicolumn{1}{c}{0.0}   & 0.0 & 0.0 & 0.0 & 0.0 & 0.0 & 0.0 & 0.0 & 33.33 & 0.0 & 0.0 & 33.33 & 0.0 & 0.0 & 66.67 \\
                             &                            & CoT                     & \multicolumn{1}{c}{0.0}  & \multicolumn{1}{c}{0.0}   & \multicolumn{1}{c}{0.0}   & \multicolumn{1}{c}{0.0}   & 50 & 0.0 & 0.0 & 0.0 & 0.0 & 0.0 & 0.0 & 0.0 & 50.0 & 0.0 & 50.0 & 0.0 & 0.0 & 100 \\
                             &                            & ToT                     & \multicolumn{1}{c}{44.44}   & \multicolumn{1}{c}{11.11}   & \multicolumn{1}{c}{0.0}   & \multicolumn{1}{c}{44.44}   & 11.11 & 0.0 & 0.0 & 0.0 & 0.0 & 11.11 & 0.0 & 55.56 & 33.33 & 0.0 & 55.56 & 0.0 & 0.0 & 100 \\
                             &                            & GToT                    & \multicolumn{1}{c}{0.0}  & \multicolumn{1}{c}{0.0}   & \multicolumn{1}{c}{0.0}   & \multicolumn{1}{c}{50.0}   & 50.0 & 50.0 & 0.0 & 0.0 & 0.0 & 0.0 & 0.0 & 0.0 & 50.0 & 0.0 & 100 & 0.0 & 0.0 & 100 \\ \hline
\multirow{2}{*}{EvoSuite}    & \multicolumn{2}{c}{SF110} & 50.78 & 0.0 & 0.0 & 6.90 & 93.10 & 0.0 & 4.44 & 0.0 & 0.0 & 17.53 & 0.0 & 54.67 & 76.94 & 0.07 & 6.90 & 0.0 & 3.96 & 100 \\ 
                             & \multicolumn{2}{c}{Defects4J} & 38.24 & 0.0 & 0.0 & 6.90 & 93.10 & 0.0 & 4.44 & 0.0 & 0.0 & 10.71 & 0.0 & 62.95 & 82.18 & 6.73 & 0.38 & 0.0 & 4.34 & 100 \\
\bottomrule
\end{tabular}}
\begin{tablenotes}
\tiny
\item[1]$^*$ AR: Assertion Roulette, CLT: Conditional Logic Test, CI: Constructor Initialization, EM: Empty Test,  EH: Exception Handling, GF: General Fixture, MG: Mistery Guest, RP: Redundant Print, RA: Redundant Assertion, SE: Sensitive Equality, ST: Sleepy Test, EA: Eager Test, LT: Lazy Test, DA: Duplicate Assert, UT: Unknown Test, IT: Ignored Test, RO: Resource Optimism, MNT: Magic Number Test, ZSL: Zero-shot Learning, FSL: Few-shot Learning, CoT: Chain-of-Thought, ToT: Tree-of-Thoughts, GToT: Guided Tree-of-Thoughts.
\end{tablenotes}
\vspace{3mm}
\end{table*}

\vspace{0.5em}
\noindent\colorbox{gray!20}{{\parbox{0.98\linewidth}{
\textbf{Finding 24:}

Across our benchmarks, LLM-generated tests tend to exhibit simpler structures, showing lower rates of complexity-related smells such as General Fixture (GF: 0.0–35.96\%), Redundant Assertion (RA: 0.0–6.17\%), and Conditional Logic Test (CLT: 0.0–14.34\%) compared to EvoSuite. Conversely, EvoSuite consistently shows higher rates of Exception Handling (EH: 93.10\%), Lazy Test (LT: 76.94–82.18\%), and Eager Test (EA: 54.67–62.95\%) under our configuration, which is consistent with SBST’s emphasis on structural coverage rather than stylistic simplicity.}
}}
\vspace{0.5em}

\noindent\textbf{\underline{Dataset-Level Trends in Test Smell Prevalence.}}

Test smell distribution reveals dataset-specific trends. Defects4J exhibits the highest Assertion Roulette rates (73.15\% GPT-3.5-Turbo/FSL), indicating challenges in generating precise assertions for real-world bugs. Our findings partially align with \citet{siddiq2024using}'s observations, but expand the analysis by comparing SF110, Defects4J, and CMD datasets. SF110 shows elevated General Fixture (93.10\% in EvoSuite) and 100\% Magic Number Test, while CMD demonstrates the lowest Assertion Roulette (6.4\% GPT-3.5-Turbo/CoT). The analysis reveals that LLM-generated tests tend to be stateless and isolated, lacking the inter-method or state-dependent sequences typical of EvoSuite tests. This comprehensive examination illuminates how project context and software evolution influence test smell profiles across different datasets.

\vspace{0.5em}
\noindent\colorbox{gray!20}{{\parbox{0.98\linewidth}{
\textbf{Finding 25:}
LLM-generated CMD tests show limited capturing of shared object states (GF: 0.0–35.96\%, MG: 0.0\%) compared to EvoSuite (4.44\%), revealing challenges in modeling stateful interactions. Defects4J tests display high AR rates (27.97–79.35\%), especially in FSL, underlining difficulties in generating precise assertions for for real-world defects localization.
}}}
\vspace{0.5em}

\noindent\textbf{\underline{Impact of Prompt Engineering on Test Quality.}}

Test smell prevalence varies across LLMs, with CoT and GToT prompts improving quality. GPT-4 (CMD) reduces Assertion Roulette to 14.56\%, and GPT-3.5-Turbo achieves the lowest AR at 6.4\% (CoT on CMD), yet both still show nearly 100\% Magic Number Test. Mistral 7B and Mixtral 8x7B yield more smells, with Mistral 7B reaching 27.97\% AR on Defects4J. EvoSuite, though competitive, exhibits 100\% MNT and notable AR rates (38.24–50.78\%). These findings align with \citet{siddiq2024using}'s report of universally high MNT and variable AR. Our results extend their analysis by evaluating more models and prompts, demonstrating how reasoning-based prompting reduces AR. This prompt-model-smell interplay emphasizes that improvements depend on structured prompting strategies, not just model capabilities. MNT's persistence confirms its status as a core limitation in current test generation methods.

\vspace{0.5em}
\noindent\colorbox{gray!20}{{\parbox{0.98\linewidth}{
\textbf{Finding 26:}
CoT and GToT produce the cleanest tests, notably reducing AR smells (GPT-3 CoT: 6.4–40.98\%, GPT-3 GToT: 19.86–38.23\%, GPT-4 CoT: 14.56\%) compared to FSL (18.45–79.35\%) and EvoSuite (38.24–50.78\%). However, all strategies consistently struggle with Magic Number smells (85.64–100\%), highlighting ongoing challenges with numeric literals.
}}}
\vspace{0.5em}

\paragraph{\textbf{Relation to prior work.}}

\citet{siddiq2024using} present the first systematic analysis of test smells in LLM-generated tests, applying TsDetect on HumanEval and SF110. They report near-universal Magic Number (MNT) and Lazy Test (LT) smells, with Assertion Roulette (AR) and Eager Test (EA) also prevalent, while EvoSuite exhibited higher rates of Exception Handling (EH) and LT. Their study focuses on three models (Codex, GPT-3.5, StarCoder) and two benchmarks, providing an important initial view of how smells manifest in generated tests. Our work extends this line by covering a broader spectrum of LLM families and prompting strategies across three benchmarks (Defects4J, SF110, CMD), and by contrasting them against EvoSuite. We confirm the persistence of MNT, but show that reasoning-based prompting substantially reduces AR while EvoSuite continues to exhibit higher EH, LT, and EA, highlighting complementary strengths and weaknesses between SBST and LLM-based approaches.

\vspace{0.5em}
\noindent\highlight{%
Summary of \textbf{RQ6:}
LLM-generated tests show distinct smell profiles compared to EvoSuite. Magic Number Tests remain pervasive (85–100\% across all settings), but reasoning-based prompts (CoT, GToT) cut Assertion Roulette rates to as low as 6.4\% (vs.\ up to 79.3\% with FSL and 38–51\% with EvoSuite). EvoSuite, in turn, exhibits high Lazy Test (77–82\%) and Exception Handling (93\%) smells, underscoring its coverage-oriented bias. Overall, LLMs yield structurally simpler tests with fewer stateful smells (e.g., GF $\geq36\%$ vs.\ 93\% in EvoSuite), and prompt design emerges as a decisive factor in mitigating assertion-related issues. These results extend \citet{siddiq2024using} by quantifying prompt-driven shifts in smell prevalence across multiple datasets and models.
}

\section{Discussion}
\label{discussion}
This section discusses the potential threats to validity, limitations of our study, and suggestions for future work.

\subsection{Threats to Validity}
\label{threat-to-validity}
This section addresses the threats to the validity of this study.

\noindent\textbf{External Validity.} 

The generalizability of our findings faces several limitations. Though we used three datasets (Defects4J, SF110, and CMD), these may not fully represent real-world software complexity. CMD's small size of just two projects restricts our observations, particularly regarding the unexpected cases where GPT-3.5-Turbo outperformed GPT-4, suggesting model performance varies significantly with dataset characteristics and prompting approaches. Further experimentation on a broader dataset is necessary to confirm our observations.

Despite our efforts to mitigate training data contamination through CMD, LLMs may still have encountered similar test patterns during pretraining, potentially overestimating their true capabilities—a persistent challenge in LLM-based software engineering research. Our evaluation of  four models from OpenAI and MistralAI further limits the breadth of our conclusions. Future work should explore a wider range of models, including open-source LLMs trained specifically for software engineering tasks.

Model knowledge cutoff dates also likely influenced our results, with Mistral 7B's older knowledge base (2021-2022) compared to Mixtral 8x7B's more recent release (December 2023) creating performance disparities. Our findings reflect these models' capabilities during our study period (November 2023–April 2024), and newer versions may demonstrate different results.

\noindent\textbf{Internal Validity.} 

Several internal threats could affect our results' reliability. LLM output randomness introduces variability in generated tests, which we mitigated by executing each prompt 30 times per test suite and performing statistical analyses. Similarly, we ran 30 iterations to address EvoSuite's variability from its evolutionary search process.

Test smell detection methodology presents another challenge. Recent studies~\citep{panichella2022test} highlight concerns about detection tools' reliability, noting high false-positive and false-negative rates. Some smells like Assertion Roulette and Eager Test don't always indicate poor maintainability in LLM-generated tests. To mitigate this, we manually validated a subset of test cases to better assess whether detected smells genuinely impact maintainability, though this doesn't eliminate all possible biases.

Finally, prompting strategy effectiveness represents another concern. While GToT consistently improves test structure and reliability, prompt designs may need further refinement for different models and datasets. Future research should explore model-specific prompt tuning to optimize LLM-generated test quality.

\subsection{Limitations and Future Work}

\subsubsection{Limitations.}
Our study does not assess fault detection. Specifically, we did not implement (i) regression testing on paired buggy and fixed versions (i.e., generating tests on the fixed version and executing them on the buggy counterpart, as in~\citet{shamshiri2015automatically}), (ii) the exception-oriented fault-detection setup used in recent LLM-vs-SBST comparisons (e.g.,~\citet{tang2024chatgpt}), which was introduced to sidestep unreliable automatically generated assertions but inherently captures only faults that manifest as runtime exceptions, or (iii) mutation-based analyses. Extending our evaluation to fault detection requires dependable oracles at scale and datasets instrumented for paired versions or mutation loops, which are outside the scope of the present study and outlined in Future Work.
We also limited our comparison to EvoSuite as a single SBST baseline to maintain a clear experimental scope focused on LLM-generated test suites. Other automated test generation tools such as fuzzers, random testing, or symbolic execution frameworks were not included. We acknowledge the value of these alternatives and consider their integration an important direction for future work.

\subsubsection{Future Work.}

\noindent\textbf{Regression testing on buggy/fixed pairs.} 
We will re-instrument our benchmarks to enable a principled regression protocol (generate tests on the fixed version and execute them on the buggy counterpart) so that fault-revealing capability can be assessed with a controlled oracle, following established practice on Defects4J.

\noindent\textbf{Principled, mutation-guided pipelines.} 
Rather than treating mutation score as a raw end-metric, we will adopt mutation-guided loops in which surviving mutants drive subsequent test improvements (prompt augmentation/repair), with pre-filters ensuring tests are executable and assertive before mutation. This direction is supported by MuTAP(\citet{dakhel2024effective}), which demonstrates that leveraging mutants as feedback can substantially raise fault-revealing effectiveness.

\noindent\textbf{Hybrid SBST + LLM with full-fledged refactoring.} 
We will combine SBST (for structural coverage and input diversity) with LLM-based refactoring beyond identifier naming (e.g., clarifying fixtures and test scaffolding, restructuring helpers, strengthening oracles, reducing brittle mocks or magic values, and improving assertion intent) so that readability and maintainability gains translate into developer effectiveness. \citet{deljouyi2024leveraging} already show that LLM-enhanced SBST tests can improve bug-fix performance (up to +33\% bugs fixed, $-20$\% time) and that semantics-preserving renaming can make generated tests as readable as developer-written tests. Our next step is to generalize these benefits to structural refactoring of test code (layout, fixtures, oracles), not only surface-level names or comments.

\noindent\textbf{Controlled oracles at scale.} 
Finally, we will explore non-LLM-derived oracles (e.g., property- or metamorphic-style checks, or contract-based assertions) to reduce ambiguity in execution outcomes and make exception-only detection unnecessary, addressing a key limitation observed in recent LLM-vs-SBST comparisons.

\noindent\textbf{Extended baselines.}
We also plan to broaden our evaluation by incorporating other automated test generation tools such as fuzzers, random test generators, and symbolic/concolic engines. This will enable a more comprehensive assessment of LLM-based test generation in relation to diverse testing paradigms.

\section{Related work}
\label{relatedwork}

\subsection{LLM-Based Test Generation: Strengths and Limitations}

Recent studies have assessed LLMs for unit test generation along multiple dimensions, not limited to static properties. For instance, \citet{siddiq2024using} evaluate syntactic and compilation correctness, coverage, and the presence of test smells across HumanEval and SF110, showing that while coverage can be competitive in some cases, validity and smell-related issues remain frequent. \citet{yuan2024evaluating} conduct both a quantitative analysis and a user study, reporting that although many LLM-generated tests fail to compile or execute, the passing ones can achieve coverage and readability close to human-written tests; moreover, iterative refinement (ChatTester) improves compilability and assertion correctness. Comparative works have also positioned LLMs against SBST tools: \citet{tang2024chatgpt} examine correctness, readability, coverage, and bug detection versus EvoSuite; \citet{yang2024evaluation} provide a broad empirical evaluation across different LLMs and datasets, concluding that LLMs still underperform EvoSuite in coverage, largely due to validity and dependency resolution issues.

Overall, prior work has already examined runtime behavior (e.g., execution and coverage) and aspects of maintainability (e.g., readability, smells). Our study complements this body of evidence by broadening scale and scope, covering three datasets with class-level generation, multiple LLM families, and five prompting strategies and by structuring failures into a fine-grained taxonomy that highlights hallucination-driven issues. We explicitly avoid exclusive claims and instead position our results as an extension and consolidation across settings and prompts.

\vspace{-3mm}
\subsection{The Impact of Prompt Engineering in Test Generation}

The impact of prompt engineering on LLM-based test generation has been explored in recent research. \citet{yuan2024evaluating} and \citet{chen2024chatunitest} examined the effects of different prompting techniques on test correctness, demonstrating that Few-Shot Learning (FSL) improves correctness but does not fully mitigate hallucinated dependencies. Similarly, \citet{yang2024evaluation} introduced Chain-of-Thought (CoT) prompting, highlighting its ability to enhance logical reasoning in test generation. 

Building upon these insights, our study systematically compares five prompting strategies namely ZSL, FSL, CoT, ToT, and GToT across multiple LLMs and datasets, focusing on structural correctness, compilability, coverage, and maintainability to assess their practical effectiveness in automated unit test generation.

\vspace{-3mm}
\subsection{Readability, Maintainability, and Test Smell Analysis}

Aspects of maintainability have been discussed in prior work, though often in limited scope. \citet{siddiq2024using} quantify the prevalence of smells such as duplicated asserts and empty tests in LLM-generated suites, reporting nontrivial rates particularly on SF110. \citet{yuan2024evaluating} complement this with a user study indicating that passing LLM-generated tests can achieve readability close to human-written ones, while also noting frequent failures and incorrect assertions in the broader population. Beyond Java, \citet{schafer2023empirical} show that LLM-generated JavaScript tests can include meaningful oracles and competitive coverage, though such results depend strongly on runtime and language-specific contexts.

Building on these observations, our study provides a broader and more systematic assessment of maintainability. We analyze test smells using TsDetect, evaluate readability with a specialized deep learning model for JUnit tests \citep{scalabrino2018comprehensive}, and incorporate structural measures such as Match Success Rate (MSR) and Code Extraction Success Rate (CSR). In contrast to prior evaluations, we scale this analysis to three datasets with class-level test generation, multiple LLM families, and five prompting strategies, explicitly examining how reasoning-based prompting modulates maintainability indicators. This comprehensive approach uncovers structural and readability challenges in LLM-generated tests and provides actionable insights for prompt design and post-processing aimed at improving long-term usability.

\vspace{-3mm}
\subsection{Advancing LLM-Based Unit Test Improvement}

Recent studies have explored LLM-driven unit test improvement techniques, primarily focusing on iterative refinement. \citet{gu2024testart} introduced TestArt, a co-evolutionary approach that iteratively refines LLM-generated tests through repair iterations, enhancing correctness over multiple steps. Similarly, \citet{zhang2025citywalk} proposed CITYWALK, which employs iterative refinement guided by contextual cues to improve generated tests, with a focus on increasing functional correctness. \citet{alshahwan2024automated} investigated LLM-driven test improvements at Meta, focusing on refining correctness but without addressing key aspects such as maintainability, readability, or test smells. These studies demonstrate progress in test improvement but remain limited to correctness-focused refinements.

In contrast, our work provides a multi-dimensional assessment that goes beyond correctness-focused refinements. We complement prior improvement techniques by evaluating structure- and maintenance-oriented qualities: extractability (MSR/CSR), syntactic correctness, compilability, coverage, readability, and test smells. Moreover, we introduce readability and maintainability metrics to assess whether generated tests are understandable and practically usable in real-world software projects.

Our findings indicate that while iterative refinement can enhance the syntactic and structural quality of LLM-generated tests, it does not fully address maintainability and structural robustness: compilability remains fragile and test smells persist across datasets and prompting strategies. This suggests the need for hybrid approaches that combine Search-Based Software Testing (SBST) with LLM-based generation to jointly target correctness and maintainability.

\vspace{-3mm}
\subsection{Key Takeaways and Research Gap}

Prior works on LLM-based test generation primarily focus on syntactic correctness and code coverage, leaving key aspects of readability, maintainability, and structural reliability underexplored. Our study fills this gap by providing a comprehensive evaluation framework that goes beyond static correctness checks to assess multiple dimensions of test quality.  Specifically, we evaluate test readability and maintainability using deep learning-based readability scoring, aligned with human evaluation guidelines.

Unlike previous studies that explore only one or two prompting techniques, we systematically compare five strategies—ZSL, FSL, CoT, ToT, and GToT. We also integrate mutation testing, a missing component in prior research, to evaluate fault detection capabilities. Beyond correctness and structural coverage (line, instruction, method), we analyze test smells, code extraction reliability (MSR \& CSR), compilation errors, and runtime failures, uncovering key challenges in dependency resolution and API usage.
Unlike previous studies that explore only one or two prompting techniques, we systematically compare five strategies(ZSL, FSL, CoT, ToT, and GToT) across four LLMs and three datasets. Beyond structural correctness and coverage (line, instruction, and method), we analyze code extractability (MSR \& CSR), compilation success, and test smells to uncover recurring issues in dependency resolution and API usage.

A key differentiator of our work is the diversity of datasets: SF110 (academic), Defects4J (real-world bugs), and CMD (industrial codebase). This design ensures that our findings generalize across heterogeneous software environments. By bridging structural reliability, readability, maintainability, and dataset diversity, our study establishes a broad, data-driven foundation for understanding the practical effectiveness and current limitations of LLM-generated test suites.

\section{Discussion of Findings and Implications}
\label{discussion-of-findings-and-implications}

Our study reveals that reasoning-based prompting (CoT, GToT) improves the structural correctness and maintainability of LLM-generated tests, yet reliability and integration robustness remain major challenges. While many tests compile successfully, they often exhibit dependency mismatches or redundant logic, and test smells continue to affect maintainability. Hybrid strategies combining Search-Based Software Testing (SBST) with LLM-driven generation and targeted human oversight appear essential for achieving higher robustness. Although prompt engineering enhances structural reliability, hallucination-induced errors persist, highlighting the need for stronger contextual grounding and improved handling of dependencies and encapsulation. These findings underscore the importance of adaptive test generation pipelines that integrate automated validation, dependency resolution, and iterative refinement to improve the practical usability of LLM-generated unit tests.

\vspace{-3mm}
\subsection{Implications for Researchers} 

Our findings emphasize reasoning-based prompting's significant role in improving LLM-generated test suites. Unlike prior studies~\citep{siddiq2024using, tang2024chatgpt, yang2024evaluation}, which did not extensively analyze prompt engineering, we systematically evaluate ZSL, FSL, CoT, ToT, and GToT. Results show that structured reasoning significantly enhances test adherence, compilability, and structural reliability, with GToT consistently outperforming other strategies.

Our multi-level coverage analysis reveals that LLM-generated tests achieve moderate coverage but remain inferior to EvoSuite, reinforcing the need for hybrid approaches that integrate SBST techniques with LLM-based generation. Our findings also challenge the common assumption that larger models inherently produce better tests: GPT-3.5-turbo occasionally outperforms GPT-4 on CMD, highlighting the need to better understand how architectural and dataset characteristics influence performance.

Beyond correctness and coverage, our readability and maintainability analysis reveals that while LLMs produce syntactically correct and readable tests, readability alone does not guarantee maintainability. Test smells—particularly Magic Number Tests and Assertion Roulette—persist across models and prompting strategies, increasing debugging complexity. These results highlight the need for more refined prompting strategies and automated post-processing techniques to reduce test smells and improve long-term test reliability.

\vspace{-3mm}
\subsection{Implications for Developers}

Our study highlights the practical value of LLM-generated test suites when readability and maintainability are prioritized over raw coverage. Unlike EvoSuite, which maximizes coverage at the expense of readability, LLM-generated tests generally exhibit clearer structure, more meaningful naming, and higher comprehensibility—particularly when reasoning-based prompting techniques such as GToT and CoT are applied. This makes these tests more accessible for long-term maintenance and collaborative development.

However, developers must remain aware of reliability concerns. While LLMs can generate large numbers of tests rapidly, issues such as partial compilability, redundant assertions, or unverified logic can still arise. 
Persistent test smells—including redundant or fragile assertions, flaky behavior, and inconsistent naming—emphasize the importance of human oversight. LLMs thus serve best as \quotes{assistive tools} rather than fully autonomous test generators. 
Careful prompt refinement, automated filtering, and lightweight post-processing validation can maximize their practical utility, ensuring that generated tests meaningfully contribute to software quality and long-term maintainability.

\section{Conclusion}
\label{conclusion}

This study presents a large-scale evaluation of Large Language Models (LLMs) for automated unit test generation, systematically analyzing how structured prompting strategies influence test quality, readability, and maintainability across multiple datasets. 
By comparing five prompting techniques (ZSL, FSL, CoT, ToT, and GToT) and four LLMs, our results show that reasoning-based prompting especially GToT and CoT substantially improves test structure, compilability, and readability compared to zero-shot prompting.
However, LLM-generated tests remain limited in robustness and reliability. Although they exhibit cleaner code and higher readability than traditional SBST-generated tests, persistent issues such as incomplete imports, redundant assertions, and recurring test smells (e.g., Magic Number Tests, Assertion Roulette) reduce their long-term maintainability. Coverage analysis further reveals that while LLMs achieve moderate coverage, they still underperform EvoSuite in exploring program structure.
Overall, our findings indicate that LLMs can indeed assist in automated test generation, but they are not yet ready to replace traditional testing tools. Their greatest strength lies in producing human-readable, maintainable test scaffolds that can complement existing SBST or developer-written tests. To unlock their full potential, future research should focus on hybrid methodologies that integrate LLM-based generation with automated validation, dependency resolution, and search-based refinement. Such approaches could enable LLMs to evolve from assistive code generators into reliable components of software quality assurance pipelines.

\section*{Declarations}

\subsection*{Funding}

This research was funded in whole, or in part, by the Luxembourg National Research Fund (FNR), grant reference AFR PhD bilateral, project reference 17185670. This work was also supported by the European Research Council (ERC) under the European Union’s Horizon 2020 research and innovation program (grant agreement No. 949014). For the purpose of open access, and in fulfilment of the obligations arising from the grant agreement, the author has applied a Creative Commons Attribution 4.0 International (CC BY 4.0) license to any Author Accepted Manuscript version arising from this submission.

\subsection*{Conflict of Interest}
The authors declare that they have no conflict of interest.

\subsection*{Ethical Approval}
This article does not contain any studies with human participants or animals performed by any of the authors.

\subsection*{Informed Consent}
Not applicable.

\subsection*{Author Contributions}
All authors contributed equally to this work. They jointly conceived the study, conducted the experiments, analyzed the results, and wrote the manuscript. All authors reviewed and approved the final version of the manuscript.

\subsection*{Data Availability Statement}
The datasets and code used in the present study are available in our repository: \href{https://anonymous.4open.science/r/LLM4TS-0F76/}{https://anonymous.4open.science/r/LLM4TS-0F76/}

\subsection*{Clinical Trial Number}
Not applicable.

\bibliographystyle{spbasic.bst}       
\bibliography{references}

\end{document}